**Coping with the delineation of emerging fields: Nanoscience and Nanotechnology as a case study.**

Teresa Muñoz-Écija, Benjamín Vargas-Quesada, Zaida Chinchilla Rodríguez[*]

**Abstract**

Proper field delineation plays an important role in scientometric studies, although it is a tough task. Based on an emerging and interdisciplinary field —nanoscience and nanotechnology— this paper highlights the problem of field delineation. First we review the related literature. Then, three different approaches to delineate a field of knowledge were applied at three different levels of aggregation: subject category, publication level, and journal level. Expert opinion interviews served to assess the data, and precision and recall of each approach were calculated for comparison. Our findings confirm that field delineation is a complicated issue at both the quantitative and the qualitative level, even when experts validate results.

**Keywords:** Field delineation; Science classification; Research topics; Bibliometrics; Scientometrics

**Highlights:**

This study offers an updated literature review of NST as a delineated field. We provide an extended and unified NST search strategy suitable for two databases, Scopus and Web of Science. After formulating a conceptual framework as to how to employ different approaches described in the literature, we tried to delineate the NST field at the journal level through a seven-step procedure. Three approaches —at subject category, publication and journal levels— were adopted to explore and analyze these two databases. The results are compared, and we offer a detailed explanation of the topics related to the journals included in each level. At the publication level, we compare the potential of the micro-field classification system developed by Waltman and van Eck (2012) with the other two approaches. Finally, we examine our findings in the light of NST expert opinions to assess the reliability of the results. The findings of this survey confirm certain problems inherent to field delineation at the quantitative and qualitative levels, especially when dealing with interdisciplinary fields (Huang, Notten, & Rasters, 2011).

[*] Corresponding author at CSIC, Institute of Public Goods and Policies (CSIC-IPP), Albasanz 26-28, 28037 Madrid, Spain. E-mail address: zaida.chinchilla@csic.es (Zaida Chinchilla-Rodríguez)

## 1. Introduction

Field delineation[†], which consists of assigning journals or publications to fields, tends to be the first stage undertaken in bibliometric or scientometric studies of fields (Zhao, 2009). Proper field delineation is important for researchers who study the structure and dynamics of a field, and it is fundamental for information retrieval in the research areas of bibliometrics and scientometrics. It is also needed when designing reliable and solid tools for rankings and domain analysis, highly valuable in science policy design and science evaluation processes (Gómez-Núñez et al., 2014). Indeed, the relevance of such studies will depend on the quality of their delineation. Specific challenges lie in the process of delineating emerging and/or interdisciplinary fields, for example, in the retrieval of relevant data with high precision and high (hefty) recall, and the construction of a core journal or publication set to comprise a field (Milanez, Noyons, & de Faria, 2016).

Several methods may be used to delineate fields. Subject-classification schemes, made up mainly of scientific journals, are one information retrieval strategy that can help to collect a relevant document set for field delineation. This is the classification system used by the top two multidisciplinary bibliographical databases, Scopus and Web of Science (WoS). They assign scientific journals to categories taking into account journal title and scope, as well as journal citation analysis. Although subject-classification schemes may define a broad field, this type of delineation encounters limitations regarding the level of detail, multidisciplinary journals, and interdisciplinary research published in a variety of journals, defying disciplinary boundaries (Zhao, 2009); in such cases, the retrieved document set contains considerable noise affecting both precision and recall (Glänzel, 2015). In order to offset these limitations, an alternative method may be used to delineate by assigning not only journals but also publications to categories (Glänzel & Schubert, 2003).

The most common information retrieval archetype used for field delineation entails the translation of an information need into a query or a set of queries. The application of complex search strategies based on lexical terms (keywords and phrases) is most widely used to collect a document set related to a field/domain, especially in emerging and interdisciplinary fields, because it can effectively cover most aspects of the field (Glänzel, 2015). This method has been progressively fine-tuned, and can now be considered a hybrid lexical-citation method. At the seminal level, it may be used to appraise a field of literature, or core journal publications, through retrieval by lexical queries. The initial seed set is then extended by citation analysis among publications using different parameters. However, this process becomes problematic when, for instance, the use of common terms bears a negative influence on precision yet a positive one on recall. Total recall can also be affected because the borders of fields are sometimes difficult to define; and the help of experts in the field to design lexical queries can produce a specialization bias, leading to a skewed and superficial overview of the field/domain (Zitt, 2015).

The delineation of scientific fields previous to developing disciplinary subject classification schemes may involve empirical and pragmatic techniques, or else automated procedures based on statistics and computerized methods. Among the latter, clustering analysis is a valuable and popular method applicable at journal and publication level, in a wide variety of scientific fields (Zitt & Bassecoulard, 2006; Bassecoulard, Lelu, & Zitt, 2007, Gómez-Núñez et al., 2014). Boyack and Klavans (2010, 2013) use hybrid methods that combine different types of citation relations and conclude that the approach generates accurate clusters.

The use of multidisciplinary classification systems at the publication level, as established by Waltman & van Eck (2012) is based on direct citation relations between publications from WoS, used to cluster them into research areas organized in a three-level hierarchical structure. Delineating at the publication level more accurately matches the current structure of scientific research in a field. Accordingly, the classification system of publications could help define a journal level system, but it might prove problematic the other way around. Field delineation at the level of publication is one possible solution when facing the limitations involved in classification at the level of journal.

---

[†] The terms field delineation, subject classification, subject category delineation, subject delineation, thematic delineation, and discipline delineation are used interchangeably throughout this text.

Nowadays, Nanoscience and Nanotechnology (NST) is an area holding vast technological and social potential for the community, presenting advancements for industry, health, the environment, and security. It therefore attracts great policy interest. Thus, NST has been included as a strategic area with an innovative and economic potential in many research and development plans, even worldwide, like the EU Research and Innovation Programme known as Horizon 2020‡, the National Science Foundation§ and the National Nanotechnology Initiative**. Previous efforts have delineated NST from different perspectives (Zitt & Bassecoulard, 2006; Leydesdorff & Zhou, 2007; Mogoutov & Kahane, 2007; Grieneisen & Zhang, 2011; Arora et al. 2013, 2014), but to date no comparison of these methods has been performed. The current study compares three different approaches to delineate the emerging field of NST, shedding new light on the key problems that underlie emerging field delineation.

*1.1. Delineation of Nanoscience & Nanotechnology (NST)*

Nanoscience & Nanotechnology (NST) constitutes an interdisciplinary and emerging domain that embraces physics, chemistry, materials science, engineering, and more. Since the 1990s, many works have attempted to delineate NST by using different perspectives, e.g. thematic delineation or bibliometric techniques.

Various lexical queries have been proposed for gathering the data set to study NST, including:

1) The prefix *nano*, to harvest all publications that include it within their title (Tolles, 2001; Meyer, Persson & Power, 2001; Dunn and Whatmore, 2002);
2) A combination of *nano* prefix and representative keywords of NST, excluding those terms that include the *nano* prefix but have nothing to do with NST (Glänzel et al. 2003; Noyons et al., 2003; Mogoutov & Kahane, 2007; Porter et al., 2008).

The drawback of searching based on the nano* prefix alone is that publications without a prefix are not retrieved (for instance, biotechnology publications). In addition, some NST keywords such as fullerenes or graphene are excluded. The second search strategy is supported by experts' opinion to include or exclude keywords, yet it tends toward bias, because the experts may include keywords associated with their particular field of knowledge (Huang et al., 2011).

An early work proposing a lexical query to study NST is that by Braun, Schubert & Zsindely (1997). These authors harvested articles with the prefix *nano* from 1986 to 1995 in order to analyze the growth of nanotechnology. Some years later, Kostoff and colleagues (2006a) studied NST to identify its thematic structure. They recreated the genealogy of NST in terms of the most important works behind the development of NST based on the assumption that they would be highly cited. Zitt & Bassecoulard (2006) delineate NST literature by combining lexical queries and citation analysis, additionally using a set of parameters to highlight relevance and reduce noise. A worldwide view of the technical structure and infrastructure of scientific production in NST drawn up by Kostoff, Koytcheff and Lau (2007) introduced us to the most productive and highly impacting countries.

---

One important finding two decades ago is that journals focusing on physics published the greatest number of NST publications, followed by interdisciplinary journals (Meyer & Persson, 1998). The search involved articles with the prefix *nano* in their titles between 1991 and 1996 retrieved from the Science Citation Index, distributing them according to the journal classification scheme proposed by Hicks & Katz (1996).

Schummer (2004) ran a co-author analysis of over 600 articles published in *nano* journals in 2002 and 2003. The results showed that most NST journals include publications by authors with just one disciplinary affiliation, especially by physicists. He furthermore explored the speed at which the scientists and engineers from different disciplines, institutions and countries take part in NST research, and whether the development of NST involves new means or levels of multi and interdisciplinary research, or institutional and geographical collaboration patterns. Results indicated that there are no particular patterns or stable levels of multi- or interdisciplinary research —while interaction exists, it shares little more than the prefix *nano*.

Bassecoulard, Lelu, & Zitt (2007) looked into NST as a multidisciplinary field of knowledge by means of a classification of publications in thematic clusters connected by the similarity of their references. With an NST data set from 1999 to 2003, they evidenced a moderate level of multidisciplinary, though just a few publications in physics and chemistry were at the root of that multidisciplinary.

The NST journals with the *nano* prefix in their titles included in the Journal Citation Report (JCR) from 25 countries were studied by Andrievski and Klyuchavera (2011). New journals in NST were found to come from fields such as nanobiology, nanomedicine, nanobiomedicine, and nanobiotechnology, whereas the bulk of scientific production in NST was published in classic natural-science and engineering journals.

Huang and his collaborators (2011) conducted a review of the nanotechnology literature that had analyzed publications and patents. They identified pros and cons of the different search strategies carried out for nanotechnology delineation: keyword queries and all their improvements, citation analyses, and the use of a core journal set to identify articles. As most of the search strategies for nanotechnology delineation shared a core keyword set, the results obtained in terms of main areas and journals in nanotechnology proved similar. Their lexical queries were designed in view of the search strategy suggested by Glänzel et al. (2003). The other search strategy was based on the use of the top 10 journals in nanotechnology; however, it did not provide a solid delineation because many nanotechnology papers are published in multidisciplinary journals.

A number of attempts to improve keyword queries for the retrieval of NST publications (Zucker et al. 2007, Kostoff et al., 2006a; Kostoff, Koytcheff y Lau, 2006b ; Mogoutov & Kahane, 2007; Maghrebi et al., 2011) rely on a data set harvested using an initial keyword search strategy and a selection of keywords based on the relevance of those terms.

On the one hand, Maghrebi et al. (2011) used keyword queries proposed by Warris (2004) to study precision and recall of the keywords by analyzing the articles of 2008 in WoS. On the other hand, the combination of keyword queries and citation analysis used to delineate NST can be seen as a highly effective approach, since it increases precision and recall, reducing the noise of information retrieval (Zitt & Bassecoulard, 2006). Although keyword queries are still used, subjectivity is reduced because expert consultation is not necessarily required (Huang et al., 2011).

Another search strategy described in the literature on NST delineation involves the use of core journals: Leydesdorff & Zhou (2007) identified a set of 10 relevant journals in NST using betweenness centrality as a measure of the interdisciplinary of scientific journals, with visualization techniques based on citation analysis of journal titles. In addition, they analyzed Nanotechnology patents from the U.S. Patent and Trade Office (USPTO) to determine whether the patents contain references to non-patent literature. Such references were found to be too general to define a set of key journals, and NST appeared to arise from the interrelation between Physics and Chemistry. The drawback of this method, as opposed to retrieval based on keyword queries and citation analyses, is that only a small portion of the NST literature (thematic corpus) is covered.

The present contribution stems from a search strategy combining several approaches previously described for the recovery of NST documents, launched in the Scopus database. Our objective was to derive an NST journal classification in the SCImago Journal and Country Rank (SJR) and in the SCImago Institutions Rankings (SIR) portals based on the Scopus data. All NST articles, conference papers, and reviews contained in Scopus were recovered for the year 2010. Subsequently, a citation analysis of the most cited journals was carried out based on the references of the analyzed documents. The journals that reflected over 1% of total citation were selected (A), excluding multidisciplinary journals. In order to identify the journals that contained the prefix *nano* in the title covered in Scopus, another search was launched in the database, taking only those journals with a *nano* prefix that had received at least one citation in 2010 (B). The sum of journal sets A and B determined the NST core journals (Muñoz-Ecija et al., 2013).

## 2. Objectives

The overall motivation behind this study was to optimize and update the NST journal classification in the framework of the SCImago Journal & CountryRank (SJR) portal. As there is no unique and infallible means of field delineation, our main aim was to highlight certain issues involved in delineating emerging fields, using NST as a study case. We explore and compare three approaches to show how field delineation varies under the different methods, and aspire to the following specific objectives:

- To update the literature review of field delineation in NST;
- To unify search strategies when harvesting the NST collection;
- To set up a framework for combining approaches at the journal level through a seven-step procedure;
- To use two distinct value data sources, Scopus and WoS, to delineate an emerging field;
- To include validation by experts in the NST field.

## 3. Material and Method

Two different databases were used to harvest the data set: Scopus and WoS Core Collection (SCI-Expanded, SSCI, and A&HCI indexes). These two databases are widely held to be the most formal data sources in terms of publications because they cover a huge journal collection, and they store publication activities including various reliable indicators (Park, Yoon, & Leydesdorff, 2016). Moreover, two separate classification journal systems were employed to obtain journal lists: Journal Citation Reports (JCR) and SCImago Journal & Country Rank (SJR) in their 2016 versions.

Likewise, two different periods of time were used: for Scopus, 2008-2015, and 2000-2016 for WoS. The first time period, 2008-2015, was chosen because it marked a certain stability in NST scientific output (Muñoz-Écija, Vargas-Quesada, & Chinchilla-Rodríguez, 2017). For the second period (2000-2016) we used the in-house version of WoS kept at the Centre for Science and Technology Studies (CWTS) of Leiden University; it incorporates the publication-level science classification system developed by Waltman and Van Eck (2012), which comprises more than 4,000 micro-level fields.

In analyzing these data sets, three alternative approaches were applied: 1) at category level ($A_1$); 2) at publication level ($A_2$); and 3) at journal level ($A_3$).

### 3.1. Approach 1: Delineation at category level ($A_1$)

To date, WoS is the only multidisciplinary database that includes Nanoscience & Nanotechnology as a subject category. Assignation of journals to subject categories in WoS is the result of a heuristic method based on citation data, long regarded as perhaps *"the best way"* to delineate fields in journal terms (Pudovkin and Garfield, 2002). Under this approach, 88 journals contained in the WoS Nanoscience and Nanotechnology subject category were selected, excluding ceased journals ([Appendix 1a](#)). Because Scopus does not include the subject category Nanoscience and Nanotechnology among their subject areas, field delineation for NST was drawn in the SCImago Journal and Country Rank portal using subject classification in conjunction with citation- and query-based approaches (Muñoz-Écija et al. 2013). A total of 87 journals contained in SJR in the year 2016 were selected for used in approach $A_3$, excluding ceased/discontinued journals (Appendix 1b).

### 3.2 Approach2: Delineation at publication level ($A_2$)

This second approach was applied for field delineation in terms of publications. The search strategy relied on the in-house version of WoS from the CWTS of Leiden University ([Appendix 2](#)) to obtain an NST data set. The strategy evolves from one used earlier to construct the NST category in SJR and SIR portals (Muñoz-Écija et al., 2013), based on lexical queries combining the *nano* prefix and a list of keywords related to *nano* by means of Boolean operators. All the keywords queries were designed taking into account lexical queries reported elsewhere (Kostoff, Koytcheff, & Lau, 2006b; Grieneisen & Zhang, 2011; Maghrebi et al., 2011; Arora et al., 2013). For the period 2000 to 2016, the number of articles and reviews retrieved amounted to 1,055,801.

This total of 1,005,801 publications were spread out into 3,433 micro-level fields. The percentage of overlapped publications was calculated by dividing the number of matched publications per micro-level fields by the total number of publications in each micro-level field. We believe that the occurrence of publication in relevant micro-fields that come to represent a field can be regarded as a publication overlap having a minimum frequency of 60% in each cluster. Consequently, only micro-level fields with a threshold equal to or greater than 0.6 % of overlapped publication were selected. Afterwards, the number of publications per journal covering the 35-selected micro-cluster were calculated (3,851 journals and 332,739 publications in total), as was the total number of publications per journal in the WoS collection for the same time period. The percentage of publications per journal is thus approximated by dividing the two values. Journals over the 0.2 % threshold of publication per journal and over 50 publications were extracted. A list of 78 journals was made, but two turned out to be the same journal that had changed its name. Also, one "ceased" journal was identified using this approach. The final number of journals was therefore 76.

### 3.3. Approach 3: Delineation at journal level ($A_3$) / Lexical query

This approach builds on a previous work (Muñoz-Écija et al., 2013), but improves upon the design for an optimal lexical query to harvest the NST data set and implements a statistical method to estimate unknown parameters.

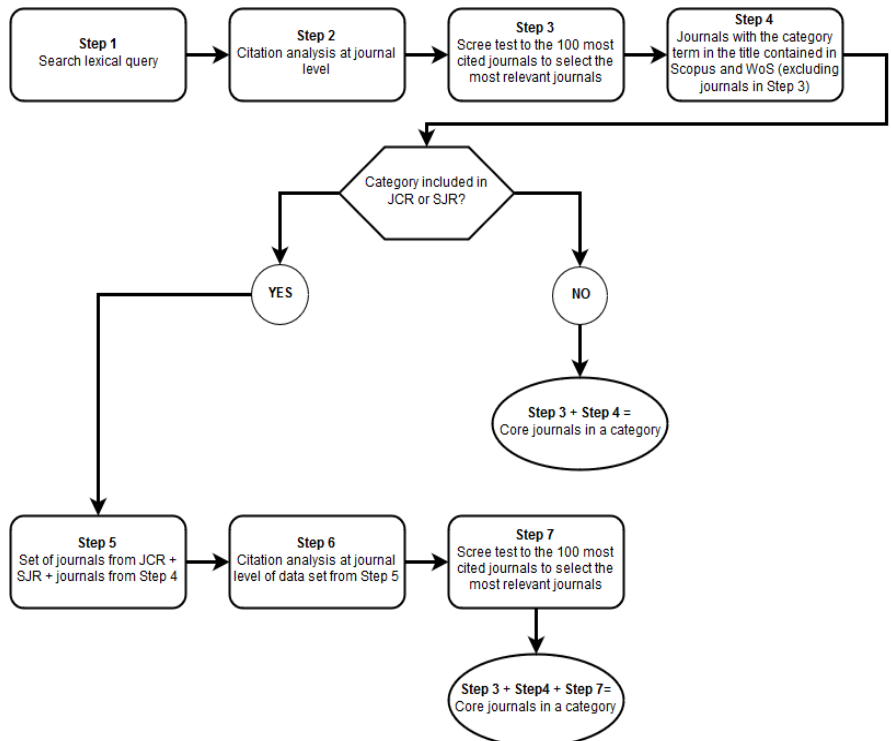

*Figure 1. Steps in field delineation at the journal level*

### Step 1. Searching lexical query

Figure 1 shows a sequence of steps at the journal level. It starts with retrieving the data set from Scopus and WoS databases by applying the NST search strategy (Appendix 2). All document types published from 2008 to 2015 were considered for both databases (Table 1).

*Table 1. Data retrieved by Approach 3 ($A_3$)*

| Database | Period | Document type | Number of publications |
|---|---|---|---|
| Scopus | 2008-2015 | All | 902,082 |
| WoS | 2008-2015 | All | 711,464 |

### Step 2. Citation analysis

We determine the set of cited journals in NST publications retrieved in each database. Before that, the multidisciplinary cited journals included in the Multidisciplinary subject category of Journal Citation Report (JCR) were removed, along with those in the General subject category of SCImago Journal Rank (SJR). This step was based on the assumption that multidisciplinary journals publish articles of many different fields and cannot be considered as journals pertaining to a specific field of knowledge (Narin, 1976). The multidisciplinarity of journals might also be studied as an indicator in the *citing* dimension rather than in the *cited* one (Leydesdorff, 2007).

### Step 3. Scree test

After calculating the percentage of citation for each cited journal, the statistical method known as the scree test (Cattell, 1966) was used to choose the most representative journals. We selected the 100 most cited journals according to the two data sets. Given that the variable "percentage of citation" is not linear, this test allows one to identify the breaking point between a steep slope and the leveling off in terms of citation analysis. Journals with a value equal to or above the breaking point in the citation analysis graph are likely to be *nano* core journals. The scree test should be combined with Ordinary Least-Square regression (OLS) to derive the most predictive breaking point, minimizing the sum of squared errors (Figure 2 and 3). In our analysis, the breaking point served to set off 13 journals from the NST Scopus data set and 15 journals from the NTS WoS data set.

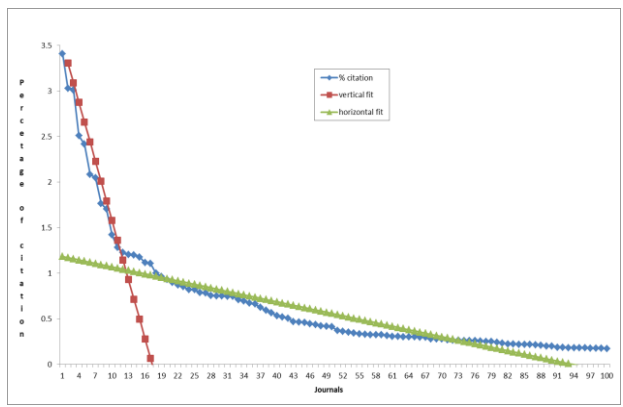

*Figure 2. Scree test of 100 most-cited journals from NTS Scopus data set. Red line minimizes the error between the observed and predicted value of axis y (Percentage of citation); green line minimizes the error of axis x (100 most cited journal)*

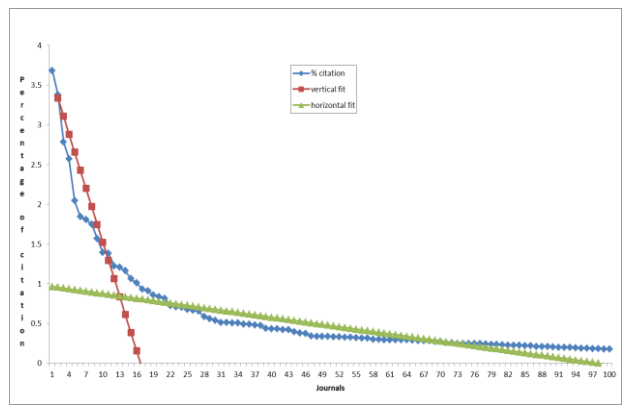

*Figure 3. Scree test of 100 most-cited journals from NST WoS data set. Red line minimizes the error between the observed and predicted value of axis y (Percentage of citation); green line minimizes the error of axis x (100 most cited journal)*

### *Step 4. Nano-prefix journals*
In the fourth step, we search into Scopus and WoS databases by source/publication field using the nano* prefix for term search. All retrieved items were checked so as to choose only journals that were not ceased. By selecting only in-press journals/active journal titles in both databases (Appendix 3), we came to a total number of 91 journals. To avoid duplicates, retrieved journals with *nano* prefix in their titles that had been previously identified by scree test were removed from this group (3 in Scopus and 1 in WoS). A total of 88 journals were thereby selected for Scopus, and 90 for WoS.

### *Step 5. Journal compilation from JCR and SJR plus Nano-prefix journals*
We selected all journals contained in the Nanoscience & Nanotechnology category under JCR and SJR versions 2016 (87 and 90 journals, respectively). After adding the journals with *nano* prefix retrieved in step 4 to detect new journals (not identified in the previous steps), we went on to denote this group of journals as NST A Scopus and NST B WoS. In order to avoid data duplication, the journals previously selected by employing a scree test (step 3) were deleted from the datasets, giving 123 journals for NST A, and also 123 for NST B.

### *Step 6. Citation analysis (from step 5)*
Publications from NST A and B were retrieved, and a separate citation analysis was accomplished. As in step 2, we determined which journals had been cited, the number of received citations, and the percentage of citation per journal.

### *Step 7. Scree test (from step 6)*
Like in Step 3, the Scree test method is used to choose the breaking point of the variable "percentage of citation", taking into account the 100 most cited journals from NST A (Figure 4) and NST B (Figure 5). The optimal breakpoint predicted, in view of both data sets, is 17. All journals placed beyond point 17, including point 17, are therefore considered the core nano-journals from NST A and B. Only seven previously unidentified journals from Scopus were detected, and three from WoS.

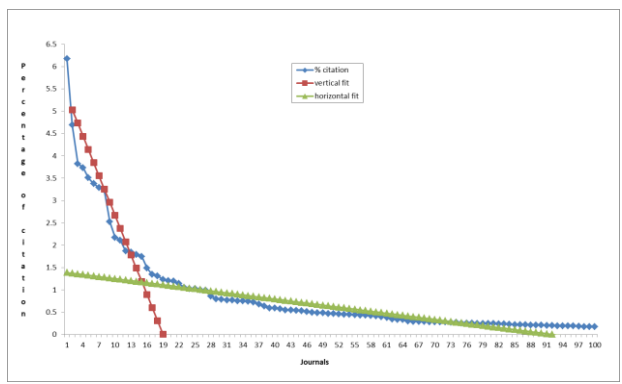 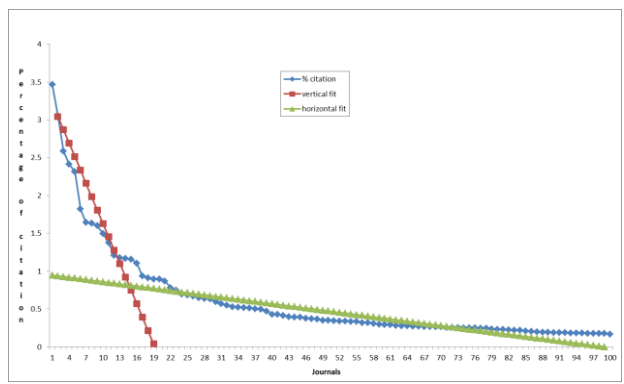

*Figure 4. Scree test 100 most-cited journals from NST data set A. Red line minimizes the error between the observed and predicted value of axis y (Percentage of citation); green line minimizes the error of axis x (100 mos- cited journal)*

*Figure 5. Scree test 100 most-cited journals from NST data set B. Red line minimizes the error between the observed and predicted value of axis y (Percentage of citation); green line minimizes the error of axis x (100 most cited journal)*

In the final step, we sum up all those journals identified throughout steps 3, 4 and 7 for each database: respectively, 108 and 110 journals[††] for Scopus and WoS. These two core journal lists do not show significant differences. Three journals were identified by the Scopus data set but not by WoS, and five journals were identified by WoS data yet not by Scopus. Hence, we consider the aggregation of both datasets —that is, 113 journals— to be the nano-core journals that could delineate the field of NST.

*Precision and recall of the three approaches*

3.4. Precision
To compare the three approaches, we studied the precision of each at the journal level and at the publication level, following the study of Maghrebi et al. (2011). In this study, the precision is defined taking into account the definition from the National Nanotechnology Initiative of the United States (NNI): "Nanotechnology is science, engineering, and technology conducted at the nanoscale, which is about 1 to 100 nanometers. Nanoscience and nanotechnology are the study and application of extremely small things and can be used across all the other science fields, such as chemistry, biology, physics, materials science, and engineering."[‡‡]

*3.4.1. Precision by scope*

In light of the "Aim and scope" and "Overview" sections, we classified journals as high, medium, or low precision based on how well they match the definition. Ultimately, we drew up the following guidelines for denoting and interpreting the results at this level:

- Value 1 = High precision. Journals that cover only NST topics
- Value 0.5 = Medium precision. Journals that cover NST topics and other topics
- Value 0 = Low precision. Journals not covering NST topics

---

[††] At first 108 were selected, but one journal had split into three different titles, for which reason all three titles were selected, making the final number of journals 110.
[‡‡] http://www.nano.gov/nanotech-101/what/definition

*3.4.2. Precision by interviews*

We sent an online questionnaire managed by Qualtrics software to NST experts in Portugal, Spain, and The Netherlands. The experts (only Ph.D. researchers) were classified through two filters: their general background field and their particular background in NST (subfield). The questionnaire consisted of a random sample of 50 articles and reviews under different approaches and from the initial seed data: specifically, 15 articles and reviews from each approach, plus five from the seminal data (Appendix 6). Experts in NST had to indicate whether an item was relevant or not, taking into account the NNI definition. This closed question survey included a remark section per item, in case one wished to elaborate or comment on their answers. Lastly, we denoted as relevant 1 and as irrelevant 0 for the purposes of calculating precision.

3.5. Recall

Recall, at the journal level, is derived from the assumption that the total number of journals identified by the three approaches denotes total recall, meaning total recall equals 163. In addition, we measured recall by counting journals with a high or medium level of precision. At the publication level, recall is calculated assuming that all articles and reviews identified by the three approaches should represent total recall. Thus, we considered the random sample as total recall, and considered relevant publications to be all those validated by experts whose precision was equal to or greater than 0.5.

**4. Results**

*4.1. Comparing approaches*

*4.1.1 Comparing approaches at the journal level* ($A_1 \cap A_2 \cap A_3$)
The total number of journals retrieved was 163. [Figure 6](#) provides the number of journals obtained under the three approaches: 44 journals were matched by the three.

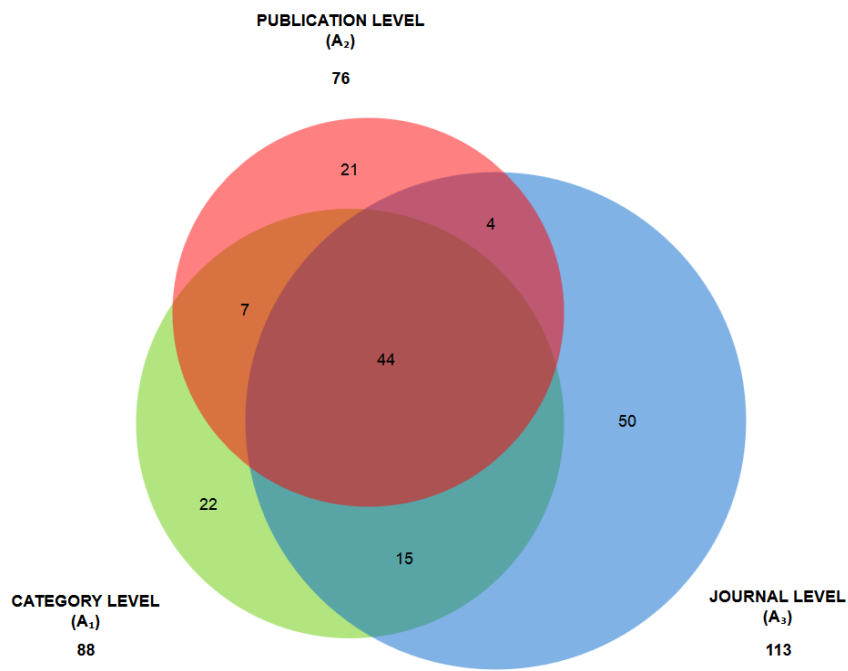

*Figure 6. Comparison of the three approaches to delineate a field at the journal level. Numbers denote the number of journals*

Altogether, 41 journals had the *nano* prefix in their titles, 22 of them covering a vast range of topics in NST, while the remaining 19 dealt only with specific NST topics (Carbon nanomaterials [1]; Nanomaterials [4]; Nanoenergy [1]; Nanoenvironment [1]; Nanomedicine and Nanobiotechnology [4]; Nanoparticles [1]; Nanophotonics [3]; Nanopatents [1]; Nanoscale communication and networking [1]; Nanotoxicology [2]). Meanwhile, three journals without the *nano* prefix covered topics related to Nanomaterials [2] and Nanochemistry [1], but were not solely related to NST (Appendix 4).

This group of journals displays a high level of precision given the fact that all topics covered by the journals are relevant in NST.

*4.1.2. Matching journals under two approaches*

The intersection between $A_1$ and $A_2$ ($A_1 \cap A_2$) displayed seven journals; five of them focus on NST or a specific topic in NST, and the other two cover adjacent areas. As seen in Appendix 4, journals residing at this intersection may be classified under the topics: Nanomaterials; Nanoparticles; Nanophotonics; NST; Science (interdisciplinary), or Physical chemistry.

The next intersection, of $A_1$ and $A_3$, took in 15 journals. What is interesting here (Appendix 4) is that 13 journals shared the characteristic of the *nano* prefix in their title. In contrast, one features the prefix "micro" in its title. The journals shown in Appendix 4 cover specific areas in NST: Nanochemistry [1]; Nanoengineering [5]; Nanofluidics [1]; Nanomaterials [2]; Nanomedicine and nanobiotechnology [4]; Nanophotonics [1]; and Nanophysics [1].

The final intersection between two approaches, $A_2 \cap A_3$, presented four journals, one of them specializing in computational nanoscience. The rest might be considered as journals that cover basic/pure knowledge applicable for research development in NST, e.g. materials science, or more specifically energy, medicine & biology, and optics & electronics.

Regarding the precision of the journals, those journals of intersection between sets $A_1 \cap A_2$ and $A_1 \cap A_3$ showed a high level of precision, focusing on NST. Journals in the intersection $A_2$ with $A_3$ showed a lower level of precision, as they covered other knowledge fields in addition to NST-related topics.

*4.1.3. Journals outside the intersections*

Journals that are not found to occupy any intersection are listed in Appendix 5. $A_1$ identified 22 journals outside an intersection, meaning they could only be identified under this particular approach (Appendix 5). A peculiar characteristic is shown by nine of these journals: the prefix *micro* in their titles. Although microscale studies ($10^{-6}$) use a low scale, it is still larger if we take into account the definition of nanoscale ($10^{-9}$). Five of these journals make reference to both scales: micro and nano; three of them make no reference to nano. Another covers a specific topic in NST, namely nanoporous materials. Besides, among the 14 remaining journals, 12 encompass *nano*-related topics, but these journals are not merely focused on *nano* research. The other two are strictly NST journals. Hence, these journals gave a high level of precision in six cases, and a medium to low level of precision in the rest.

$A_2$ detected 21 journals (Appendix 5), four focused on carbon materials, and four on electrochemistry. These electrochemistry journals may provide basic/pure knowledge for the study of nanoelectrochemistry, but they are not specialized journals in NST. Just one of them refers to NST in its scope. The same happens with two of the journals focused on condensed matter & materials physics. One focuses on the study of gold and fabricated materials. The rest of the journals identified here cover some topics close to NST or even other field categories. Six of these journals showed a high level of precision, while others displayed a medium or low level of precision.

$A_3$ distinguished 50 journals, though 31 journals are only included in the Scopus database; the other 19 journals are included in both databases (Appendix 5). Those identified journals here may be classified into two groups: journals of NST per se (general and specific) and journals connected to NST. The first group contains 34 journals on the following *nano* topics: Nanochemistry [1]; Nanoengineering [9]; Nanoenvironment [1]; Nanoethics [1]; Nanomedicine & nanobiotechnology [7]; Nanomaterials [3]; Nanoscience & Nanotechnology [6]; Nanoparticles [1]; Nanopharmachology [1]; and Nanotoxicology [1]. The second one contains 16 journals not specializing in an NST topic, but rather topics quite close to NST. Some of them even include the *nano* prefix or a keyword related to NST in the section 'Aim and scope', but they better match other fields of knowledge. Consequently, the precision of journals in $A_1$ outside the intersection presented a medium to low level of precision, although the 34 journals with the *nano* prefix in their titles showed a high level of precision.

*4.1.4. Precision and recall at the journal level*

The range of values for precision was not large. Journal approach scored the highest rank and the CWTS approach scored the lowest ([Table 2](#)). In other words, all the journals retrieved might be considered as relevant journals in NST because the values are closer to 1; just a few journals delineated by each approach might be considered non-useful journals for the delineation of NST.

Recall scored lower values than precision values (Table 2). The journal approach achieved the highest values, followed by Wos category approach and CWTS. Subsequently, comparing precision and recall of the three approaches reveals a trade-off of sorts, with a slight improvement in recall for the journal approach. In sum, precision and recall performed better for the journal approach than for the WoS category and CWTS approaches, but the variation is minor.

*Table 2. Precision and recall at the journal level*

| Approach | Total number of journals | Number of high-precise journals | Number of medium-precise journals | Number of low-precise journals | Precision | Recall |
|---|---|---|---|---|---|---|
| $A_1$ | 88 | 73 | 7 | 8 | 0.87 | 0.49 |
| $A_2$ | 76 | 58 | 14 | 4 | 0.86 | 0.44 |
| $A_3$ | 113 | 95 | 10 | 8 | 0.89 | 0.64 |

\* Total recall = 163

*4. 2. Comparing approaches at the publication level*

We use the in-house WoS database of CTWS containing 1,005,801 publications spread out in 3433 micro-level fields. ($A_0$). Figure 7 shows the number of publications in the journals identified by each approach (2000-2016), as well as the number of publication intersections between them.

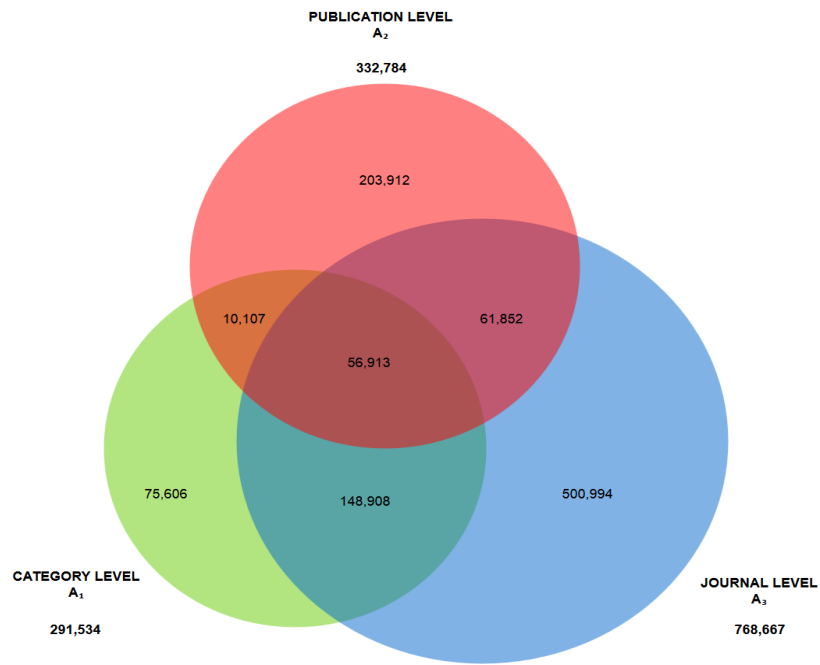

***Figure 7. Comparison of three approaches to delineate a field category at the publication level. Numbers denote the number of publications and reviews***

*4.2.1. Precision at the publication level*

As mentioned above, we conducted a questionnaire with a random sample to be studied by NST experts. The number of completed questionnaires was 98 (Appendix 7). According to the experts' answers, the CWTS classification approach achieved the top level of average precision with 0.62, followed by journal approach (0.56). The WoS classification approach secured the lowest value of precision (0.47) (Table 3). In addition, the seed data ($A_0$) obtained 0.56 precision. Journal approach scored the highest level of recall with a compensated precision value; WoS category approach displayed the same pattern. By contrast, CWTS approach, as well as the data seed, ranked high values of precision but limited recall. In other words, the higher the precision is, the lower recall is. These outcomes indicate that the journal approach may be considered an alternative approach to delineate a field, in terms of relevant publications, because of the trade-off between precision and recall.

*Table 3. Precision of recall at publication level*

| Approach | Precision | Recall |
|---|---|---|
| $A_0$ | 0.56 | 0.38 |
| $A_1$ | 0.47 | 0.34 |
| $A_2$ | 0.62 | 0.34 |
| $A_3$ | 0.56 | 0.46 |

Most of the questionnaires were completed by physicists, chemists, biologists and engineers. Taking into account the background field of each interviewee (Appendix 8), all fields except materials science considered that the approach with the highest level of precision was CWTS, closely followed by the journal approach.

The comparison of precision outcomes at the publication level and at the journal level showed distinctions. Precision at journal level identified the journal approach as the most precise, although there is no significant difference with regard to the other approaches. Nevertheless, precision at the publication level ranked CWTS classification as the most precise approach. It so happens that many journals publish *nano* publications, but at the same time, they continue to put out publications close to *nano* yet not specifically NST. This could be why there is a distinction when comparing the two approaches: studying precision at the publication level, we see it in greater detail, allowing for a more objective assessment of precision.

In general, a very substantial level of disagreement was detected in terms of the background field of each interviewee and the perceived relevance or irrelevance of an item according to the responses. We should point out that no publication received a 100% answer as relevant or not according to the experts.

In order to analyze/evaluate the level/reliability of agreement among experts, we used Fleiss's kappa statistic (Fleiss, 1971) for interrater reliability. This value ranges from -1 to +1, where 0 represents agreement expected at random, and 1 represents perfect agreement among experts. In addition, we used the Landis and Koch (1977) guidelines for interpreting Fleiss's kappa values. The resulting level/reliability of agreement among all experts was 0.27 (Table 4). Some disagreement among the experts might be traced to their different backgrounds, hence their beliefs about what is relevant or not in the field.

Despite the fact that the random sample involved different approaches, publications with a 0.8 of precision or more were selected, 13 publications in total. After matching these publications to the three approaches, the journal approach was found to match 10 publications (76.92%) and CWTS classification approach matched seven (53.84%). This could mean that although the average of precision in the journal approach is less than under the CWTS classification, there are many relevant publications in the journal approach. For example, three of the four publications with 0.9 (or over) percentage of precision match the journal approach; only once matches the CWTS classification approach.

*Table 4. Fleiss's Kappa values*

|  | Fleiss's kappa |
|---|---|
| Total | 0.27 |

| Field | Fleiss's kappa |
|---|---|
| Physics | 0.31 |
| Chemistry | 0.26 |
| Engineering | 0.18 |
| Biology | 0.09 |

| Subfield | Fleiss's kappa |
|---|---|
| Nanotheoretical physics | 0.42 |
| Nanobiosystems | 0.40 |
| Nanomagnetism | 0.34 |
| Nanoelectronics | 0.29 |
| Nanomaterials | 0.26 |
| Nanobiomedicine | 0.24 |
| Nanoprocesses | 0.21 |
| Nanoenergy | 0.19 |
| Nanoengineering | 0.19 |
| Nanophotonics | 0.19 |
| Nanotoxicology & Sustainability | 0.16 |
| Nanofabrication | 0.07 |

The results obtained after the validation process by experts determined whether journal publications were relevant or not relevant. In view of the validation process, it was unexpectedly found that journals putting out relevant publications also put out irrelevant publications. It may be that many publications tend to be classified as *nano* publications, but they are not NST per se. Our random sample set is too small to compare results at the journal level, that is, taking into account journal title, with the journal included in the sample.

*4.2. Visualization maps at the publication level*

Visualization of the data was achieved through VOSviewer software (van Eck & Waltman, 2010) and shows the journals in the field of NST under the three approaches (Figure 8).

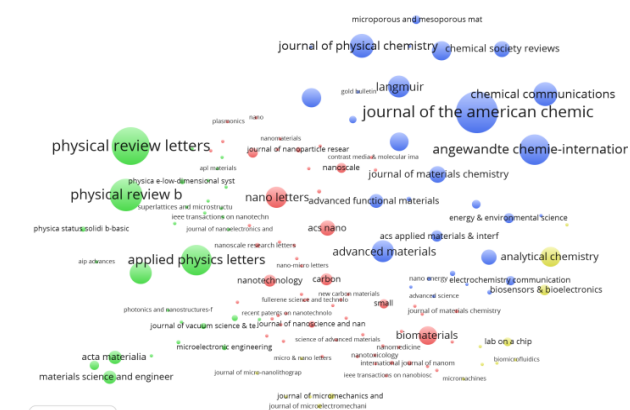

*Figure 8. NST journals citation network*

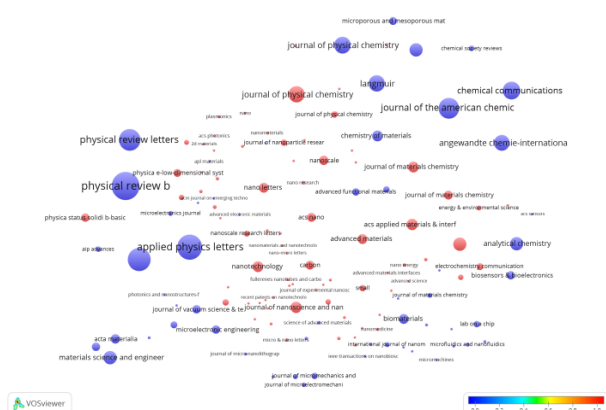

*Figure 9. Overlap map number of publications per journal by CWTS approach ($A_2$)*

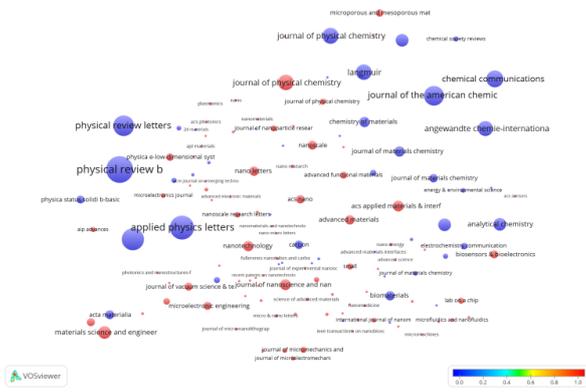 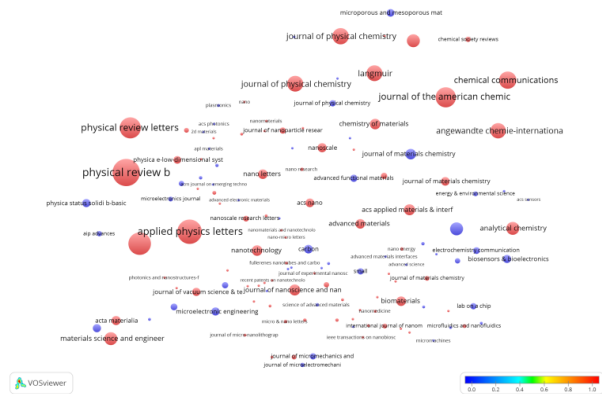

*Figure 10. Overlap map of number of publications per journal by WoS approach ($A_1$)*

*Figure 11. Overlap map of number of publications per journal by journal approach ($A_3$)*

Each circle represents a journal. The size of each journal represents the number of citations received. Journals located close to each other are more related than journals located far away. Color indicates that clusters, as well as journals in each cluster, are strongly related to each other. Four clusters are seen, three of them large and one small. The group of journals for every cluster can be summed up as: journals focusing on physics in the middle and lower left (green); journals focusing on areas in NST in the middle (red); journals focusing on chemistry toward the upper and middle right; and journals focusing on micro and nanoengineering to the lower right (yellow).

The red cluster consists of journals that publish the scientific output in NST, covering journals publishing all areas of NST, and nanomaterials-physical and mechanical characteristics (upper middle) and nanomaterials-structure and properties (lower middle). The green cluster holds journals publishing nanophysics research, although journals on physics are included as well. The lower area of this cluster displays journals focusing on micro and nanoengineering process (the physics, engineering, and materials of the devices). The blue one consists of journals on chemistry research. Lastly, the yellow cluster consists of journals focused on micro and nanoengineering, but applied to biology, medicine, and chemistry.

Overlapped visualization based on a number of publications by the journal in each approach illustrates that most journals with a high level of overlap in $A_2$ are located in the middle area, and no many of them are in the outer periphery ([Figure 9](#)). In a similar vein, many of the journals from $A_1$ are located in the middle area, although many others are in the outer periphery ([Figure 10](#)). By contrast, journals from $A_3$ are located all over the visualization ([Figure 11](#)). As is clearly seen, many of the journals distinguished by this approach are located in the outer area, and many of the journals located in the middle areas are not selected by this approach ($A_3$). That is, many of the journal located in middle areas of the map are not been identified by the journal approach ($A_3$). Instead, they have been identified by the use of approaches $A_1$ or $A_2$.

## 5. Discussion and Conclusion

The overall advancement of research means that new fields are emerging, and they may be hard to define. Yet a proper field delineation is essential for information retrieval purposes and designing reliable and solid tools, especially in the framework of science policy design and science evaluation processes. In this paper, we deal with the problem that researchers face when attempting to delineate emergent and interdisciplinary fields such as NST.

Evidence shows that the use of different multidisciplinary and international bibliographical data sources —such as Scopus and WoS— for collecting data and comparing approaches can contribute to accurate field delineation (Glänzel, 2015), although the completeness and quality of the data set might be skewed (Mongeon & Paul-Hus, 2016). The inexistence, thus far, of an international gold-standard classification system for blibliometric fields has been underlined in previous studies (Gomez& Bordons, 1996; Waltman & Van Eck, 2012). It is well known that field delineation is problematic, even when different data sources are used, new methods are applied, citation relations are measured, precision and recall are calculated, and the results are validated by experts. While validation by experts in the field may enhance the quality of the delineation process, a lack of agreement may prevail, with opinions partly conditioned by the experts´ particular field of work.

Depending on the field to be delineated and the purpose at hand, each of the methods suggested in this study could be suitable (Zitt, 2015). All the approaches have their pros and cons. Each vision reflects nuances that contribute to field delineation in some way or another. Because no method can be singled out as the best one, this study aimed to contribute empirically to the questionable area of using search strategies and citation analysis at the journal level to delineate research fields. Key elements for the success of such a process are the application and replicability of methods used, and the design and development of classifications, indicators and/or tools relying on scientific references from databases. A comparative and reproducible method is essential for arriving at consistent and reliable results. The approach presented here entails a seven-step procedure that can be replicated in other research areas and emergent fields in large databases characterized by a very robust and detailed framework.

Our line of research has become a series of stepping-stones (Gómez-Núñez et al., 2011; 2014: Muñoz-Écija et al, 2013) toward the goal of optimizing and boosting the SJR journal classification system and subsequent journal assignment. In this study itself, taking journals as units of analysis did not guarantee proper identification of the publications: some publications belonged to other subject categories, some journals that had published relevant output in the field were not included under the corresponding subject category, etc. Notwithstanding, the journal classification system can be said to display "a robust poor health". It is still used by the main bibliographical databases such as Scopus and WoS. The classifications are of widespread interest and utility when it comes to evaluating science and research, and a valuable resource for decision- and policy-makers (e.g., when hiring, promoting and funding research assessment exercises). We do not proclaim that this classification proposal is definitive or exclusively appropriate. Rather, within the context of classifying emerging and interdisciplinary fields like NST, we conclude it is necessary to keep striving to combine techniques and units of analysis in the search for a process that will lead us closer to consensus among the scientific community.

**Acknowledgments**

We thank Nees Jan van Eck and Ludo Waltman for their generous and valuable advice in the early stage of this work. We also thank Ed Noyons for his comments and helpful suggestions on the design of the questionnaire. Thanks are likewise due to all the NST experts who were consulted and offered feedback, in a spirit of community, to enhance analysis of the results.

**Funding**

This study was funded by the Ph.D. International Mobility Programme of the University of Granada and by the Ministerio de Economía y Competitividad of Spain through the State Programme of Research, Development and Innovation oriented to the Challenges of Society [NANOMETRICS, CSO2014-57770-R].

**Author's contribution**

**Teresa Muñoz-Écija**: Conceived and designed the analysis; Collected the data; Contributed data or analysis tools; Performed the analysis; Wrote the paper.
**Benjamín Vargas-Quesada**: Contributed data or analysis tools; Wrote the paper.
**Zaida Chinchilla-Rodríguez**: Conceived and designed the analysis; Wrote the paper.

**Appendix 1a. Journals list of Wos category Nanoscience & Nanotechnology (ceased journals excluded)**

| *Journals list of WoS category N&N* |
|---|
| ACM JOURNAL ON EMERGING TECHNOLOGIES IN COMPUTING SYSTEMS |
| ACS APPLIED MATERIALS & INTERFACES |
| ACS ENERGY LETTERS |
| ACS NANO |
| ACS PHOTONICS |
| ACS SENSORS |
| ADVANCED ELECTRONIC MATERIALS |
| ADVANCED FUNCTIONAL MATERIALS |
| ADVANCED HEALTHCARE MATERIALS |
| ADVANCED MATERIALS |
| ADVANCED SCIENCE |
| ADVANCES IN NATURAL SCIENCES-NANOSCIENCE AND NANOTECHNOLOGY |
| AIP ADVANCES |
| APL MATERIALS |
| APPLIED NANOSCIENCE |
| BEILSTEIN JOURNAL OF NANOTECHNOLOGY |
| BIOMEDICAL MICRODEVICES |
| BIOMICROFLUIDICS |
| BIOSENSORS & BIOELECTRONICS |
| CHEMNANOMAT |
| CURRENT NANOSCIENCE |
| DIGEST JOURNAL OF NANOMATERIALS AND BIOSTRUCTURES |
| ENVIRONMENTAL SCIENCE-NANO |
| FULLERENES NANOTUBES AND CARBON NANOSTRUCTURES |
| IEEE TRANSACTIONS ON NANOBIOSCIENCE |
| IEEE TRANSACTIONS ON NANOTECHNOLOGY |
| IET NANOBIOTECHNOLOGY |
| INORGANIC AND NANO-METAL CHEMISTRY |
| INTERNATIONAL JOURNAL OF NANOMEDICINE |
| INTERNATIONAL JOURNAL OF NANOTECHNOLOGY |
| JOURNAL OF BIOMEDICAL NANOTECHNOLOGY |
| JOURNAL OF EXPERIMENTAL NANOSCIENCE |
| JOURNAL OF LASER MICRO NANOENGINEERING |
| JOURNAL OF MICROELECTROMECHANICAL SYSTEMS |
| JOURNAL OF MICROMECHANICS AND MICROENGINEERING |
| JOURNAL OF MICRO-NANOLITHOGRAPHY MEMS AND MOEMS |
| JOURNAL OF NANO RESEARCH |
| JOURNAL OF NANOBIOTECHNOLOGY |
| JOURNAL OF NANOELECTRONICS AND OPTOELECTRONICS |
| JOURNAL OF NANOMATERIALS |
| JOURNAL OF NANOPARTICLE RESEARCH |
| JOURNAL OF NANOPHOTONICS |
| JOURNAL OF NANOSCIENCE AND NANOTECHNOLOGY |
| JOURNAL OF PHYSICAL CHEMISTRY C |
| JOURNAL OF PHYSICAL CHEMISTRY LETTERS |

| *Journals list of WoS category N&N* |
|---|
| **JOURNAL OF VACUUM SCIENCE & TECHNOLOGY B** |
| **LAB ON A CHIP** |
| **MATERIALS EXPRESS** |
| **MATERIALS SCIENCE AND ENGINEERING A-STRUCTURAL MATERIALS PROPERTIES MICROSTRUCTURE AND PROCESSING** |
| **MICRO & NANO LETTERS** |
| **MICROELECTRONIC ENGINEERING** |
| **MICROELECTRONICS JOURNAL** |
| **MICROELECTRONICS RELIABILITY** |
| **MICROFLUIDICS AND NANOFLUIDICS** |
| **MICROMACHINES** |
| **MICROPOROUS AND MESOPOROUS MATERIALS** |
| **MICROSYSTEM TECHNOLOGIES-MICRO-AND NANOSYSTEMS-INFORMATION STORAGE AND PROCESSING SYSTEMS** |
| **NANO** |
| **NANO COMMUNICATION NETWORKS** |
| **NANO ENERGY** |
| **NANO LETTERS** |
| **NANO RESEARCH** |
| **NANO TODAY** |
| **NANOMATERIALS** |
| **NANOMATERIALS AND NANOTECHNOLOGY** |
| **NANOMEDICINE** |
| **NANOMEDICINE-NANOTECHNOLOGY BIOLOGY AND MEDICINE** |
| **NANO-MICRO LETTERS** |
| **NANOPHOTONICS** |
| **NANOSCALE** |
| **NANOSCALE AND MICROSCALE THERMOPHYSICAL ENGINEERING** |
| **NANOSCALE RESEARCH LETTERS** |
| **NANOSCIENCE AND NANOTECHNOLOGY LETTERS** |
| **NANOTECHNOLOGY** |
| **NANOTECHNOLOGY REVIEWS** |
| **NANOTOXICOLOGY** |
| **NATURE NANOTECHNOLOGY** |
| **PARTICLE & PARTICLE SYSTEMS CHARACTERIZATION** |
| **PHOTONICS AND NANOSTRUCTURES-FUNDAMENTALS AND APPLICATIONS** |
| **PHYSICA E-LOW-DIMENSIONAL SYSTEMS & NANOSTRUCTURES** |
| **PLASMONICS** |
| **PRECISION ENGINEERING-JOURNAL OF THE INTERNATIONAL SOCIETIES FOR PRECISION ENGINEERING AND NANOTECHNOLOGY** |
| **RECENT PATENTS ON NANOTECHNOLOGY** |
| **REVIEWS ON ADVANCED MATERIALS SCIENCE** |
| **SCIENCE OF ADVANCED MATERIALS** |
| **SCRIPTA MATERIALIA** |
| **SMALL** |
| **WILEY INTERDISCIPLINARY REVIEWS-NANOMEDICINE AND NANOBIOTECHNOLOGY** |

**Appendix 1b. Journals list of SJR category Nanoscience & Nanotechnology (ceased and discountinued published journals excluded)**

| |
|---|
| **ACM JOURNAL ON EMERGING TECHNOLOGIES IN COMPUTING SYSTEMS** |
| **ACS APPLIED MATERIALS & INTERFACES** |
| **ACS NANO** |
| **ADVANCED FUNCTIONAL MATERIALS** |
| **ADVANCED MATERIALS** |
| **AIP ADVANCES** |
| **BEILSTEIN JOURNAL OF NANOTECHNOLOGY** |
| **BIOMATERIALS** |
| **BIOMEDICAL MICRODEVICES** |
| **BIOMICROFLUIDICS** |
| **BIOSENSORS AND BIOELECTRONICS** |
| **CURRENT NANOSCIENCE** |
| **DIGEST JOURNAL OF NANOMATERIALS AND BIOSTRUCTURES** |
| **E-JOURNAL OF SURFACE SCIENCE AND NANOTECHNOLOGY** |
| **FULLERENES NANOTUBES AND CARBON NANOSTRUCTURES** |
| **IEEE NANOTECHNOLOGY MAGAZINE** |
| **IEEE TRANSACTIONS ON NANOBIOSCIENCE** |
| **IEEE TRANSACTIONS ON NANOTECHNOLOGY** |
| **IET NANOBIOTECHNOLOGY** |
| **INTERNATIONAL JOURNAL OF NANO AND BIOMATERIALS** |
| **INTERNATIONAL JOURNAL OF NANOMANUFACTURING** |
| **INTERNATIONAL JOURNAL OF NANOMECHANICS SCIENCE AND TECHNOLOGY** |
| **INTERNATIONAL JOURNAL OF NANOMEDICINE** |
| **INTERNATIONAL JOURNAL OF NANOPARTICLES** |
| **INTERNATIONAL JOURNAL OF NANOSCIENCE** |
| **INTERNATIONAL JOURNAL OF NANOTECHNOLOGY** |
| **JOURNAL OF BIOMEDICAL NANOTECHNOLOGY** |

**JOURNAL OF BIONANOSCIENCE**

**JOURNAL OF COMPUTATIONAL AND THEORETICAL NANOSCIENCE**

**JOURNAL OF EXPERIMENTAL NANOSCIENCE**

**JOURNAL OF LASER MICRO NANOENGINEERING**

**JOURNAL OF MICRO/ NANOLITHOGRAPHY, MEMS, AND MOEMS**

**JOURNAL OF MICROMECHANICS AND MICROENGINEERING**

**JOURNAL OF MICRO-BIO ROBOTIC**

**JOURNAL OF NANO RESEARCH**

**JOURNAL OF NANOBIOTECHNOLOGY**

**JOURNAL OF NANOELECTRONICS AND OPTOELECTRONICS**

**JOURNAL OF NANOMATERIALS**

**JOURNAL OF NANOPARTICLE RESEARCH**

**JOURNAL OF NANOPHOTONICS**

**JOURNAL OF NANOSCIENCE AND NANOTECHNOLOGY**

**JOURNAL OF PHYSICAL CHEMISTRY C**

**JOURNAL OF PHYSICAL CHEMISTRY LETTERS**

**JOURNAL OF SURFACE INVESTIGATION**

**LAB ON A CHIP - MINIATURISATION FOR CHEMISTRY AND BIOLOGY**

**MATERIALS EXPRESS**

**MATERIALS SCIENCE & ENGINEERING A: STRUCTURAL MATERIALS: PROPERTIES, MICROSTRUCTURE AND PROCESSING**

**MICRO AND NANO LETTERS**

**MICRO AND NANOSYSTEMS**

**MICROELECTRONIC ENGINEERING**

**MICROELECTRONICS AND RELIABILITY**

**MICROELECTRONICS JOURNAL**

**MICROFLUIDICS AND NANOFLUIDICS**

**MICROPOROUS AND MESOPOROUS MATERIALS**

**MICROSYSTEM TECHNOLOGIES**

**SCIENCE OF ADVANCED MATERIALS**

**SCRIPTA MATERIALIA**

**SMALL**

**INORGANIC AND NANO METAL CHEMISTRY**

**WILEY INTERDISCIPLINARY REVIEWS NANOMEDICINE AND NANOBIOTECHNOLOGY**

# Appendix 2. Lexical queries for Scopus and WoS

**Lexical queries for Scopus**

#1

TITLE-ABS-KEY(nano* )

#2

(TITLE-ABS-KEY ( nanoa ) OR TITLE-ABS-KEY ( nanoacalles ) OR TITLE-ABS-KEY ( nanoagraylea ) OR TITLE-ABS-KEY ( nanoalga* ) OR TITLE-ABS-KEY ( nanoapiculatum ) OR TITLE-ABS-KEY ( nanoarchaeaor ) OR TITLE-ABS-KEY ( nanoarchaeota ) OR TITLE-ABS-KEY ( nanoarchaeum ) OR TITLE-ABS-KEY ( nanoastegotherium ) OR TITLE-ABS-KEY ( "nano*aryote*" ) OR TITLE-ABS-KEY ( nanobacteri* ) OR TITLE-ABS-KEY ( nanobagrus ) OR TITLE-ABS-KEY ( nanobalcis ) OR TITLE-ABS-KEY ( nanobaris ) OR TITLE-ABS-KEY ( nanobates ) OR TITLE-ABS-KEY ( nanobatinae ) OR TITLE-ABS-KEY ( nanobius ) OR TITLE-ABS-KEY ( nanobryaceae ) OR TITLE-ABS-KEY ( nanobryoides ) OR TITLE-ABS-KEY ( nanobuthus ) OR TITLE-ABS-KEY ( nano-bible ) OR TITLE-ABS-KEY ( nanocalcar ) OR TITLE-ABS-KEY ( nanocambridgea ) OR TITLE-ABS-KEY ( nanocapillare ) OR TITLE-ABS-KEY ( nanocapillary ) OR TITLE-ABS-KEY ( nanocarpa ) OR TITLE-ABS-KEY ( nanocarpum ) OR TITLE-ABS-KEY ( nanocarpus ) OR TITLE-ABS-KEY ( nanocassiope ) OR TITLE-ABS-KEY ( nanocavia ) OR TITLE-ABS-KEY ( nanocephalum ) OR TITLE-ABS-KEY ( nanocheirodon ))

#3

(TITLE-ABS-KEY ( nanochilina ) OR TITLE-ABS-KEY ( nanochilus ) OR TITLE-ABS-KEY ( nanochitina ) OR TITLE-ABS-KEY ( nanochlaenius ) OR TITLE-ABS-KEY ( nanochlorum ) OR TITLE-ABS-KEY ( nanochoerus ) OR TITLE-ABS-KEY ( nanochromis ) OR TITLE-ABS-KEY ( nanochromosom* ) OR TITLE-ABS-KEY ( nanochrysopa ) OR TITLE-ABS-KEY ( nanochthonius ) OR TITLE-ABS-KEY ( nanocixius ) OR TITLE-ABS-KEY ( nanocladius ) OR TITLE-ABS-KEY ( nanoclarelia ) OR TITLE-ABS-KEY ( nanoclavelia ) OR TITLE-ABS-KEY ( nanoclimacium ) OR TITLE-ABS-KEY ( nanoclymenia ) OR TITLE-ABS-KEY ( nanocnide ) OR TITLE-ABS-KEY ( nanocochlea ) OR TITLE-ABS-KEY ( nanocolletes ) OR TITLE-ABS-KEY ( nanocondylodesmus ) OR TITLE-ABS-KEY ( nanocopia ) OR TITLE-ABS-KEY ( nanocoquimba ) OR TITLE-ABS-KEY ( nanocrinus ) OR TITLE-ABS-KEY ( nanoctenus ) OR TITLE-ABS-KEY ( nanocthispa ) OR TITLE-ABS-KEY ( nanocuridae ) OR TITLE-ABS-KEY ( nanocurie ) OR TITLE-ABS-KEY ( nano-curie ) OR TITLE-ABS-KEY ( nanocuris ) OR TITLE-ABS-KEY ( nanocyclopia ) OR TITLE-ABS-KEY ( nanocynodon ) OR TITLE-ABS-KEY ( nanocythere ))

#4

(TITLE-ABS-KEY ( nanodacna ) OR TITLE-ABS-KEY ( nanodactylus ) OR TITLE-ABS-KEY ( nanodamon ) OR TITLE-ABS-KEY ( nanodea ) OR TITLE-ABS-KEY ( nanodealbata ) OR TITLE-ABS-KEY ( nanodectes ) OR TITLE-ABS-KEY ( nanodella ) OR TITLE-ABS-KEY ( nanodelphys ) OR TITLE-ABS-KEY ( nanodendron ) OR TITLE-ABS-KEY ( nanodes ) OR TITLE-ABS-KEY ( nanodiaparsis ) OR TITLE-ABS-KEY ( nanodiaptomus ) OR TITLE-ABS-KEY ( nanodidelphys ) OR TITLE-ABS-KEY ( nanodiella ) OR TITLE-ABS-KEY ( nanodiodes ) OR TITLE-ABS-KEY ( nanodiplosis ) OR TITLE-ABS-KEY ( nanodiscus ) OR TITLE-ABS-KEY ( nanodisticha ) OR TITLE-ABS-KEY ( nanodromia ) OR TITLE-ABS-KEY ( nanodynerus ) OR TITLE-ABS-KEY ( nanoesi ) OR TITLE-ABS-KEY ( nanofauna* ) OR TITLE-ABS-KEY ( nanofila ) OR TITLE-ABS-KEY ( "nano-filt*" ) OR TITLE-ABS-KEY ( nanofilt* ) OR TITLE-ABS-KEY ( nanoflagel* ) OR TITLE-ABS-KEY ( nanoflow ) OR TITLE-ABS-KEY ( "nano-flow" ) OR TITLE-ABS-KEY ( nanofilidae ) OR TITLE-ABS-KEY( Nanog ) OR TITLE-ABS-KEY( Nanog1 ))

#5

(TITLE-ABS-KEY( Nanog2 ) OR TITLE-ABS-KEY( Nanogalathea ) OR TITLE-ABS-KEY( Nanogeterotroph* ) OR TITLE-ABS-KEY( Nanoglobum ) OR TITLE-ABS-KEY( Nanoglossa ) OR TITLE-ABS-KEY( Nanognathia ) OR TITLE-ABS-KEY( Nanognathus ) OR TITLE-ABS-KEY( Nanogomphodon ) OR TITLE-ABS-KEY( Nanogona ) OR TITLE-ABS-KEY( Nanogonalos ) OR TITLE-ABS-KEY( Nanogorgon ) OR TITLE-ABS-KEY( Nanogram* ) OR TITLE-ABS-KEY("nano-gram*" ) OR TITLE-ABS-KEY( Nanogramma ) OR TITLE-ABS-KEY( Nanograptus ) OR TITLE-ABS-KEY( Nanogyra ) OR TITLE-ABS-KEY (Nanogyrini ) OR TITLE-ABS-KEY( Nanohalus ) OR TITLE-ABS-KEY( Nanohammus ) OR TITLE-ABS-KEY( Nanohemicera ) OR TITLE-ABS-KEY( Nanoheterotroph* ) OR TITLE-ABS-KEY( Nanohystrix ) OR TITLE-ABS-KEY( Nanoides ) OR TITLE-ABS-KEY( Nanoini ) OR TITLE-ABS-KEY( Nanojapyx ) OR TITLE-ABS-KEY( Nanokerala ) OR TITLE-ABS-KEY( Nanokelvin* ) OR TITLE-ABS-KEY( Nanokermes ) OR TITLE-ABS-KEY( Nanola ) OR TITLE-ABS-KEY( Nanolachesilla ) OR TITLE-ABS-KEY( Nanolania ) OR TITLE-ABS-KEY( Nanolauthia ) OR TITLE-ABS-KEY( NanoLC* ) OR TITLE-ABS-KEY("nano-LC*" ) OR TITLE-ABS-KEY( Nanolestes ) OR TITLE-ABS-KEY( Nanolichus ) OR TITLE-ABS-KEY("nano-liter*" ) OR TITLE-ABS-KEY("nano-litre*" ) OR TITLE-ABS-KEY( Nanoliter* ) OR TITLE-ABS-KEY( Nanolitre* ) OR TITLE-ABS-KEY( Nanolobus ) OR TITLE-ABS-KEY( Nanoloricida ))

#6

(TITLE-ABS-KEY( Nanolpium ) OR TITLE-ABS-KEY( Nanolumen ) OR TITLE-ABS-KEY( Nanomaja ) OR TITLE-ABS-KEY( Nanomantinae ) OR TITLE-ABS-KEY( Nanomantini ) OR TITLE-ABS-KEY( Nanomantis ) OR TITLE-ABS-KEY( Nanomeli* ) OR TITLE-ABS-KEY( Nanomelon ) OR TITLE-ABS-KEY( Nanomermis ) OR TITLE-ABS-KEY( Nanomerus ) OR TITLE-ABS-KEY( Nanomeryx ) OR TITLE-ABS-KEY( Nanometa ) OR TITLE-ABS-KEY("nano-meter*" ) OR TITLE-ABS-KEY( Nanometidae ) OR TITLE-ABS-KEY( Nanometinae ) OR TITLE-ABS-KEY( Nanometra ) OR TITLE-ABS-KEY("nano-metre*" ) OR TITLE-ABS-KEY("nm" ) OR TITLE-ABS-KEY( Nanometer* ) OR TITLE-ABS-KEY( Nanometre* ) OR TITLE-ABS-KEY( Nanomia ) OR TITLE-ABS-KEY( Nanomias ) OR TITLE-ABS-KEY( Nanomicrophyes ) OR TITLE-ABS-KEY( Nanomilleretta ) OR TITLE-ABS-KEY( Nanomimus ) OR TITLE-ABS-KEY( Nanomis ) OR TITLE-ABS-KEY( Nanomitra ) OR TITLE-ABS-KEY( Nanomitriella ) OR TITLE-ABS-KEY( Nanomitriopsis ) OR TITLE-ABS-KEY( Nanomitus ) OR TITLE-ABS-KEY( Nanomol* ) OR TITLE-ABS-KEY("nano-molar" ) OR TITLE-ABS-KEY("nano-mole*" ) OR TITLE-ABS-KEY( Nanomutilinae ) OR TITLE-ABS-KEY( Nanomutilla ) OR TITLE-ABS-KEY( Nanomyces ) OR TITLE-ABS-KEY( Nanomyina ) OR TITLE-ABS-KEY( Nanomyrmacyba ) OR TITLE-ABS-KEY( Nanomyrme ) OR TITLE-ABS-KEY( Nanomys ) OR TITLE-ABS-KEY( Nanomysis ) OR TITLE-ABS-KEY( Nanomysmena ))

#7

(TITLE-ABS-KEY( Nanonaucoris ) OR TITLE-ABS-KEY( Nanonavis ) OR TITLE-ABS-KEY( Nanoneis ) OR TITLE-ABS-KEY( Nanonemoura ) OR TITLE-ABS-KEY( Nanonocticolus ) OR TITLE-ABS-KEY( Nanonycteris ) OR TITLE-ABS-KEY("Nanook of the North" ) OR TITLE-ABS-KEY ( nanopachyiulus )  OR  TITLE-ABS-KEY ( nanopagurus )  OR  TITLE-ABS-KEY ( nanopareia )  OR  TITLE-ABS-KEY ( nanoparia )  OR  TITLE-ABS-KEY ( nanopatula )  OR  TITLE-ABS-KEY ( nanopennatum )  OR  TITLE-ABS-KEY ( nanoperla )  OR  TITLE-ABS-KEY ( nanophareus )  OR  TITLE-ABS-KEY ( nanophemera )  OR  TITLE-ABS-KEY ( nanophtalm* )  OR  TITLE-ABS-KEY ( nanophya )  OR  TITLE-ABS-KEY ( nanophydes )  OR  TITLE-ABS-KEY ( nanophydinae )  OR  TITLE-ABS-KEY ( nanophydini )  OR  TITLE-ABS-KEY ( nanophyes )  OR  TITLE-ABS-KEY ( nanophyetinae )  OR  TITLE-ABS-KEY ( nanophyetus )  OR  TITLE-ABS-KEY ( nanophyidae )  OR  TITLE-ABS-KEY ( nanophyinae )  OR  TITLE-ABS-KEY ( nanophyini )  OR  TITLE-ABS-KEY ( nanophylla )  OR  TITLE-ABS-KEY ( nanophylliini )  OR  TITLE-ABS-KEY ( nanophyllium )  OR  TITLE-ABS-KEY ( nanophyllum )  OR  TITLE-ABS-KEY ( nanophyllus )  OR  TITLE-ABS-KEY ( nanophytes  ))

#8

(TITLE-ABS-KEY ( nanophyti ) OR TITLE-ABS-KEY ( nanophyto* ) OR TITLE-ABS-KEY ( nanopilumnus ) OR TITLE-ABS-KEY ( nanopitar ) OR TITLE-ABS-KEY ( nanoplagia ) OR TITLE-ABS-KEY ( nanoplax ) OR TITLE-ABS-KEY ( nanoplaxes ) OR TITLE-ABS-KEY ( nanoplectrus ) OR TITLE-ABS-KEY ( nanoplinthisus ) OR TITLE-ABS-KEY ( nanopodella ) OR TITLE-ABS-KEY ( nanopodellus ) OR TITLE-ABS-KEY ( nanopolymorphum ) OR TITLE-ABS-KEY ( nanopolystoma ) OR TITLE-ABS-KEY ( nanopria ) OR TITLE-ABS-KEY ( nanoprotist* ) OR TITLE-ABS-KEY ( nanops ) OR TITLE-ABS-KEY ( nanopsallus ) OR TITLE-ABS-KEY ( nanopsis ) OR TITLE-ABS-KEY ( nanopsocetae ) OR TITLE-ABS-KEY ( nanopsocus ) OR TITLE-ABS-KEY ( nanopterodectes ) OR TITLE-ABS-KEY ( nanopterum ) OR TITLE-ABS-KEY ( nanoptilium ) OR TITLE-ABS-KEY ( nanopus ) OR TITLE-ABS-KEY ( nanopyxis ) OR TITLE-ABS-KEY ( nanoqia ) OR TITLE-ABS-KEY ( nanoqsunquak ) OR TITLE-ABS-KEY ( nanor ) OR TITLE-ABS-KEY ( nanorafonus ) OR TITLE-ABS-KEY ( nanorana ) OR TITLE-ABS-KEY ( nanoraphidia ) OR TITLE-ABS-KEY ( nanorchestes ) OR TITLE-ABS-KEY ( nanorchestidae ) OR TITLE-ABS-KEY ( nanorhamphus ))

#9

(TITLE-ABS-KEY ( nanorhathymus ) OR TITLE-ABS-KEY ( nanorhopaea ) OR TITLE-ABS-KEY ( nanorrhacus ) OR TITLE-ABS-KEY ( nanorrhynchus ) OR TITLE-ABS-KEY( Nanorthidae ) OR TITLE-ABS-KEY( Nanorthis ) OR TITLE-ABS-KEY( Nanos ) OR TITLE-ABS-KEY( Nanosalicium ) OR TITLE-ABS-KEY( Nanosatellite* ) OR TITLE-ABS-KEY( Nanosauridae ) OR TITLE-ABS-KEY( Nanosaurus ) OR TITLE-ABS-KEY( Nanoschema ) OR TITLE-ABS-KEY( Nanoschetus ) OR TITLE-ABS-KEY( Nanoscydmus ) OR TITLE-ABS-KEY( Nanoscypha ) OR TITLE-ABS-KEY("nano-second*" ) OR TITLE-ABS-KEY( Nanosecond* ) OR TITLE-ABS-KEY( Nanosella ) OR TITLE-ABS-KEY( Nanosellini ) OR TITLE-ABS-KEY( Nanoserranus ) OR TITLE-ABS-KEY( Nanosesarma ) OR TITLE-ABS-KEY( Nanosetus ) OR TITLE-ABS-KEY( Nanosilene ) OR TITLE-ABS-KEY("nano-SIMS" ) OR TITLE-ABS-KEY( Nanosiren ) OR TITLE-ABS-KEY( Nanosius ) OR TITLE-ABS-KEY( Nanosmia ) OR TITLE-ABS-KEY( Nanosmilus ) OR TITLE-ABS-KEY( Nanosomus ) OR TITLE-ABS-KEY( Nanospadix ) OR TITLE-ABS-KEY( Nanospathulatum ) OR TITLE-ABS-KEY( Nanospira ) OR TITLE-ABS-KEY( Nanospondylus ) OR TITLE-ABS-KEY( Nanospora ) OR TITLE-ABS-KEY( Nanospray* ) OR TITLE-ABS-KEY("nano-spray*" ) OR TITLE-ABS-KEY( Nanosteatoda ) OR TITLE-ABS-KEY( Nanostellata ) OR TITLE-ABS-KEY( Nanostictis ))

#10

(TITLE-ABS-KEY( Nanostoma ) OR TITLE-ABS-KEY( Nanostomus ) OR TITLE-ABS-KEY( Nanostrangalia ) OR TITLE-ABS-KEY( Nanostrea ) OR TITLE-ABS-KEY( Nanostreptus ) OR TITLE-ABS-KEY( Nanosura ) OR TITLE-ABS-KEY( Nanosylvanella ) OR TITLE-ABS-KEY( Nanotagalus ) OR TITLE-ABS-KEY( Nanotanaupodus ) OR TITLE-ABS-KEY( Nanotaphus ) OR TITLE-ABS-KEY( Nanotermitodius ) OR TITLE-ABS-KEY( Nanothamnus ) OR TITLE-ABS-KEY( Nanothecioidea ) OR TITLE-ABS-KEY( Nanothecium ) OR TITLE-ABS-KEY( Nanothinophilus ) OR TITLE-ABS-KEY( Nanothrips ) OR TITLE-ABS-KEY( Nanothyris ) OR TITLE-ABS-KEY( Nanotitan ) OR TITLE-ABS-KEY( Nanotitanops ) OR TITLE-ABS-KEY( Nanotopsis ) OR TITLE-ABS-KEY( Nanotragulus ) OR TITLE-ABS-KEY( Nanotragus ) OR TITLE-ABS-KEY( Nanotrema ) OR TITLE-ABS-KEY( Nanotrephes ) OR TITLE-ABS-KEY( Nanotrigona ) OR TITLE-ABS-KEY( Nanotriton ) OR TITLE-ABS-KEY( Nanotrombium ) OR TITLE-ABS-KEY( Nanotyrannus ) OR TITLE-ABS-KEY( Nanoviridae ) OR TITLE-ABS-KEY( Nanovirus ) OR TITLE-ABS-KEY( Nanowana ) OR TITLE-ABS-KEY( Nanowestratia ) OR TITLE-ABS-KEY( Nanoxylocopa ) OR TITLE-ABS-KEY( Nano2 ) OR TITLE-ABS-KEY( Nano3 ) OR TITLE-ABS-KEY(plankton* ) OR TITLE-ABS-KEY( "N*plankton*" ) OR TITLE-ABS-KEY("m*plankton*" ) OR TITLE-ABS-KEY("b*plankton*" ) OR TITLE-ABS-KEY("p*plankton*" ) OR TITLE-ABS-KEY("z*plankton*" ))

#11= #1 AND NOT (#2 OR #3 OR #4 OR #5 OR #6 OR #7 OR #8 OR #9 OR #10)

#12

( ( ( TITLE-ABS-KEY ( gadonano* )  OR  TITLE-ABS-KEY ( glyconanopartic* )  OR  TITLE-ABS-KEY ( heteronano* )  OR  TITLE-ABS-KEY ( immunonanogold )  OR  TITLE-ABS-KEY ( immunonanopartic* )  OR  TITLE-ABS-KEY ( polynanocrystal* )  OR  TITLE-ABS-KEY ( subnanocluster* )  OR  TITLE-ABS-KEY ( ultrananocrystal* )  OR  TITLE-ABS-KEY ( ultrananodiamond* ) ) )  OR  ( ( ( TITLE-ABS-KEY ( "self assembl*" )  OR  TITLE-ABS-KEY ( "self organiz*" )  OR  TITLE-ABS-KEY ( "directed assembl*" ) ) )  AND  ( ( TITLE-ABS-KEY ( monolayer* )  OR  TITLE-ABS-KEY ( "mono-layer*" )  OR  TITLE-ABS-KEY ( film* )  OR  TITLE-ABS-KEY ( quantum* )  OR  TITLE-ABS-KEY ( multilayer* )  OR  TITLE-ABS-KEY ( "multi-layer*" )  OR  TITLE-ABS-KEY ( array )  OR  TITLE-ABS-KEY ( molecul* )  OR  TITLE-ABS-KEY ( polymer* )  OR  TITLE-ABS-KEY ( "co-polymer*" )  OR  TITLE-ABS-KEY ( copolymer* )  OR  TITLE-ABS-KEY ( mater* )  OR  TITLE-ABS-KEY ( biolog* )  OR  TITLE-ABS-KEY ( supramolecul* ) ) ) )  AND NOT  ( TITLE-ABS-KEY ( nano* ) )

#13

( ( TITLE-ABS-KEY ( "quantum dot*" )  OR  TITLE-ABS-KEY ( "quantum well*" )  OR  TITLE-ABS-KEY ( "quantum wire*" )  OR  TITLE-ABS-KEY ( "molecul* motor*" )  OR  TITLE-ABS-KEY ( "molecul* ruler*" )  OR  TITLE-ABS-KEY ( "molecul* wir*" )  OR  TITLE-ABS-KEY ( "molecul* devic*" )  OR  TITLE-ABS-KEY ( "molecular engineering" )  OR  TITLE-ABS-KEY ( "molecular electronic*" )  OR  TITLE-ABS-KEY ( "single molecul*" )  OR  TITLE-ABS-KEY ( fulleren* )  OR  TITLE-ABS-KEY ( buckyball )  OR  TITLE-ABS-KEY ( buckminsterfullerene )  OR  TITLE-ABS-KEY ( "C-60" )  OR  TITLE-ABS-KEY ( c70 )  OR  TITLE-ABS-KEY ( c120 )  OR  TITLE-ABS-KEY ( swcnt )  OR  TITLE-ABS-KEY ( mwcnt )  OR  TITLE-ABS-KEY ( "coulomb blockad*" )  OR  TITLE-ABS-KEY ( bionano* )  OR  TITLE-ABS-KEY ( "Langmuir-blodgett" )  OR  TITLE-ABS-KEY ( coulombstaircase* )  OR  TITLE-ABS-KEY ( "coulomb-staircase" )  OR  TITLE-ABS-KEY ( "PDMS stamp*" )  OR  TITLE-ABS-KEY ( graphene )  OR  TITLE-ABS-KEY ( "dye-sensitized solar cell" )  OR  TITLE-ABS-KEY ( dssc )  OR  TITLE-ABS-KEY ( ferrofluid* )  OR  TITLE-ABS-KEY ( fullerid* )  OR  TITLE-ABS-KEY ( fullerinol* )  OR  TITLE-ABS-KEY ( fullerite* )  OR  TITLE-ABS-KEY ( fullerol* )  OR  TITLE-ABS-KEY ( fulleropyrrolidin* )  OR  TITLE-ABS-KEY ( "core-shell" )  OR  TITLE-ABS-KEY ( nems )  OR  TITLE-ABS-KEY ( "atom* scale" )  OR  TITLE-ABS-KEY ( "ballistic transport*" )  OR  TITLE-ABS-KEY ( "DNA comput*" )  OR  TITLE-ABS-KEY ( "porous silicon" )  OR  TITLE-ABS-KEY ( "a*fulleren*" )  OR  TITLE-ABS-KEY ( "b*fulleren*" )  OR  TITLE-ABS-KEY ( "c*fulleren*" )  OR  TITLE-ABS-KEY ( "d*fulleren*" )  OR  TITLE-ABS-KEY ( "e*fulleren*" )  OR  TITLE-ABS-KEY ( "f*fulleren*" )  OR  TITLE-ABS-KEY ( "g*fulleren*" )  OR  TITLE-ABS-KEY ( "h*fulleren*" )  OR  TITLE-ABS-KEY ( "i*fulleren*" )  OR  TITLE-ABS-KEY ( "j*fulleren*" )  OR  TITLE-ABS-KEY ( "k*fulleren*" )  OR  TITLE-ABS-KEY ( "l*fulleren*" )  OR  TITLE-ABS-KEY ( "m*fulleren*" )  OR  TITLE-ABS-KEY ( "n*fulleren*" )  OR  TITLE-ABS-KEY ( "o*fulleren*" )  OR  TITLE-ABS-KEY ( "p*fulleren*" )  OR  TITLE-ABS-KEY ( "q*fulleren*" )  OR  TITLE-ABS-KEY ( "r*fulleren*" )  OR  TITLE-ABS-KEY ( "s*fulleren*" )  OR  TITLE-ABS-KEY ( "t*fulleren*" )  OR  TITLE-ABS-KEY ( "u*fulleren*" )  OR  TITLE-ABS-KEY ( "v*fulleren*" )  OR  TITLE-ABS-KEY ( "w*fulleren*" )  OR  TITLE-ABS-KEY ( "x*fulleren*" )  OR  TITLE-ABS-KEY ( "y*fulleren*" )  OR  TITLE-ABS-KEY ( "z*fulleren*" ) )  OR  ( ( TITLE-ABS-KEY ( c60 ) )  AND NOT  ( TITLE-ABS-KEY ( c60 )  AND  TITLE-ABS-KEY ( steel ) ) ) )  AND NOT  ( TITLE-ABS-KEY ( nano* ) )

#14

( ( TITLE-ABS-KEY ( tem ) OR TITLE-ABS-KEY ( stm ) OR TITLE-ABS-KEY ( edx ) OR TITLE-ABS-KEY ( afm ) OR TITLE-ABS-KEY ( hrtem ) OR TITLE-ABS-KEY ( sem ) OR TITLE-ABS-KEY ( eels ) OR TITLE-ABS-KEY ( sers ) OR TITLE-ABS-KEY ( mfm ) OR TITLE-ABS-KEY ( uv-vis ) OR TITLE-ABS-KEY ( xps ) OR TITLE-ABS-KEY ( nsom ) OR TITLE-ABS-KEY ( "atom* force microscop*" ) OR TITLE-ABS-KEY ( "tunnel* microscop*" ) OR TITLE-ABS-KEY ( "scanning probe microscop*" ) OR TITLE-ABS-KEY ( "transmission electron microscop*" ) OR TITLE-ABS-KEY ( "scanning electron microscop*" ) OR TITLE-ABS-KEY ( "energy dispersive X-ray" ) OR TITLE-ABS-KEY ( "xray photoelectron*" ) OR TITLE-ABS-KEY ( "x-ray photoelectron" ) OR TITLE-ABS-KEY ( "electron energy loss spectroscop*" ) OR TITLE-ABS-KEY ( "enhanced raman-scattering" ) OR TITLE-ABS-KEY ( "surface enhanced raman scattering" ) OR TITLE-ABS-KEY ( "single molecule microscopy" ) OR TITLE-ABS-KEY ( "focused ion beam" ) OR TITLE-ABS-KEY ( "ellipsometry" ) OR TITLE-ABS-KEY ( "magnetic force microscopy" ) OR TITLE-ABS-KEY ( "UV-Visible Spectroscop*" ) OR TITLE-ABS-KEY ( "Ultraviolet-visible spectroscop*" ) OR TITLE-ABS-KEY ( "near field scanning optical microscop*" ) OR TITLE-ABS-KEY ( pebbles ) OR TITLE-ABS-KEY ( quasicrystal* ) OR TITLE-ABS-KEY ( "quasi-crystal*" ) ) AND ( TITLE-ABS-KEY ( monolayer* ) OR TITLE-ABS-KEY ( "mono-layer*" ) OR TITLE-ABS-KEY ( film* ) OR TITLE-ABS-KEY ( quantum* ) OR TITLE-ABS-KEY ( multilayer* ) OR TITLE-ABS-KEY ( "multi-layer*" ) OR TITLE-ABS-KEY ( array* ) ) ) AND NOT ( TITLE-ABS-KEY ( nano* ) )

#15

( ( TITLE-ABS-KEY ( biosensor* ) OR TITLE-ABS-KEY ( nems ) OR TITLE-ABS-KEY ( "sol gel*" ) OR TITLE-ABS-KEY ( "solgel*" ) OR TITLE-ABS-KEY ( dendrimer* ) OR TITLE-ABS-KEY ( "soft lithograph*" ) OR TITLE-ABS-KEY ( "molecular simul*" ) OR TITLE-ABS-KEY ( "molecular machin*" ) OR TITLE-ABS-KEY ( "molecular imprinting" ) OR TITLE-ABS-KEY ( "quantum effect*" ) OR TITLE-ABS-KEY ( "molecular sieve*" ) OR TITLE-ABS-KEY ( "surface energy" ) OR TITLE-ABS-KEY ( "mesoporous material*" ) OR TITLE-ABS-KEY ( "mesoporous silica" ) OR TITLE-ABS-KEY ( "porous silicon" ) OR TITLE-ABS-KEY ( "zeta potential" ) OR TITLE-ABS-KEY ( "epitax*" ) OR TITLE-ABS-KEY ( cnt ) OR TITLE-ABS-KEY ( "electron beam lithography" ) OR TITLE-ABS-KEY ( "e-beam lithography" ) OR TITLE-ABS-KEY ( "quantum size effect" ) OR TITLE-ABS-KEY ( "quantum device" ) OR TITLE-ABS-KEY ( "chemical vapor deposition" ) OR TITLE-ABS-KEY ( cvd ) OR TITLE-ABS-KEY ( "chemical vapour deposition" ) OR TITLE-ABS-KEY ( "surface plasmon resonance" ) OR TITLE-ABS-KEY ( "differential scanning calorimetry" ) ) AND ( TITLE-ABS-KEY ( monolayer* ) OR TITLE-ABS-KEY ( "mono-layer*" ) OR TITLE-ABS-KEY ( film* ) OR TITLE-ABS-KEY ( quantum* ) OR TITLE-ABS-KEY ( multilayer* ) OR TITLE-ABS-KEY ( "multi-layer*" ) OR TITLE-ABS-KEY ( array* ) ) ) AND NOT ( TITLE-ABS-KEY ( nano* ) )

FINAL QUERY (#11 OR #12 OR #13 OR #14 OR #15)

**Lexical queries for WOS**

#1

TS=nano*

#2

TS=((gadonano* OR glyconanopartic* OR heteronano* OR immunonanogold OR immunonanopartic*) OR (("self assembl*" OR "self organiz*" OR "directed assembl*") AND (monolayer* OR "mono-layer*" OR film* OR quantum* OR multilayer* OR "multi-layer*" OR array OR molecul* OR polymer* OR "co-polymer*" OR copolymer* OR mater* OR biolog* OR supramolecul*)))

#3

TS=(( polynanocrystal* OR subnanocluster* OR ultrananocrystal* OR ultrananodiamond*) OR (("self assembl*" OR "self organiz*" OR "directed assembl*") AND (monolayer* OR "mono-layer*" OR film* OR quantum* OR multilayer* OR "multi-layer*" OR array OR molecul* OR polymer* OR "co-polymer*" OR copolymer* OR mater* OR biolog* OR supramolecul*)))

#4

TS=("quantum dot*" OR "quantum well*" OR "quantum wire*" OR "molecul* motor*" OR "molecul* ruler*" OR "molecul* wir*" OR "molecul* devic*" OR "molecular engineering" OR "molecular electronic*" OR "single molecul*")

#5

TS=(fulleren* OR buckyball OR buckminsterfullerene OR "C-60" OR (c60 NOT(c60 AND steel)) OR c70 OR c120 OR swcnt OR mwcnt)

#6

TS=("coulomb blockad*" OR bionano* OR "Langmuir-blodgett" OR coulombstaircase* OR "coulomb-staircase" OR "PDMS stamp*" OR graphene OR "dye-sensitized solar cell" OR dssc OR ferrofluid*)

#7

TS=(fullerid* OR fullerinol* OR fullerite* OR fullerol* OR fulleropyrrolidin* OR "core-shell" OR nems OR "atom* scale" OR "ballistic transport*" OR "DNA comput*")

#8

TS=("porous silicon" OR "a*fulleren*" OR "b*fulleren*" OR "c*fulleren*" OR "d*fulleren*" OR "e*fulleren*" OR "f*fulleren*" OR "g*fulleren*" OR "h*fulleren*" OR "i*fulleren*")

#9

TS=("j*fulleren*" OR "k*fulleren*" OR "l*fulleren*" OR "m*fulleren*" OR "n*fulleren*" OR "o*fulleren*" OR "p*fulleren*" OR "q*fulleren*" OR "r*fulleren*" OR "s*fulleren*")

#10

TS=("t*fulleren*" OR "u*fulleren*" OR "v*fulleren*" OR "w*fulleren*" OR "x*fulleren*" OR "y*fulleren*" OR "z*fulleren*")

#11

TS=((tem OR stm OR edx OR afm OR hrtem OR sem) AND (monolayer* OR "mono-layer*" OR film* OR quantum* OR multilayer* OR "multi-layer*" OR array*))

#12

TS=((eels OR sers OR mfm OR uv-vis OR xps OR nsom) AND (monolayer* OR "mono-layer*" OR film* OR quantum* OR multilayer* OR "multi-layer*" OR array*))

#13

TS=(("atom* force microscop*" OR "tunnel* microscop*" OR "scanning probe microscop*" OR "transmission electron microscop*" OR "scanning electron microscop*") AND (monolayer* OR "mono-layer*" OR film* OR quantum* OR multilayer* OR "multi-layer*" OR array*))

#14

TS=(("energy dispersive X-ray" OR "xray photoelectron*" OR "x-ray photoelectron" OR "electron energy loss spectroscop*") AND (monolayer* OR "mono-layer*" OR film* OR quantum* OR multilayer* OR "multi-layer*" OR array*))

#15

TS=(("enhanced raman-scattering" OR "surface enhanced raman scattering" OR "single molecule microscopy" OR "focused ion beam") AND (monolayer* OR "mono-layer*" OR film* OR quantum* OR multilayer* OR "multi-layer*" OR array*))

#16

TS=(("ellipsometry" OR "magnetic force microscopy" OR "UV-Visible Spectroscop*") AND (monolayer* OR "mono-layer*" OR film* OR quantum* OR multilayer* OR "multi-layer*" OR array*))

#17

TS=(("Ultraviolet-visible spectroscop*" OR "near field scanning optical microscop*" OR pebbles OR quasicrystal* OR "quasi-crystal*") AND (monolayer* OR "mono-layer*" OR film* OR quantum* OR multilayer* OR "multi-layer*" OR array*))

#18

TS=((biosensor* OR nems OR "sol gel*" OR "solgel*" OR dendrimer*) AND (monolayer* OR "mono-layer*" OR film* OR quantum* OR multilayer* OR "multi-layer*" OR array*))

#19

TS=(("soft lithograph*" OR "molecular simul*" OR "molecular machin*" OR "molecular imprinting" OR "quantum effect*") AND (monolayer* OR "mono-layer*" OR film* OR quantum* OR multilayer* OR "multi-layer*" OR array*))

#20

TS=(("molecular sieve*" OR "surface energy" OR "mesoporous material*" OR "mesoporous silica" OR "porous silicon") AND (monolayer* OR "mono-layer*" OR film* OR quantum* OR multilayer* OR "multi-layer*" OR array*))

#21

TS=(("zeta potential" OR "epitax*" OR cnt OR "electron beam lithography" OR "e-beam lithography") AND (monolayer* OR "mono-layer*" OR film* OR quantum* OR multilayer* OR "multi-layer*" OR array*))

#22

TS=(("quantum size effect" OR "quantum device" OR "chemical vapor deposition" OR cvd OR "chemical vapour deposition" OR "surface plasmon resonance" OR "differential scanning calorimetry") AND (monolayer* OR "mono-layer*" OR film* OR quantum* OR multilayer* OR "multi-layer*" OR array*))

#23

TS=(nanoa OR nanoacalles OR  nanoagraylea OR  nanoalga* OR nanoapiculatum OR nanoarchaeaor OR nanoarchaeota OR nanoarchaeum OR nanoastegotherium OR "nano*aryote*")

#24

TS=(nanobacteri* OR nanobagrus OR nanobalcis OR nanobaris OR nanobates OR nanobatinae OR nanobius OR nanobryaceae OR nanobryoides OR nanobuthus)

#25

TS=(nano-bible OR nanocalcar OR nanocambridgea OR nanocapillare OR  nanocapillary OR nanocarpa OR nanocarpum OR nanocarpus OR nanocassiope OR nanocavia OR nanocephalum OR nanocheirodon)

#26

TS=(nanochilina OR nanochilus OR nanochitina OR nanochlaenius OR nanochlorum OR nanochoerus OR nanochromis OR nanochromosom* OR nanochrysopa OR nanochthonius)

#27

TS=(nanocixius OR nanocladius OR nanoclarelia OR nanoclavelia OR nanoclimacium OR nanoclymenia OR nanocnide OR nanocochlea OR nanocolletes OR nanocondylodesmus)

#28

TS=(nanocopia OR nanocoquimba OR nanocrinus OR nanoctenus OR nanocthispa OR nanocuridae OR nanocurie OR nano-curie OR nanocuris OR nanocyclopia OR nanocynodon OR nanocythere)

#29

TS=(nanodacna OR nanodactylus OR nanodamon OR nanodea OR nanodealbata OR nanodectes OR nanodella OR nanodelphys OR nanodendron OR nanodes)

#30

TS=(nanodiaparsis OR nanodiaptomus OR nanodidelphys OR nanodiella OR nanodiodes OR nanodiplosis OR nanodiscus OR nanodisticha OR nanodromia OR nanodynerus)

#31

TS=(nanoesi OR nanofauna* OR nanofila OR "nano-filt*" OR nanofilt* OR nanoflagel* OR nanoflow OR "nano-flow" OR nanofilidae OR Nanog OR Nanog1)

#32

TS=(Nanog2 OR Nanogalathea OR Nanogeterotroph* OR Nanoglobum OR Nanoglossa OR Nanognathia OR Nanognathus OR Nanogomphodon OR Nanogona OR Nanogonalos)

#33

TS=(Nanogorgon OR Nanogram* OR "nano-gram*" OR Nanogramma OR Nanograptus OR Nanogyra OR Nanogyrini OR Nanohalus OR Nanohammus OR Nanohemicera)

#34

TS=(Nanoheterotroph* OR Nanohystrix OR Nanoides OR Nanoini OR Nanojapyx OR Nanokerala OR Nanokelvin* OR Nanokermes OR Nanola OR Nanolachesilla)

#35

TS=(Nanolania OR Nanolauthia OR NanoLC* OR "nano-LC*" OR Nanolestes OR Nanolichus OR "nano-liter*" OR "nano-litre*" OR Nanoliter* OR Nanolitre* OR Nanolobus OR Nanoloricida)

#36

TS=(Nanolpium OR Nanolumen OR Nanomaja OR Nanomantinae OR Nanomantini OR Nanomantis OR Nanomeli* OR Nanomelon OR Nanomermis OR Nanomerus OR Nanomeryx OR Nanometa)

#37

TS=("nano-meter*" OR Nanometidae OR Nanometinae OR Nanometra OR "nano-metre*" OR "nm" OR Nanometer* OR Nanometre* OR Nanomia OR Nanomias OR Nanomicrophyes)

#38

TS=(Nanomilleretta OR Nanomimus OR Nanomis OR Nanomitra OR Nanomitriella OR Nanomitriopsis OR Nanomitus OR Nanomol* OR "nano-molar" OR "nano-mole*")

#39

TS=(Nanomutilinae OR Nanomutilla OR Nanomyces OR Nanomyina OR Nanomyrmacyba OR Nanomyrme OR Nanomys OR Nanomysis OR Nanomysmena)

#40

TS=(Nanonaucoris OR Nanonavis OR Nanoneis OR Nanonemoura OR Nanonocticolus OR Nanonycteris OR "Nanook of the North" OR nanopachyiulus OR nanopagurus OR nanopareia)

#41

TS=(nanoparia OR nanopatula OR nanopennatum OR nanoperla OR nanophareus OR nanophemera OR nanophtalm* OR nanophya OR nanophydes OR nanophydinae)

#42

TS=(nanophydini OR nanophyes OR nanophyetinae OR nanophyetus OR nanophyidae OR nanophyinae OR nanophyini OR nanophylla OR nanophylliini OR nanophyllium OR nanophyllum OR nanophyllus OR nanophytes)

#43

TS=(nanophyti OR nanophyto* OR nanopilumnus OR nanopitar OR nanoplagia OR nanoplax OR nanoplaxes OR nanoplectrus OR nanoplinthisus OR nanopodella OR nanopodellus OR nanopolymorphum)

#44

TS=(nanopolystoma OR nanopria OR nanoprotist* OR nanops OR nanopsallus OR nanopsis OR nanopsocetae OR nanopsocus OR nanopterodectes OR nanopterum)

#45

TS=(nanoptilium OR nanopus OR nanopyxis OR nanoqia OR nanoqsunquak OR nanor OR nanorafonus OR nanorana OR nanoraphidia OR nanorchestes OR nanorchestidae OR nanorhamphus)

#46

TS=(nanorhathymus OR nanorhopaea OR nanorrhacus OR nanorrhynchus OR Nanorthidae OR Nanorthis OR Nanos OR Nanosalicium OR Nanosatellite* OR Nanosauridae OR Nanosaurus)

#47

TS=(Nanoschema OR Nanoschetus OR Nanoscydmus OR Nanoscypha OR "nano-second*" OR Nanosecond* OR Nanosella OR Nanosellini OR Nanoserranus OR Nanosesarma OR Nanosetus)

#48

TS=(Nanosilene OR "nano-SIMS" OR Nanosiren OR Nanosius OR Nanosmia OR Nanosmilus OR Nanosomus OR Nanospadix OR Nanospathulatum OR Nanospira)

#49

TS=(Nanospondylus OR Nanospora OR Nanospray* OR "nano-spray*" OR Nanosteatoda OR Nanostellata OR Nanostictis)

#50

TS=(Nanostoma OR Nanostomus OR Nanostrangalia OR Nanostrea OR Nanostreptus OR Nanosura OR Nanosylvanella OR Nanotagalus OR Nanotanaupodus OR Nanotaphus)

#51

TS=(Nanotermitodius OR Nanothamnus OR Nanothecioidea OR Nanothecium OR Nanothinophilus OR Nanothrips OR Nanothyris OR Nanotitan OR Nanotitanops)

#52

TS=(Nanotopsis OR Nanotragulus OR Nanotragus OR Nanotrema OR Nanotrephes OR Nanotrigona OR Nanotriton OR Nanotrombium OR Nanotyrannus OR Nanoviridae)

#53

TS=(Nanovirus OR Nanowana OR Nanowestratia OR Nanoxylocopa OR Nano2 OR Nano3 OR plankton* OR "N*plankton*" OR "m*plankton*" OR "b*plankton*" OR "p*plankton*" OR "z*plankton*")

FINAL QUERY (#1 OR #2 OR #3 OR #4 OR #5 OR #6 OR #7 OR #8 OR #9 OR #10 OR #11 OR #12 OR #13 OR #14 OR #15 OR #16 OR #17 OR #18 OR #19 OR #21 OR #22) NOT (#23 OR #24 OR #25 OR #26 OR #27 OR #28 OR #29 OR #30 OR #31 OR #32 OR #33 OR #34 OR #35 OR #36 OR #37 OR #38 OR #39 OR #40 OR #41 OR #42 OR #43 OR #44 OR #45 OR #46 OR #47 OR #48 OR #49 OR #50 OR #51 OR #52 OR #53)

**Appendix 3. Journals list with nano prefix in their titles retrieved from WoS and Scopus**

| *Journals list with nano prefix in their titles* |
| --- |
| ACS NANO* |
| ADVANCES IN NATURAL SCIENCES NANOSCIENCE AND NANOTECHNOLOGY |
| APPLIED NANOSCIENCE |
| ARTIFICIAL CELLS NANOMEDICINE AND BIOTECHNOLOGY |
| BEILSTEIN JOURNAL OF NANOTECHNOLOGY |
| BIOINSPIRED BIOMIMETIC AND NANOBIOMATERIALS |
| CANCER NANOTECHNOLOGY |
| CHEMNANOMAT |
| CURRENT NANOSCIENCE |
| DIGEST JOURNAL OF NANOMATERIALS AND BIOSTRUCTURES |
| E JOURNAL OF SURFACE SCIENCE AND NANOTECHNOLOGY |
| ENVIRONMENTAL NANOTECHNOLOGY MONITORING AND MANAGEMENT |
| ENVIRONMENTAL SCIENCE NANO |
| FULLERENES NANOTUBES AND CARBON NANOSTRUCTURES |
| IEEE NANOTECHNOLOGY MAGAZINE |
| IEEE TRANSACTIONS ON NANOBIOSCIENCE |
| IEEE TRANSACTIONS ON NANOTECHNOLOGY |
| IET NANOBIOTECHNOLOGY |
| INORGANIC AND NANO METAL CHEMISTRY |
| INTEGRATIVE BIOLOGY – QUANTITATIVE BIOSCIENCE FROM NANO TO MACRO |
| INTERNATIONAL JOURNAL OF BIOMEDICAL NANOSCIENCE AND NANOTECHNOLOGY |
| INTERNATIONAL JOURNAL OF NANO AND BIOMATERIALS |
| INTERNATIONAL JOURNAL OF NANOELECTRONICS AND MATERIALS |
| INTERNATIONAL JOURNAL OF NANOMANUFACTURING |
| INTERNATIONAL JOURNAL OF NANOMEDICINE |
| INTERNATIONAL JOURNAL OF NANOPARTICLES |
| INTERNATIONAL JOURNAL OF NANOSCIENCE |
| INTERNATIONAL JOURNAL OF NANOTECHNOLOGY |
| INTERNATIONAL JOURNAL OF SMART AND NANO MATERIALS |
| JOURNAL OF BIOMEDICAL NANOTECHNOLOGY |
| JOURNAL OF BIONANOSCIENCE |
| JOURNAL OF COMPUTATIONAL AND THEORETICAL NANOSCIENCE |
| JOURNAL OF EXPERIMENTAL NANOSCIENCE |
| JOURNAL OF LASER MICRO NANOENGINEERING |
| JOURNAL OF MICRO NANOLITHOGRAPHY MEMS AND MOEMS |
| JOURNAL OF MICRO-BIO ROBOTICS |
| JOURNAL OF NANO AND ELECTRONIC PHYSICS |
| JOURNAL OF NANO RESEARCH |
| JOURNAL OF NANOBIOTECHNOLOGY |
| JOURNAL OF NANOELECTRONICS AND OPTOELECTRONICS |
| JOURNAL OF NANOMATERIALS |
| JOURNAL OF NANOMECHANICS AND MICROMECHANICS |
| JOURNAL OF NANOMEDICINE AND NANOTECHNOLOGY |
| JOURNAL OF NANOPARTICLE RESEARCH |

| *Journals list with nano prefix in their titles* |
|---|
| **JOURNAL OF NANOPHOTONICS** |
| **JOURNAL OF NANOSCIENCE AND NANOTECNOLOGY** |
| **JOURNAL OF NANOTECHNOLOGY** |
| **JOURNAL OF VACUUM SCIENCE AND TECHNOLOGY B NANOTECHNOLOGY AND MICROELECTRONICS** |
| **MICRO AND NANO LETTERS** |
| **MICRO AND NANOSYSTEMS** |
| **MICROFLUIDICS AND NANOFLUIDICS** |
| **MICROSYSTEM TECHNOLOGIES-MICRO-AND NANOSYSTEMS-INFORMATION STORAGE AND PROCESSING SYSTEMS** |
| **NAMI JISHU YU JINGMI GONGCHENG NANOTECHNOLOGY AND PRECISION ENGINEERING** |
| **NANO** |
| **NANO BIOMEDICINE** |
| **NANO BIOMEDICINE AND ENGINEERING** |
| **NANO COMMUNICATION NETWORKS** |
| **NANO ENERGY** |
| **NANO LETTER\* °** |
| **NANO LIFE** |
| **NANO MICRO LETTERS** |
| **NANO RESEARCH** |
| **NANO STRUCTURES AND NANO OBJECTS** |
| **NANO TODAY** |
| **NANOETHICS** |
| **NANOIMPACT** |
| **NANOMATERIALS** |
| **NANOMATERIALS AND NANOTECHNOLOGY** |
| **NANOMEDICINE** |
| **NANOMEDICINE NANOTECHNOLOGY BIOLOGY AND MEDICINE** |
| **NANOPHOTONICS** |
| **NANOSCALE AND MICROSCALE THERMOPHYSICAL ENGINEERING** |
| **NANOSCALE RESEARCH LETTERS** |
| **NANOSCALE\*** |
| **NANOSCIENCE AND NANOTECHNOLOGY ASIA** |
| **NANOSCIENCE AND NANOTECHNOLOGY LETTERS** |
| **NANOSCIENCE AND TECHNOLOGY: AN INTERNATIONAL JOURNAL (INTERNATIONAL JOURNAL OF NANOMECHANICS SCIENCE AND TECHNOLOGY** |
| **NANOTECHNOLOGIES IN RUSSIA** |
| **NANOTECHNOLOGY** |
| **NANOTECHNOLOGY PERCEPTIONS** |
| **NANOTECHNOLOGY REVIEWS** |
| **NANOTECHNOLOGY SCIENCE AND APPLICATIONS** |
| **NANOTOXICOLOGY** |
| **NATURE NANOTECHNOLOGY** |
| **OPENNANO** |
| **PHOTONICS AND NANOSTRUCTURES FUNDAMENTALS AND APPLICATIONS** |
| **PHYSICA E LOW DIMENSIONAL SYSTEMS AND NANOSTRUCTURES** |
| **PRECISION ENGINEERING-JOURNAL OF THE INTERNATIONAL SOCIETIES FOR PRECISION** |

| *Journals list with nano prefix in their titles* |
| --- |
| **ENGINEERING AND NANOTECHNOLOGY** |
| **PROCEEDINGS OF THE INSTITUTION OF MECHANICAL ENGINEERS, PART N: JOURNAL OF NANOMATERIALS, NANOENGENEIRING AND NANOSYSTEMS** |
| **RECENT PATENTS ON NANOTECHNOLOGY** |
| **WILEY INTERDISCIPLINARY REVIEWS NANOMEDICINE AND NANOBIOTECHNOLOGY** |

° Identified by scree test from NST WoS data set
*Identified by scree test from NST Scopus data set

**Appendix 4. Matching journals between all intersections**

| Title | ISSN | Thematic area | A₁ | A₂ | A₃ | Scopus | Wos | Precision |
|---|---|---|---|---|---|---|---|---|
| ACS APPLIED MATERIALS & INTERFACES | 1944-8244 | Materials science (Nanomaterials) | x | x | x | Y | Y | 1 |
| ACS NANO | 1936-0851 | Nanoscience & Nanotechnology | x | x | x | Y | Y | 1 |
| ACS PHOTONICS | 2330-4022 | Nanophotonics | x | x |   | Y | Y | 1 |
| ADVANCED MATERIALS | 0935-9648 | Materials science (Nanomaterials) | x | x | x | Y | Y | 1 |
| ADVANCED SCIENCE | 2198-3844 | Interdisciplinary (Materials science / Physics / Chemistry /Medicine / Engineering / Life Science) | x | x |   | Y | Y | 1 |
| ADVANCES IN NATURAL SCIENCES-NANOSCIENCE AND NANOTECHNOLOGY | 2043-6262 | Nanoscience & Nanotechnology | x | x | x | Y | Y | 1 |
| APPLIED NANOSCIENCE | 2190-5509 | Nanoscience & Nanotechnology | x | x | x | N | Y | 1 |
| BEILSTEIN JOURNAL OF NANOTECHNOLOGY | 2190-4286 | Nanoscience & Nanotechnology | x | x | x | Y | Y | 1 |
| CHEMNANOMAT | 2199-692X | Materials science (Nanomaterials) | x | x | x | N | Y | 1 |
| CURRENT NANOSCIENCE | 1573-4137 | Nanoscience & Nanotechnology | x | x | x | Y | Y | 1 |
| DIGEST JOURNAL OF NANOMATERIALS AND BIOSTRUCTURES | 1842-3582 | Nanomaterials | x |   | x | Y | Y | 1 |
| ENVIRONMENTAL SCIENCE-NANO | 2051-8153 | Nanoenvironment | x | x | x | Y | Y | 1 |
| FULLERENES NANOTUBES AND CARBON NANOSTRUCTURES | 1536-383X | Carbon materials | x | x | x | Y | Y | 1 |
| IEEE TRANSACTIONS ON NANOBIOSCIENCE | 1536-1241 | Nanomedicine | x | x | x | Y | Y | 1 |
| IEEE TRANSACTIONS ON NANOTECHNOLOGY | 1536-125X | Nanoscience & Nanotechnology | x | x | x | Y | Y | 1 |
| IET NANOBIOTECHNOLOGY | 1751-8741 | Nanobiotechnology | x |   | x | Y | Y | 1 |
| INORGANIC AND NANO-METAL CHEMISTRY | 1553-3174 | Nanochemistry (Inorganic & nuclear chemistry) | x |   | x | Y | Y | 1 |
| INTERNATIONAL JOURNAL OF NANOMEDICINE | 1176-9114 | Nanomedicine | x |   | x | Y | Y | 1 |
| INTERNATIONAL JOURNAL OF NANOTECHNOLOGY | 1475-7435 | Nanoscience & Nanotechnology | x | x | x | Y | Y | 1 |
| JOURNAL OF BIOMEDICAL NANOTECHNOLOGY | 1550-7033 | Nanomedicine & Nanobiotechnology | x | x | x | Y | Y | 1 |
| JOURNAL OF COMPUTATIONAL AND THEORETICAL NANOSCIENCE | 1546-1955 | Computational nanoscience |   | x | x | Y | Y | 1 |
| JOURNAL OF EXPERIMENTAL NANOSCIENCE | 1745-8080 | Nanoscience & Nanotechnology | x | x | x | Y | Y | 1 |
| JOURNAL OF LASER MICRO NANOENGINEERING | 1880-0688 | Nanoengineering | x |   | x | Y | Y | 1 |
| JOURNAL OF MATERIALS CHEMISTRY A | 2050-7488 | Materials science (Energy) |   | x | x | Y | Y | 0.5 |
| JOURNAL OF MATERIALS CHEMISTRY B | 2050-7518 | Materials science (Medicine and biology) |   | x | x | Y | Y | 0.5 |
| JOURNAL OF MATERIALS CHEMISTRY C | 2050-7526 | Materials science (Optic, magnetic and electronic devices) |   | x | x | Y | Y | 0.5 |
| JOURNAL OF MICRO-NANOLITHOGRAPHY MEMS AND MOEMS | 1932-5150 | Nanoengineering (Micro and nano and micro-opto | x |   | x | Y | Y | 1 |

| Title | ISSN | Thematic area | A₁ | A₂ | A₃ | Scopus | Wos | Precision |
|---|---|---|---|---|---|---|---|---|
| | | electromechanical systems) | | | | | | |
| JOURNAL OF NANO RESEARCH | 1662-5250 | Nanoscience & Nanotechnology | x | x | x | Y | Y | 1 |
| JOURNAL OF NANOBIOTECHNOLOGY | 1477-3155 | Nanobiotechnology | x | x | x | Y | Y | 1 |
| JOURNAL OF NANOELECTRONICS AND OPTOELECTRONICS | 1555-130X | Nanophotonics | x | x | x | Y | Y | 1 |
| JOURNAL OF NANOMATERIALS | 1687-4110 | Nanomaterials | x | x | x | Y | Y | 1 |
| JOURNAL OF NANOPARTICLE RESEARCH | 1388-0764 | Nanoparticles | x | x | x | Y | Y | 1 |
| JOURNAL OF NANOPHOTONICS | 1934-2608 | Nanophotonics | x | x | x | Y | Y | 1 |
| JOURNAL OF NANOSCIENCE AND NANOTECHNOLOGY | 1533-4880 | Nanoscience & Nanotechnology | x | x | x | Y | Y | 1 |
| JOURNAL OF PHYSICAL CHEMISTRY C | 1932-7447 | Nanochemistry (Physical chemistry) | x | x | x | Y | Y | 1 |
| JOURNAL OF PHYSICAL CHEMISTRY LETTERS | 1948-7185 | Chemistry (Physical chemistry) | x | x | | Y | Y | 1 |
| JOURNAL OF VACUUM SCIENCE & TECHNOLOGY B | 2166-2746 | Nanophysics | x | | x | Y | Y | 1 |
| MATERIALS EXPRESS | 2158-5849 | Nanomaterials | x | x | | Y | Y | 1 |
| MATERIALS SCIENCE AND ENGINEERING A-STRUCTURAL MATERIALS PROPERTIES MICROSTRUCTURE AND PROCESSING | 0921-5093 | Materials science (Nanomaterials) | x | | x | Y | Y | 1 |
| MICRO & NANO LETTERS | 1750-0443 | Nanoscience & Nanotechnology | x | x | x | Y | Y | 1 |
| MICROFLUIDICS AND NANOFLUIDICS | 1613-4982 | Nanofluidics | x | | x | Y | Y | 1 |
| MICROSYSTEM TECHNOLOGIES-MICRO- AND NANOSYSTEMS-INFORMATION STORAGE AND PROCESSING SYSTEMS | 0946-7076 | Nanoengineering | x | | x | Y | Y | 1 |
| NANO | 1793-2920 | Nanoscience & Nanotechnology | x | x | x | Y | Y | 1 |
| NANO COMMUNICATION NETWORKS | 1878-7789 | Nanoscale communication and networking | x | x | x | Y | Y | 1 |
| NANO ENERGY | 2211-2855 | Nanoenergy | x | x | x | Y | Y | 1 |
| NANO LETTERS | 1530-6984 | Nanoscience & Nanotechnology | x | x | x | Y | Y | 1 |
| NANO RESEARCH | 1998-0124 | Nanoscience & Nanotechnology | x | x | x | Y | Y | 1 |
| NANO TODAY | 1748-0132 | Nanoscience & Nanotechnology | x | x | x | Y | Y | 1 |
| NANO-MICRO LETTERS | 2150-5551 | Nanoscience & Nanotechnology | x | x | x | Y | Y | 1 |
| NANOMATERIALS | 2079-4991 | Nanomaterials | x | x | x | N | Y | 1 |
| NANOMATERIALS AND NANOTECHNOLOGY | 1847-9804 | Nanomaterials | x | x | x | Y | Y | 1 |
| NANOMEDICINE | 1743-5889 | Nanomedicine | x | x | x | Y | Y | 1 |
| NANOMEDICINE-NANOTECHNOLOGY BIOLOGY AND MEDICINE | 1549-9634 | Nanomedicine & Nanobiotechnology | x | | x | Y | Y | 1 |
| NANOPHOTONICS | 2192-8614 | Nanophotonics | x | x | x | Y | Y | 1 |
| NANOSCALE | 2040-3364 | Nanoscience & Nanotechnology | x | x | x | Y | Y | 1 |
| NANOSCALE AND MICROSCALE THERMOPHYSICAL ENGINEERING | 1556-7265 | Nanoengineering | x | | x | Y | Y | 1 |
| NANOSCALE RESEARCH LETTERS | 1931-7573 | Nanoscience & Nanotechnology | x | x | x | Y | Y | 1 |

| Title | ISSN | Thematic area | A₁ | A₂ | A₃ | Scopus | Wos | Precision |
|---|---|---|---|---|---|---|---|---|
| NANOSCIENCE AND NANOTECHNOLOGY LETTERS | 1941-4900 | Nanoscience & Nanotechnology | x | x | x | Y | Y | 1 |
| NANOTECHNOLOGY | 0957-4484 | Nanoscience & Nanotechnology | x | x | x | Y | Y | 1 |
| NANOTECHNOLOGY REVIEWS | 2191-9089 | Nanoscience & Nanotechnology | x | x | x | Y | Y | 1 |
| NANOTOXICOLOGY | 1743-5390 | Nanotoxicology | x | x | x | Y | Y | 1 |
| NATURE NANOTECHNOLOGY | 1748-3387 | Nanoscience & Nanotechnology | x | x | x | Y | Y | 1 |
| PARTICLE & PARTICLE SYSTEMS CHARACTERIZATION | 0934-0866 | Nanoparticles | x | x |  | Y | Y | 1 |
| PHOTONICS AND NANOSTRUCTURES-FUNDAMENTALS AND APPLICATIONS | 1569-4410 | Nanophotonics | x |  | x | Y | Y | 1 |
| PHYSICA E-LOW-DIMENSIONAL SYSTEMS & NANOSTRUCTURES | 1386-9477 | Nanomaterials | x | x | x | Y | Y | 1 |
| PLASMONICS | 1557-1955 | Nanophotonics | x | x |  | Y | Y | 1 |
| PRECISION ENGINEERING-JOURNAL OF THE INTERNATIONAL SOCIETIES FOR PRECISION ENGINEERING AND NANOTECHNOLOGY | 0141-6359 | Nanoengineering | x |  | x | Y | Y | 1 |
| RECENT PATENTS ON NANOTECHNOLOGY | 1872-2105 | Nanopatens | x | x | x | Y | Y | 1 |
| SMALL | 1613-6810 | Nanoscience & Nanotechnology | x | x |  | Y | Y | 1 |
| WILEY INTERDISCIPLINARY REVIEWS-NANOMEDICINE AND NANOBIOTECHNOLOGY | 1939-0041 | Nanomedicine & Nanobiotechnology | x | x | x | Y | Y | 1 |

x = Intersection (cover in the approach)
Y = Included
N = No included

# Appendix 5. Journals with null intersection

| Title | ISSN | Thematic area | A₁ | A₂ | A₃ | Scopus | Wos | Precision |
|---|---|---|---|---|---|---|---|---|
| 2D MATERIALS | 2053-1583 | Nanomaterials (2D materials) | | x | | Y | Y | 0.5 |
| ACM JOURNAL ON EMERGING TECHNOLOGIES IN COMPUTING SYSTEMS | 1550-4832 | Computational nanoscience | x | | | Y | Y | 1 |
| ACS ENERGY LETTERS | 2380-8195 | Energy | x | | | N | Y | 0.5 |
| ACS SENSORS | 2379-3694 | Sensor science | x | | | N | Y | 0 |
| ACTA MATERIALIA | 1359-6454 | Materials science | | | x | Y | Y | 0.5 |
| ADVANCED ELECTRONIC MATERIALS | 2199-160X | Materials science | x | | | Y | Y | 0 |
| ADVANCED ENERGY MATERIALS | 1614-6832 | Materials science (Energy) | | x | | Y | Y | 0.5 |
| ADVANCED FUNCTIONAL MATERIALS | 1616-301X | Materials science | x | | | Y | Y | 0.5 |
| ADVANCED HEALTHCARE MATERIALS | 2192-2640 | Materials science | x | | | Y | Y | 0 |
| ADVANCED MATERIALS INTERFACES | 2196-7350 | Materials science | | x | | Y | Y | 0.5 |
| ADVANCED OPTICAL MATERIALS | 2195-1071 | Material science / Optical science | | x | | Y | Y | 0.5 |
| AIP ADVANCES | 2158-3226 | Physics | x | | | Y | Y | 0 |
| ANALYTICAL CHEMISTRY | 0003-2700 | Chemistry | | | x | Y | Y | 0 |
| ANGEWANDTE CHEMIE- | 1433-7851 | Physics | | | x | Y | Y | 0 |
| APL MATERIALS | 2166-532X | Material science | x | | | Y | Y | 1 |
| APPLIED PHYSICS LETTERS | 0003-6951 | Physics | | | x | Y | Y | 0.5 |
| ARTIFICIAL CELLS NANOMEDICINE AND BIOTECHNOLOGY | 2169-1401 | Nanomedicine & Nanobiotechnology | | | x | Y | Y | 1 |
| BIOINSPIRED BIOMIMETIC AND NANOBIOMATERIALS | 2045-9858 | Nanomaterials | | | x | Y | Y | 1 |
| BIOMATERIALS | 0142-9612 | Materials science | | | x | Y | Y | 1 |
| BIOMEDICAL MICRODEVICES | 1387-2176 | Nanoengineering (Biomedical Micro/Nano electromechanical systems) | x | | | Y | Y | 0.5 |
| BIOMICROFLUIDICS | 1932-1058 | Micro and nanofluidics | x | | | Y | Y | 0.5 |
| BIOSENSORS & BIOELECTRONICS | 0956-5663 | Sensor science | x | | | Y | Y | 0.5 |
| CANCER NANOTECHNOLOGY | 1868-6958 | Nanomedicine | | | x | Y | N | 1 |
| CARBON | 0008-6223 | Carbon materials | | x | | Y | Y | 1 |
| CARBON LETTERS | 1976-4251 | Carbon materilas | | x | | Y | Y | 1 |
| CHEMELECTROCHEM | 2196-0216 | Chemistry (Physical chemistry: | | x | | Y | Y | 0.5 |
| CHEMICAL COMMUNICATIONS | 1359-7345 | Chemistry | | | x | Y | Y | 0 |
| CHEMICAL SOCIETY REVIEWS | 0306-0012 | Chemistry | | | x | Y | Y | 0 |
| CHEMISTRY OF MATERIALS | 0897-4756 | Chemistry | | | x | Y | Y | 0 |
| CONTRAST MEDIA & MOLECULAR IMAGING | 1555-4309 | Biochemistry (Molecular biology: Magneting Resonance Imaging) | | x | | Y | Y | 1 |
| DIAMOND AND RELATED MATERIALS | 0925-9635 | Carbon materials | | x | | Y | Y | 1 |
| E JOURNAL OF SURFACE SCIENCE AND NANOTECHNOLOGY | 1348-0391 | Materials science | | | x | Y | N | 1 |
| ELECTROCHEMISTRY COMMUNICATIONS | 1388-2481 | Chemistry (Physical chemistry: | | x | | Y | Y | 0.5 |

| Title | ISSN | Thematic area | A₁ | A₂ | A₃ | Scopus | Wos | Precision |
|---|---|---|---|---|---|---|---|---|
| ELECTROCHIMICA ACTA | 0013-4686 | Chemistry (Physical chemistry: | | x | | Y | Y | 0 |
| ENERGY & ENVIRONMENTAL | 1754-5692 | Environmental science | | x | | Y | Y | 0 |
| ENVIRONMENTAL NANOTECHNOLOGY MONITORING | 2215-1532 | Nanoenvironment | | | x | Y | N | |
| GOLD BULLETIN | 0017-1557 | Materials science (Gold) | | x | | Y | Y | 0 |
| IEEE NANOTECHNOLOGY MAGAZINE | 1932-4510 | Nanoengineering | | | x | Y | N | |
| INTEGRATIVE BIOLOGY | 1757-9694 | Biology | | | x | Y | Y | 0.5 |
| INTERNATIONAL JOURNAL OF BIOMEDICAL NANOSCIENCE AND NANOTECHNOLOGY | 1756-0799 | Nanomedicine & Nanobiotechnology | | | x | Y | N | 1 |
| INTERNATIONAL JOURNAL OF NANO AND BIOMATERIALS | 1752-8933 | Nanomaterials | | | x | Y | N | 1 |
| INTERNATIONAL JOURNAL OF NANOELECTRONICS AND MATERIALS | 1985-5761 | Nanoengineering (Electronics) | | | x | Y | N | 1 |
| INTERNATIONAL JOURNAL OF NANOMANUFACTURING | 1746-9392 | Nanoengineering (Manufacturing) | | | x | Y | N | 1 |
| INTERNATIONAL JOURNAL OF NANOPARTICLES | 1753-2507 | Nanoparticles | | | x | Y | N | 1 |
| INTERNATIONAL JOURNAL OF NANOSCIENCE | 0219-581X | Nanoscience & Nanotechnology | | | x | Y | N | 1 |
| INTERNATIONAL JOURNAL OF SMART AND NANO MATERIALS | 1947-5411 | Nanomaterials | | | x | Y | N | 1 |
| JOURNAL OF APPLIED PHYSICS | 0021-8979 | Physics | | | x | Y | Y | 0.5 |
| JOURNAL OF BIONANOSCIENCE | 1557-7910 | Nanobiotechnology | | | x | Y | N | 1 |
| JOURNAL OF MICRO-BIO ROBOTICS | 2194-6418 | Engineering (Robotics) | | | x | Y | N | 1 |
| JOURNAL OF MICROELECTROMECHANICAL SYSTEMS | 1057-7157 | Nanoengineering (Micro / Nano electromechanical systems) | x | | | Y | Y | 0 |
| JOURNAL OF MICROMECHANICS AND MICROENGINEERING | 0960-1317 | Nanoengineering (Micro / Nano electromechanical systems) | x | | | Y | Y | 1 |
| JOURNAL OF NANO AND ELECTRONIC PHYSICS | 2077-6772 | Nanoengineering (Electronics) | | | x | Y | N | 1 |
| JOURNAL OF NANOMECHANICS AND MICROMECHANICS | 2153-5434 | Nanoengineering (Mechanics) | | | x | Y | N | 1 |
| JOURNAL OF NANOMEDICINE AND NANOTECHNOLOGY | 2157-7439 | Nanomedicine | | | x | Y | N | 1 |
| JOURNAL OF NANOTECHNOLOGY | 1687-9503 | Nanoscience & Nanotechnology | | | x | Y | N | 1 |
| JOURNAL OF PHYSICAL CHEMISTRY B | 1520-6106 | Chemistry (Physical | | | x | Y | Y | 0 |
| JOURNAL OF SOLID STATE ELECTROCHEMISTRY | 1432-8488 | Chemistry (Physical chemistry: | | x | | Y | Y | 0 |
| JOURNAL OF THE AMERICAN CHEMICAL SOCIETY | 0002-7863 | Chemistry | | | x | Y | Y | 0 |
| LAB ON A CHIP | 1473-0197 | Chemistry (Micro and nano electromechanical systems - Lab on a chip) | x | | | Y | Y | 1 |

| Title | ISSN | Thematic area | A₁ | A₂ | A₃ | Scopus | Wos | Precision |
|---|---|---|---|---|---|---|---|---|
| LANGMUIR | 0743-7463 | Chemistry | | | x | Y | Y | 0.5 |
| MACROMOLECULES | 0024-9297 | Chemistry (Polymer chemistry - | | | x | Y | Y | 0.5 |
| MATERIALS RESEARCH EXPRESS | 2053-1591 | Materias science | | x | | Y | Y | 1 |
| MICRO AND NANOSYSTEMS | 1876-4029 | Nanoengineering (Micro / Nano electromechanical systems) | | | x | Y | N | |
| MICROELECTRONIC ENGINEERING | 0167-9317 | Nanoengineering (Micro / Nano electromechanical systems) | x | | | Y | Y | 1 |
| MICROELECTRONICS JOURNAL | 0959-8324 | Nanoengineering (Microelectromechanical systems) | x | | | Y | Y | 0 |
| MICROELECTRONICS RELIABILITY | 0026-2714 | Nanoengineering (Micro and nano electromechanical systems) | x | | | Y | Y | 0 |
| MICROMACHINES | 2072-666X | Nanoengineering (Micro and nano electromechanical systems) | x | | | Y | Y | 1 |
| MICROPOROUS AND MESOPOROUS MATERIALS | 1387-1811 | Porous materials (Nanoscale porous) | x | | | Y | Y | 1 |
| NAMI JISHU YU JINGMI GONGCHENG NANOTECHNOLOGY AND PRECISION ENGINEERING | 1672-6030 | Nanoengineering | | | x | Y | N | 1 |
| NANO BIOMEDICINE | 1883-5198 | Nanomedicine & Nanobiotechnology | | | x | Y | N | 1 |
| NANO BIOMEDICINE AND ENGINEERING | 2150-5578 | Nanomedicine & Nanobiotechnology | | | x | Y | N | 1 |
| NANO LIFE | 1793-9844 | Nanomedicine & Nanobiotechnology | | | x | Y | N | 1 |
| NANO STRUCTURES AND NANO OBJECTS | 2352-507X | Chemistry (Physical Chemistry) | | | x | Y | N | 1 |
| NANOETHICS | 1871-4757 | Nanoethics | | | x | Y | Y | 1 |
| NANOIMPACT | 2452-0748 | Nanotoxicology | | | x | Y | N | 1 |
| NANOSCIENCE AND NANOTECHNOLOGY ASIA | 2210-6812 | Nanoscience & Nanotechnology | | | x | Y | N | 1 |
| NANOSCIENCE AND TECHNOLOGY: AN INTERNATIONAL JOURNAL (INTERNATIONAL JOURNAL OF NANOMECHANICS SCIENCE AND TECHNOLOGY) | 2572-4258 | Nanoengineering (Micro and nano electromechanical systems) | | | x | Y | N | 1 |
| NANOTECHNOLOGIES IN RUSSIA | 1995-0780 | Nanoscience & Nanotechnology | | | x | Y | N | 1 |
| NANOTECHNOLOGY PERCEPTIONS | 1660-6795 | Nanoscience & Nanotechnology | | | x | Y | N | 1 |
| NANOTECHNOLOGY SCIENCE AND APPLICATIONS | 1177-8903 | Nanoscience & Nanotechnology | | | x | Y | N | 1 |
| NEW CARBON MATERIALS | 1872-5805 | Carbon materials | | x | | Y | Y | 1 |

| Title | ISSN | Thematic area | A₁ | A₂ | A₃ | Scopus | Wos | Precision |
|---|---|---|---|---|---|---|---|---|
| NPG ASIA MATERIALS | 1884-4049 | Materials science | | x | | Y | Y | 0.5 |
| OPENNANO | 2352-9520 | Nanobiology | | | x | Y | N | |
| PARTICLE AND FIBRE TOXICOLOGY | 1743-8977 | Toxicology | | x | | Y | Y | 0.5 |
| PHYSICA STATUS SOLIDI B-BASIC SOLID STATE PHYSICS | 0370-1972 | Physics (Condensed matter & Materials physics) | | x | | Y | Y | 0.5 |
| PHYSICA STATUS SOLIDI-RAPID RESEARCH LETTERS | 1862-6254 | Physics (Condensed matter & Materials physics) | | x | | Y | Y | 0.5 |
| PHYSICAL REVIEW B | 1098-0121 | Physics (Condensed matter & Materials physics) | | | x | Y | Y | 0.5 |
| PHYSICAL REVIEW LETTERS | 0031-9007 | Physics | | | x | Y | Y | 0 |
| PROCEEDINGS OF THE INSTITUTION OF MECHANICAL ENGINEERS, PART N: JOURNAL OF NANOMATERIALS, NANOENGENEIRING AND NANOSYSTEMS | 1740-3499 | Nanoengineering (Mechanical engineering) | | | x | Y | N | 1 |
| REVIEWS ON ADVANCED MATERIALS SCIENCE | 1605-8127 | Materials science (Nanomaterials) | x | | | Y | Y | 0.5 |
| SCIENCE OF ADVANCED MATERIALS | 1947-2935 | Materials science | x | | | Y | Y | 0.5 |
| SCRIPTA MATERIALIA | 1359-6462 | Materials science | x | | | Y | Y | 0 |
| SUPERLATTICES AND MICROSTRUCTURES | 0749-6036 | Materials science (Nanomaterials) | | x | | Y | Y | 0.5 |

x = Cover in the approch
Y = Included
N = No included

# Appendix 6. Questionnaire

| Publications | Abstract | Keywords | Keywords_plus | Relevant | No relevant | Further comments |
|---|---|---|---|---|---|---|
| Todai, M., Hagihara, K., Kishida, K., Inui, H., & Nakano, T. (2016). Microstructure and fracture toughness in boron added NbSi 2 (C40)/MoSi 2 (C11 b) duplex crystals. *Scripta Materialia*, *113*(1), 236-240. doi: 10.1016/j.scriptamat.2015.11.004 | The effect of B-addition on the microstructure and fracture toughness of (Mo0.85Nb0.15)Si-2 crystals with an oriented lamellar microstructure was investigated. B-addition led to an increase in the volume fraction of the Cub phase, which possesses different orientation relationship from that of the fine lamellae, and a reduction in their precipitation rate. The fracture toughness of the B-added crystal with the varied microstructure exhibited a value more than 4.0 MPa m(1/2), that was significantly higher than that of the ternary crystal. | Transition metal silicides; Lamellar structure; Toughness; Microstructure | plastic-deformation behavior; mosi2 single-crystals; mechanical-properties; oriented lamellae; phase-field; silicides; composite; segregation; improvement; elements | | | |
| Howell, S. L., Padalkar, S., Yoon, K., Li, Q., Koleske, D. D., Wierer, J. J., Wang, G. T., & Lauhon, L. J. (2013). Spatial mapping of efficiency of GaN/InGaN nanowire array solar cells using scanning photocurrent microscopy. *Nano Letters, 13*(11), 5123-5128. doi: 10.1021/nl402331u | GaN-InGaN core shell nanowire array devices are characterized by spectrally resolved scanning photocurrent microscopy (SPCM). The spatially resolved external quantum efficiency is correlated with structure and composition inferred from atomic force microscope (AFM) topography, scanning transmission electron microscope (STEM) imaging, Raman microspectroscopy, and scanning photocurrent microscopy (SPCM) maps of the effective absorption edge. The experimental analyses are coupled with finite difference time domain simulations to provide mechanistic understanding of spatial variations in carrier generation and collection, which is essential to the development of heterogeneous novel architecture solar cell devices. | Nanowire; InGaN; SPCM; solar cell; photovoltaics | multiple-quantum-wells; light-emitting-diodes; raman-scattering; photovoltaics; devices; nanorod | | | |

| Publications | Abstract | Keywords | Keywords_plus | Relevant | No relevant | Further comments |
|---|---|---|---|---|---|---|
| Yang, P., Xu, R., Nanita, S. C., & Cooks, R. G. (2006). Thermal formation of homochiral serine clusters and implications for the origin of homochirality. *Journal of the American Chemical Society*, *128*(51), 17074-17086. doi: 10.1021/ja064617d | Spontaneous assembly of amino acids into vapor-phase clusters occurs on heating the solid compounds in air. In comparison to the other amino acids, serine forms clusters to an unusual extent, showing a magic number octamer on sublimation; this octamer can be ionized and characterized by mass spectrometry. Two isomers of the vapor-phase serine octamer are generated, the minor one at 130 degrees C and the major at 220 degrees C. The higher temperature cluster shows a strong homochiral preference, as confirmed by isotopic labeling experiments. This serine cluster, like that generated earlier from solution in electrospray ionization experiments, undergoes gas-phase enantioselective substitution reactions with other amino acids. These reactions transfer the chirality of serine to the other amino acid through enantioselective incorporation into the octamer. Other serine pyrolysis products include alanine, glycine, ethanolamine, and small dipeptides, and many of these, too, are observed to be incorporated into the thermally formed serine octamers. Chiral chromatographic analysis confirmed that L-serine sublimation produced DL-alanine, glycine, and ethanolamine, while in the presence of hydrogen sulfide, L-serine yielded L-cysteine. The data demonstrate that sublimation of serine under relatively mild conditions yields chirally enriched serine octamers and that the chiral preference of the starting serine can be transferred to other compounds through cluster-forming chemical reactions. | | ionization mass-spectrometry; sonic spray ionization; magic number clusters; phase h/d exchange; gas-phase; electrospray-ionization; amino-acids; supramolecular chirality; structure elucidation; octamer | | | |
| Bergman, L., Rosenholm, J., Öst, A. B., Duchanoy, A., Kankaanpää, P., Heino, J., & Lindén, M. (2008). On the complexity of electrostatic suspension stabilization of functionalized silica nanoparticles for biotargeting and imaging applications. *Journal of Nanomaterials*, *2008*. doi: 10.1155/2008/712514 | Different means of attaching streptavidin to surface functionalized silica particles with a diameter of 240 nm were investigated with special focus on suspension stability for electrostatically stabilized suspensions. The influence of two different fluorescent dyes covalently linked to the streptavidin on suspension stability was also studied. The results clearly show that the stability of the suspensions is crucially dependent on all functional groups present on the surface. The surface functions may either directly affect the effective surface charge if the functions contain charged groups, or indirectly by affecting the relative concentration of charged groups on the particle surface. Poly( ethylene imine)- functionalized silica particles, where the polymer is grown by surface hyperbranching polymerization, are shown to be promising candidates for bioapplications, as the zeta-potential can remain strongly positive even under biologically relevant conditions. | | gene delivery; in-vivo; surface functionalization; transfection efficiency; drug-delivery; particles; carriers; cells; polyethylenimine; glutaraldehyde | | | |

| Publications | Abstract | Keywords | Keywords_plus | Relevant | No relevant | Further comments |
|---|---|---|---|---|---|---|
| Kanamadi, C. M., Das, B. K., Kim, C. W., Cha, H. G., Ji, E. S., Kang, D. I., Jadhav, A. P., & Kang, Y. S. (2009). Template assisted growth of cobalt ferrite nanowires. *Journal of Nanoscience and Nanotechnology, 9*(8), 4942-4947. doi: 10.1166/jnn.2009.1271 | Cobalt ferrite nanowires with a diameter of about 30 nm have been prepared inside anodized aluminum oxide (AAO) templates with one end closed nanopores using a vacuum infiltration method. The ferrite phase formation was confirmed by X-ray diffraction. Field emission scanning electron microscopy (FE-SEM) and transmission electron microscopy (TEM) were employed to characterize the morphology and structure of nanowires. SQUID magnetometer measurement showed that the nanowires to have both ferrimagnetic and superparamagnetic characteristics. A model for formation of discontinuous nanowires and particle agglomeration inside the template is discussed to explain these results. | Anodized Aluminum Oxide; AAO; Nanowires; Ferrite; Ferrimagnetic | magnetic-properties; anodic alumina; beta-feooh; thin-films; arrays; nanostructures; nanoparticles; fabrication; behavior; nanotubes | | | |
| Fang, L., & Li, W. J. (2012). Hydrothermal synthesis of flake-like MnCO3 film under high gravity field and their thermal conversion to hierarchical Mn3O4. *Micro & Nano Letters, 7*(4), 353-356. doi: 10.1049/mnl.2011.0530 | A flake-like MnCO3 film has been successfully synthesised by the hydrothermal method, with high gravity field using 0.5 mol/L MnCl2 and 5 mol/L CO(NH)(2) as precursor at 120 degrees C for 30 min in the aqueous solution-bromobenzene system. The effects of the relative centrifugal field and the reaction temperature on the formation of flake-like MnCO3 film were examined. The resultant samples were characterised by X-ray diffraction, Fourier transform infrared, FE-scanning electron microscopy and thermogravimetric differential thermal analysis. The results reveal that low temperature and high gravity field are favourable for the formation of flake-like MnCO3 film. The hierarchical Mn3O4 were obtained after calcinations of the flake-like MnCO3 film at 700 degrees C for 1 h in air. | | rechargeable lithium batteries; manganese carbonate; metal carbonate; ionic liquid; nanocrystals; mn(oh)(2); system; route | | | |
| Gonzalez, J. (2008). Kohn-Luttinger superconductivity in graphene. *Physical Review B, 78*(20), 205431. doi: 10.1103/PhysRevB.78.205431 | We investigate the development of superconductivity in graphene when the Fermi level becomes close to one of the Van Hove singularities of the electron system. The origin of the pairing instability lies in the strong anisotropy of the e-e scattering at the Van Hove filling, which leads to a channel with attractive coupling when making the projection of the BCS vertex on the symmetry modes with nontrivial angular dependence along the Fermi line. We show that the scale of the superconducting instability may be pushed up to temperatures larger than 10 K, depending on the ability to tune the system to the proximity of the Van Hove singularity. | | renormalization-group approach; 2 dimensions; fermions | | | |

| Publications | Abstract | Keywords | Keywords_plus | Relevant | No relevant | Further comments |
|---|---|---|---|---|---|---|
| Xiang, C., Yang, Y., & Penner, R. M. (2009). Cheating the diffraction limit: electrodeposited nanowires patterned by photolithography. *Chemical Communications*, (8), 859-873. doi: 10.1039/b815603d | The diffraction limit, d approximate to lambda/2, constrains the resolution with which structures may be produced using photolithography. Practical limits for d are in the 100 nm range. To circumvent this limit, photolithography can be used to fabricate a sacrificial electrode that is then used to initiate and propagate the growth by electrodeposition of a nanowire. We have described a version of this strategy in which the sacrificial electrode delimits one edge of the nascent nanowire, and a microfabricated "ceiling" constrains its height during growth. The width of the nanowire is determined by the electrochemical deposition parameters (deposition time, applied potential, and solution composition). Using this method, called lithographically patterned nanowire electrodeposition (LPNE), nanowires with minimum dimensions of 11 nm (w) x 5 nm (h) have been obtained. The lengths of these nanowires can be wafer-scale. LPNE has been used to synthesize nanowires composed of bismuth, gold, silver, palladium, platinum, and lead telluride. | | conducting polymer nanowires; step-edge decoration; field-effect transistors; light-emitting diode; quartz-crystal microbalance; telluride bi2te3 nanowires; tin oxide nanowires; silicon nanowires; transport-properties; bismuth nanowires | | | |
| Sumesh, C. K., Patel, K. D., Pathak, V. M., & Srivastava, R. (2008). Twofold conduction mechanisms in molybdenum diselenide single crystals in the wide temperature range of 300K to 12K. *Chalcogenide Letters*, 5(8), 177-180. | Molybdenum dichalcogenides, a member of VIA-VIB transition metal dichalcogenides has been looked as a potential semiconducting material for electronic devices. Thispaper reports temperature variation of (12K - 300K) electrical conductivity and carrier concentration in semiconducting crystals grown by direct vapor transport technique. It is found that the conductivity behavior is extrinsic in character with three different values of activation energy in the investigated temperature range. This behavior seems to be originating from presence of defects incorporated during growth and can be explained in the frame of a model consisting of one donor level along with one shallow acceptor level. | EDAX; A1 XRD; Growth from vapour; MoSe2 single crystals; Low temperature transport properties | | | | |

| Publications | Abstract | Keywords | Keywords_plus | Relevant | No relevant | Further comments |
|---|---|---|---|---|---|---|
| Westenfelder, B., Meyer, J. C., Biskupek, J., Kurasch, S., Scholz, F., Krill III, C. E., & Kaiser, U. (2011). Transformations of carbon adsorbates on graphene substrates under extreme heat. *Nano Letters, 11*(12), 5123-5127. doi 10.1021/nl203224z | We describe new phenomena of structural reorganization of carbon adsorbates as revealed by in situ atomic-resolution transmission electron microscopy (TEM) performed on specimens at extreme temperatures. In our investigations, a graphene sheet serves as both a quasi-transparent substrate for TEM and as an in situ heater. The melting of gold nanoislands deposited on the substrate surface is used to evaluate the local temperature profile. At annealing temperatures around 1000 K, we observe the transformation of physisorbed hydrocarbon adsorbates into amorphous carbon monolayers and the initiation of crystallization. At temperatures exceeding 2000 K the transformation terminates in the formation of a completely polycrystalline graphene state. The resulting layers are bounded by free edges primarily in the armchair configuration. | Transmission electron microscopy; in situ Joule heating; graphene; graphene edges | in-situ observation; electron-microscopy; gold particles; temperature; aberration; atoms; edge; nanostructures; spectroscopy; stability | | | |
| Sherman, E. Y., Ban, Y., Gulyaev, L. V., & Khomitsky, D. V. (2012). Spin Tunneling and Manipulation in Nanostructures. *Journal of nanoscience and nanotechnology*, *12*(9), 7535-7539. doi 10.1166/jnn.2012.6554 | The results for joint effects of tunneling and spin-orbit coupling on spin dynamics in nanostructures are presented for systems with discrete and continuous spectra. We demonstrate that tunneling plays the crucial role in the spin dynamics and the abilities of spin manipulation by external electric field. This result can be important for design of nanostructures-based spintronics devices. | Spin Dynamics; Spin-Orbit Coupling; Tunneling | single-electron spin; traversal time; transport | | | |
| Shao Z. Q., Chen, J. W. ,Li Y. Q., & Pan, X.Y. (2014). Thermodynamical properties of a three-dimensional free electron gas confined in a one-dimensional harmonical potential. Acta Physica Sinica, 63(24). doi: 10.7498/aps.63.240502 | We study the thermodynamical properties of a noninteracting electron gas confined in one dimension by a harmonic-oscillator potential. The exact analytical expression for the thermodynamical potential is obtained by using a formula of contour integration. The magnetizations, magnetic susceptibilities, and the specific heats are then studied each as a function of the strength of the magnetic field in different regimes of the temperature and effective thickness. It is shown at low temperature, the magnetization, magnetic susceptibility, and the specific heat oscillate as the strength of the magnetic field increases. Especially, there exist two modes of oscillations for the specific heat in certain regimes of low temperature and effective thickness. | thermodynamical potential; magnetization; magnetic susceptibility; specific heat | diamagnetic susceptibility; magnetic-field; size corrections; systems; particles | | | |

| Publications | Abstract | Keywords | Keywords_plus | Relevant | No relevant | Further comments |
|---|---|---|---|---|---|---|
| Soylemez, S., Udum, Y. A., Kesik, M., Hizliates, C. G., Ergun, Y., & Toppare, L. (2015). Electrochemical and optical properties of a conducting polymer and its use in a novel biosensor for the detection of cholesterol. *Sensors and Actuators B: Chemical*, *212*, 425-433. doi 10.1016/j.snb.2015.02.045 | A simple and robust cholesterol biosensor was designed by immobilizing cholesterol oxidase (ChOx) onto a conducting polymer modified graphite electrode. For this purpose, monomer, (Z)-4-(4-(9H-carbazol-9-yl) benzylidene) 2 (4 nitrophenyl) oxazol-5(4H)-one (CBNP), was synthesized and electrochemically polymerized on an electrode to achieve an effective immobilization platform for enzyme immobilization. After electropolymerization of the monomer (CBNP), electrochemical and spectroelectrochemical properties were investigated. Through the presence of nitro group on the polymer backbone hydrogen-bonding between enzyme molecules and polymer was achieved. Moreover, strong pi-pi stacking between aromatic moities in the polymer and aromatic residues of the enzyme enables a sensitive and reliable biosensor by conserving the crucial structure of biological molecules during the enzymatic reaction. The efficient interaction of the enzyme with the polymer coated surface brings easy and long-life detection of the substrate, cholesterol. After successful immobilization of ChOx with the help of glutaraldehyde as the crosslinking agent, amperometric biosensor responses were recorded at -0.7 V vs Ag wire in phosphate buffer (pH 7.0). K-M(app) (37.3 mu M), I-max (3.92 mu A), LOD (0.4063 mu M) and sensitivity (1.49 pA mu M-1 cm(-2)) values were determined. Finally, the prepared biosensor was successfully applied for determination of cholesterol content in real blood samples. | Amperometric biosensor; Polymer based biosensor; Conducting polymer; Cholesterol biosensor; Cholesterol oxidase | glassy-carbon electrode; oxidase; immobilization; film; nanotubes; graphene; network; sensor; serum | | | |
| Dyshin, A. A., Eliseeva, O. V., Bondarenko, G. V., Kolker, A. M., & Kiselev, M. G. (2016). Dispersion of single-walled carbon nanotubes in dimethylacetamide and a dimethylacetamide–cholic acid mixture. *Russian Journal of Physical Chemistry A*, *90*(12), 2434-2439. doi 10.1016/j.snb.2015.02.045 | A way of dispersing single-walled carbon nanotubes in preparing stable suspensions with high concentrations of individual nanotubes in amide solvents is described. The obtained suspensions are studied via Raman spectroscopy. The dependence of the degree of single-walled carbon nanotube (SWNT) dispersion in individual and mixed amide solvents on the type of solvent, the mass of nanotubes, and the concentration of cholic acid is established. A technique for processing spectral data to estimate the diameters and chiralities of individual nanotubes in suspension is described in detail. | dimethylacetamide; cholic acid; dispersing nanotubes; single-walled carbon nanotubes; SWNT suspensions; Raman spectroscopy; decomposition of spectra; calculating SWNT diameters; estimating chirality | dissolution; dynamics | | | |

| Publications | Abstract | Keywords | Keywords_plus | Relevant | No relevant | Further comments |
|---|---|---|---|---|---|---|
| Dechow, J., Forchel, A., Lanz, T., & Haase, A. (2000). Fabrication of NMR - Microsensors for nanoliter sample volumes. *Microelectronic engineering, 53*(1), 517-519. | The fabrication of micro-sensors for NMR-spectroscopy on both glass and GaAs is presented. Planar coils with inner diameter from 50 mu m to 400 mu m including a coplanar wave-guide leading to the bonding pads were combined with a chamber for liquid samples of 200-500 mu m diameter on the backside of the substrate. The microcoil served as a receiver in a H-1-NMR experiment at 11T(500 MHz). In initial experiments, the spectrum of 60 nl-volumes of pure silicone-oil were detected by the microcoil. | | | | | |
| Hilleringmann, U., Vieregge, T., & Horstmann, J. T. (2000). A structure definition technique for 25 nm lines of silicon and related materials. *Microelectronic engineering*, *53*(1-4), 569-572. | This paper describes an interesting solution to integrate features with nanometer scale on silicon without using a special lithography tool. Simple layer deposition and etching processes fulfil the lithography demands of the SIA roadmap for the year 2012. This structure definition technique has been used to generate polysilicon gates, silicon oxide and nitride, aluminum, tungsten and titanium nitride lines down to 25 nm width with excellent homogeneity over a 100 mm silicon wafer. It is transferable to any technology line, because only standard process steps like CVD deposition, dry and wet etching, and conventional optical lithography are necessary. | | | | | |
| Chong, Y. M., Leung, K. M., Ma, K. L., Zhang, W. J., Bello, I., & Lee, S. T. (2006). Growing cubic boron nitride films at different temperatures. *Diamond and related materials, 15*(4), 1155-1160. | Cubic boron nitride (cBN) films have been synthesized by both physical and chemical vapor deposition (PVD and CVD) methods at a wide range of substrate temperature. Some works conclude that the cBN growth is insensitive to the temperature parameter while few works suggest that substrate temperature plays a considerable role at cBN deposition, furthermore, the different temperatures used for the nucleation and growth make the situation more complex. In this work, we investigated systematically the growth of cBN films on CVD diamond surfaces at variable temperatures (from 200 to 1100 degrees C) using electron cyclotron resonance microwave plasma CVD (ECR-MPCVD) and radio-frequency magnetron sputtering (RF-MS) methods. The role of substrate temperature is discussed in the view of controlling phase purity, crystallinity, growth rate, and residual stress of cBN films deposited. Under optimized conditions, several-micrometers thick films are prepared by ECR-MPCVD as demonstrated on the example of a 3-mu m thick cBN film grown at 900 degrees C with faceted surfaces. The sizes of crystallites are fairly large (similar to 0.4 mu m) to yield visible Raman spectra characteristic to cBN. | cubic boron nitrides; chemical vapor deposition; physical vapor deposition; fluorine chemistry | chemical-vapor-deposition; pulsed-laser deposition; cbn films; bn films; growth; diamond; plasma; quality; epitaxy; phase | | | |

| Publications | Abstract | Keywords | Keywords_plus | Relevant | No relevant | Further comments |
|---|---|---|---|---|---|---|
| Lin, T., Zheng, K., Wang, C. L., & Ma, X. Y. (2007). Photoluminescence study of AlGaInP/GaInP quantum well intermixing induced by zinc impurity diffusion. *Journal of Crystal Growth*, 309(2), 140-144. doi: 10.1016/j.jcrysgro.2007.09.029 | AlGaInP/GaInP quantum well intermixing phenomena induced by Zn impurity diffusion at 540 degrees C were studied using room-temperature photo luminescence (PL) spectroscopy. As the diffusion time increased from 40 to 120 min, PL blue shift taken on the AlGaInP/GaInP quantum well regions increased from 36.3 to 171.6 meV. Moreover, when the diffusion time was equal to or above 60 min, it was observed firstly that a PL red shift occurred with a PL blue shift on the samples. After detailed analysis, it was found that the red-shift PL spectra were measured on the Ga0.51In0.49P buffer layer of the samples, and the mechanism of the PL red shift and the PL blue shift were studied qualitatively. | diffusion; metalorganic vapor phase epitaxy; semiconducting III-V materials; laser diodes | low-temperature; laser-diodes; superlattices; gaas; layer; al | | | |
| Borysenko, K. M., Mullen, J. T., Barry, E. A., Paul, S., Semenov, Y. G., Zavada, J. M., Nardelli, M.& Kim, K. W. (2010). First-principles analysis of electron-phonon interactions in graphene. *Physical Review B*, 81(12). doi 10.1103/PhysRevB.81.121412 | The electron-phonon interaction in monolayer graphene is investigated using density-functional perturbation theory. The results indicate that the electron-phonon interaction strength is of comparable magnitude for all four in-plane phonon branches and must be considered simultaneously. Moreover, the calculated scattering rates suggest an acoustic-phonon contribution that is much weaker than previously thought, revealing an important role of optical phonons even at low energies. Accordingly it is predicted, in good agreement with a recent measurement, that the intrinsic mobility of graphene may be more than an order of magnitude larger than the already high values reported in suspended samples. | | | | | |
| Vikram, S. V., Phase, D. M., & Chandel, V. S. (2010). High-TC phase transition in K2Ti6O13 lead-free ceramic synthesised using solid-state reaction. *Journal of Materials Science: Materials in Electronics*, 21(9), 902-905. doi: 10.1007/s10854-009-0015-0 | Solid-state reaction synthesised K2Ti6O13 lead-free ceramic was characterized using XRD, SEM, and X- band EPR, at room temperature. EPR-spectra showed the presence of (Fe'(Ti) - V-O(center dot center dot)) O defect associate dipoles, in orthorhombic phase, responsible for the broadening of the dielectric anomaly identified in the epsilon(r) (T) plots at T-C similar to 300 degrees C. This anomaly resembled a ferroelectric-paraelectric type phase transition following Curie-Weiss type trend. Besides, dielectric loss mechanism jointly represented electrical conduction, dipole orientation, and space charge polarization. | | potassium hexatitanate; thin-films; ferroelectrics; octatitanate; whiskers; defects; matrix; na | | | |

| Publications | Abstract | Keywords | Keywords_plus | Relevant | No relevant | Further comments |
|---|---|---|---|---|---|---|
| Bora, D. K., Braun, A., Erni, R., Fortunato, G., Graule, T., & Constable, E. C. (2011). Hydrothermal treatment of a hematite film leads to highly oriented faceted nanostructures with enhanced photocurrents. *Chemistry of Materials, 23*(8), 2051-2061. doi: 10.1021/cm102826n | A simple one-pot hydrothermal method is described for conveying a dipcoatec hematite nano particle film into an array of nanorods with superimposed flowerlike structures suitable for water splitting in photoelectrochemical cells. The hydrothermal treatment of the dip-coated hematite film with FeCl3 6H(2)O and L-arginite enchances the photocurrent by a factor of 2. It has been found that hydrothermal treatment changes the optical properties of the pristenen hematite film, but the energy band gap (Eg) does not change significantly to show some electronic effect. X-ray diffractograms of pristine and hydrothermally modified films reveal evolution fo preferential orientations and textures. Electron micrographs show that the particles are more prismatic after modification with a size of around 40nm x 200 nm. The X-ray photoelectron spectroscopy valence-band spectra point at a depletion of the spectrall intensity near the Fermi energy upon hydrothermal modification. The photocurrent density of the pristine film reached 218 mu A/cm(2) after 48 h of hydrothermal treatment, and this increase was found to be due to the high specfic surface area of the modified film and changes in the optical properties of the pristene film after hydrothermal treatment. | hematite thin film; photocurrent; hydrathermal treatment; faceted nanostructures; flowerlike superstructures | biomolecule-assisted synthesis; template-free synthesis; lithium ion battery; optical-properties; water oxidation; thin-films; alpha-fe2o3 superstructures; photocatalytic properties; electronic-structure; magnetic-property | | | |

| Publications | Abstract | Keywords | Keywords_plus | Relevant | No relevant | Further comments |
|---|---|---|---|---|---|---|
| Ghalamestani, S. G., Heurlin, M., Wernersson, L. E., Lehmann, S., & Dick, K. A. (2012). Growth of InAs/InP core–shell nanowires with various pure crystal structures. *Nanotechnology, 23*(28), 285601. doi: 10.1088/0957-4484/23/28/285601 | We have studied the epitaxial growth of an InP shell on various pure InAs core nanowire crystal structures by metal-organic vapor phase epitaxy. The InP shell is grown on wurtzite (WZ), zinc-blende (ZB), and {111}- and {110}-type faceted ZB twin-plane superlattice (TSL) structures by tuning the InP shell growth parameters and controlling the shell thickness. The growth results, particularly on the WZ nanowires, show that homogeneous InP shell growth is promoted at relatively high temperatures (similar to 500 degrees C), but that the InAs nanowires decompose under the applied conditions. In order to protect the InAs core nanowires from decomposition, a short protective InP segment is first grown axially at lower temperatures (420-460 degrees C), before commencing the radial growth at a higher temperature. Further studies revealed that the InP radial growth rate is significantly higher on the ZB and TSL nanowires compared to WZ counterparts, and shows a strong anisotropy in polar directions. As a result, thin shells were obtained during low temperature InP growth on ZB structures, while a higher temperature was used to obtain uniform thick shells. In addition, a schematic growth model is suggested to explain the basic processes occurring during the shell growth on the TSL crystal structures. | | vapor-phase epitaxy; lateral growth; inp; fabrication; devices; layers | | | |
| Chiacchiarelli, L. M., Rallini, M., Monti, M., Puglia, D., Kenny, J. M., & Torre, L. (2013). The role of irreversible and reversible phenomena in the piezoresistive behavior of graphene epoxy nanocomposites applied to structural health monitoring. Composites Science and Technology, 80, 73-79. doi: 10.1016/j.compscitech.2013.03.009 | The use of graphene for the development of a strain and damage sensor was evaluated and modeled. To achieve this, a graphene epoxy reactive mixture was used as a conductive coating which was cured onto a carbon fiber reinforced composite. This methodology proved to be very effective where substantial changes in piezoresistivity (up to 400%) were found as a function of strain (up to 2%). This contributed to a very high linear gauge factor (56.7(+/- 0.69)). The role of reversible and irreversible phenomena in the sensor piezoresistivity was modeled using the concepts of tunneling currents and conduction paths. In order to predict the response at higher deformation, an irreversible component was introduced to account for the substantial increase of piezoresistivity. A model which incorporated both components was able to predict the piezoresistivity up to high deformation. | Nano composites; Polymer-matrix properties; Electrical Properties; Modelling | fiber-reinforced composites; carbon nanotubes; electrical-properties; damage; nanofibers; strain; dispersion; matrix; rods | | | |

| Publications | Abstract | Keywords | Keywords_plus | Relevant | No relevant | Further comments |
|---|---|---|---|---|---|---|
| Park, S., Lee, J., Do, I. G., Jang, J., Rho, K., Ahn, S., Maruja, L., Kim S. J., Kim, K. M., Mao, M., Oh, E., Kim, Y. J., Kim, J., & Choi, Y. L. (2014). Aberrant CDK4 amplification in refractory rhabdomyosarcoma as identified by genomic profiling. *Scientific reports, 4*, 3623. doi: 10.1038/srep03623 | Rhabdomyosarcoma (RMS) is the most commonly occurring type of soft tissue tumor in children. However, it is rare in adults, and therefore, very little is known about the most appropriate treatment strategy for adult RMS patients. We performed genomic analysis of RMS cells derived from a 27-year-old male patient whose disease was refractory to treatment. A peritoneal seeding nodule from the primary tumor, pleural metastases, malignant pleural effusion, and ascites obtained during disease progression, were analyzed. Whole exome sequencing revealed 23 candidate variants, and 10 of 23 mutations were validated by Sanger sequencing. Three of 10 mutations were present in both primary and metastatic tumors, and 3 mutations were detected only in metastatic specimens. Comparative genomic hybridization array analysis revealed prominent amplification in the 12q13-14 region, and more specifically, the CDK4 proto-oncogene was highly amplified. ALK overexpression was observed at both protein and RNA levels. However, an ALK fusion assay using NanoString technology failed to show any ALK rearrangements. Little genetic heterogeneity was observed between primary and metastatic RMS cells. We propose that CDK4, located at 12q14, is a potential target for drug development for RMS treatment. | | childrens oncology group; prognostic-factors; intergroup-rhabdomyosarcoma; metastatic rhabdomyosarcoma; adult rhabdomyosarcoma; gene-expression; nonmetastatic rhabdomyosarcoma; dedifferentiated liposarcoma; alveolar rhabdomyosarcoma; cancer | | | |
| Foos, E. E., Yoon, W., Lumb, M. P., Tischler, J. G., & Townsend, T. K. (2013). Inorganic photovoltaic devices fabricated using nanocrystal spray deposition. ACS applied materials & interfaces, 5(18), 8828-8832. doi: 10.1021/am402423f | Soluble inorganic nanocrystals offer a potential route to the fabrication of all-inorganic devices using solution deposition techniques. Spray processing offers several advantages over the more common spin- and dip-coating procedures, including reduced material loss during fabrication, higher sample throughput, and deposition over a larger area. The primary difference observed, however, is an overall increase in the film roughness. In an attempt to quantify the impact of this morphology change on the devices, we compare the overall performance of spray-deposited versus spin-coated CdTe-based Schottky junction solar cells and model their dark current-voltage characteristics. Spray deposition of the active layer results in a power conversion efficiency of 2.3 +/- 0.3% with a fill factor of 45.7 +/- 3.4%, V-oc of 0.39 +/- 0.06 V, and J(sc) of 13.3 +/- 3.0 mA/cm(2) under one sun illumination. | photovoltaic; nanocrystal; processing; CdTe; spray; modeling | colloidal quantum dots; solar-cells; films; cdse; cdte | | | |

| Publications | Abstract | Keywords | Keywords_plus | Relevant | No relevant | Further comments |
|---|---|---|---|---|---|---|
| Gu, J., Xiao, P., Huang, Y., Zhang, J., & Chen, T. (2015). Controlled functionalization of carbon nanotubes as superhydrophobic material for adjustable oil/water separation. *Journal of Materials Chemistry A, 3*(8), 4124-4128. doi: 10.1039/c4ta07173e | A robust strategy for fabricating superhydrophobic carbon nanotube (CNT)-based hybrid materials as a separation membrane through the covalent attachment of the fluorine-bearing organosilane 1H,1H,2H,2H-perfluorodecyltriethoxysilane (PFDTS) onto -OH functionalized CNTs is proposed. This method resulted in PFDTS/CNT superhydrophobic materials with controlled functionalization that could be used to effectively separate various surfactant-stabilized water-in-oil emulsions with high separation efficiency and high flux. It maintains stable superhydrophobicity and high separation efficiency under extreme conditions, including high or low temperature and strongly acidic or alkaline solutions, and shows fire-retardant properties. | | organic-solvents; surfaces; films; graphene; oils; fabrication; removal; sensors; water | | | |
| Yin, Z., Zheng, Q., Chen, S. C., Li, J., Cai, D., Ma, Y., & Wei, J. (2015). Solution-derived poly (ethylene glycol)-TiOx nanocomposite film as a universal cathode buffer layer for enhancing efficiency and stability of polymer solar cells. Nano Research, 8(2), 456-468. doi: 10.1007/s12274-014-0615-8 | Highly efficient and stable polymer solar cells (PSCs) have been fabricated by adopting solution-derived hybrid poly(ethylene glycol)-titanium oxide (PEG-TiO (x) ) nanocomposite films as a novel and universal cathode buffer layer (CBL), which can greatly improve device performance by reducing interface energy barriers and enhancing charge extraction/collection. The performance of inverted PSCs with varied bulk-heterojunctions (BHJs) based on this hybrid nanocomposite CBL was found to be much better than those of control devices with a pure TiO (x) CBL or without a CBL. An excellent power conversion efficiency up to 9.05% under AM 1.5G irradiation (100 mW center dot cm(-2)) was demonstrated, which represents a record high value for inverted PSCs with TiO (x) -based interface materials. | titanium oxide; nanocomposites; energy conversion; organic solar cells; interface engineering; device stability | power conversion efficiency; electron-transport layer; high-performance; titanium-oxide; composite film; thin-films; tio2; tandem; transparent; interlayer | | | |

| Publications | Abstract | Keywords | Keywords_plus | Relevant | No relevant | Further comments |
|---|---|---|---|---|---|---|
| Meng, D., Wang, G., San, X., Shen, Y., Zhao, G., Zhang, Y., & Meng, F. (2015). CTAB-assisted hydrothermal synthesis of WO 3 hierarchical porous structures and investigation of their sensing properties. *Journal of Nanomaterials, 16*(1), 364. doi: 10.1155/2015/393205 | WO3 hierarchical porous structures were successfully synthesized via cetyltrimethylammonium bromide-(CTAB-) assisted hydrothermal method. The structure andmorphology were investigated using scanning electron microscope, X-ray diffractometer, transmission electron microscopy, X-ray photoelectron spectra, Brunauer-Emmett-Teller nitrogen adsorption-desorption, and thermogravimetry and differential thermal analysis. The result demonstrated that WO3 hierarchical porous structures with an orthorhombic structure were constructed by a number of nanoparticles about 50-100 nm in diameters. The H-2 gas sensing measurements showed that well-defined WO3 hierarchical porous structures with a large specific surface area exhibited the higher sensitivity compared with products without CTAB at all operating temperatures. Moreover, the reversible and fast response to H-2 gas and good selectivity were obtained. The results indicated that the WO3 hierarchical porous structures are promising materials for gas sensors. | | ion batteries; solar-cells; gas; film; nanostructures; nanoparticles; mechanism; oxide | | | |
| Gong, J., & Wilkinson, A. J. (2016). Sample size effects on grain boundary sliding. *Scripta Materialia*, *114*, 17-20. doi: 10.1016/j.scriptamat.2015.11.029 | Grain boundary sliding is an important deformation mechanism that contributes to creep and superplastic forming. In tin-based lead-free solders grain boundary sliding can make significant contributions to in service performance. Novel micro-compression tests designed to isolate individual grain boundaries and assess their mechanical resistance to sliding were conducted on tin. The boundary sliding deformation was more obvious for smaller sample cross-sections and made a larger contribution to the overall deformation. As with dislocation and twinning mediated plasticity there was a significant size effect in which the resistance to grain boundary sliding increases as the sample size is reduced. | Grain boundary sliding; Tin; Micromechanics | mechanical-properties; fracture-toughness; solder joints; micron scale; copper; plasticity; crystal; deformation; strength; aluminum | | | |

| Publications | Abstract | Keywords | Keywords_plus | Relevant | No relevant | Further comments |
|---|---|---|---|---|---|---|
| Hu, D., Wang, H. Y., & Zhu, Q. F. (2016). Design of an ultra-broadband and polarization-insensitive solar absorber using a circular-shaped ring resonator. Journal of Nanophotonics, 10(2). doi: 10.1117/1.JNP.10.026021 | An ultra-broadband and polarization-independent metamaterial absorber consisting of a chromium circular-shaped ring resonator embedded in a dielectric layer and a chromium ground plane is numerically investigated in the solar spectrum. Simulation results show that the absorber can obtain the average absorption of 95.5% over most the solar spectrum, near-infrared, and short-wavelength infrared regime (400 to 2500 nm). The absorption mechanism of the ultra-broadband metamaterial absorber originates from the overlapping of two different resonance wavelengths. In particular, the electromagnetic energy is almost completely dissipated in the chromium layer, which makes it independent on the loss of the dielectric layer and widens the range of choices for potential dielectric materials. Our designed absorber has high practical feasibility and appears to be very promising for solar energy harvesting, thermal imaging, and emissivity control applications. | broadband; polarization-independent; metamaterial; absorber | perfect absorber; light-absorption; metamaterials; cloak | | | |
| Wilson, L. R., Carder, D. A., Steer, M. J., Cockburn, J. W., Hopkinson, M., Chia, C. K., ... & Airey, R. (2002). Strategies for reducing the emission wavelength of GaAs–AlAs quantum cascade lasers. *Physica E: Low-dimensional Systems and Nanostructures, 13*(2), 835-839. | We report two novel methods for reducing the emission wavelength of GaAs-AlAs quantum cascade lasers. We demonstrate that for lasing to occur electron injection into the upper laser level must proceed via Gamma states confined below the lowest X state in the injection barrier. The limit this places on the minimum operating wavelength (lambda approximate to 8 mum) is overcome by utilising a novel injection barrier design to achieve lasing at lambda = 7.2 mum. In addition, we have deposited InAs monolayers in the active regions of a GaAs-AlAs quantum cascade laser to decrease the lasing transition wavelength, The InAs monolayers have a minimal effect on the upper laser level but decrease the confinement energy of the lower laser level. Thus a significantly reduced emission wavelength (8.3 mum compared with 11.2 mum) is achieved whilst maintaining very similar laser performance. | quantum cascade laser; GaAs based; intervalley scattering | well structures; electron-transfer; operation | | | |

| Publications | Abstract | Keywords | Keywords_plus | Relevant | No relevant | Further comments |
|---|---|---|---|---|---|---|
| Du, X., & Hlady, V. (2002). Monolayer formation on silicon and mica surfaces rearranged from N-hexadecanoyl-L-alanine supramolecular structures. *The Journal of Physical Chemistry B, 106*(29), 7295-7299. doi: 10.1021/jp0209603 | The rearrangements of N-hexadecanoyl-L-alanine supramolecular structures in water to form monolayers on silicon and mica surfaces have been investigated using atomic force microscopy. It is the first time that such a monolayer with the polar groups on the solid surface and the alkyl chains up is obtained through a rational molecular design. The monolayer formation results from the strong interaction between the molecular headgroups and the surface, the intermolecular hydrogen bonding interaction via a six-membered ring structure in the case of silicon surfaces, and the additional attractive force in the case of mica surfaces. The anisotropic and dendritic growth structures, clearly observed for the monolayers rearranged on mica surfaces, are indicative of a homochiral effect. The differences in height and morphology of the monolayers on the two types of surfaces are considered to be relevant to the surface roughness and to the interactions between the molecules and the surface. | | self-assembled monolayers; langmuir-blodgett-films; chiral discrimination; ftir spectroscopy; absorption-spectroscopy; fluorescence microscopy; air/water interface; infrared reflection; aqueous-solution; metal-complex | | | |
| Gotter, R., Ruocco, A., Butterfield, M. T., Iacobucci, S., Stefani, G., & Bartynski, R. A. (2003). Angle-resolved Auger-photoelectron coincidence spectroscopy (AR-APECS) of the Ge (100) surface. *Physical Review B*, 67(3), 033303. doi: 10.1103/PhysRevB.67.033303 | We have measured the angular distribution of Ge L3M45M45 Auger electrons in coincidence with Ge 2p(3/2) core photoelectrons along the (001) azimuth of the Ge(100) surface. Intensity modulations arising from diffraction effects are suppressed in the coincidence Auger angular distribution and, when specific emission angles of the photoelectrons are considered, new features appear. We attribute the former effect to enhanced surface specificity of the coincidence technique and the latter to sensitivity of the coincidence measurement to alignment of the core hole state. | | electron diffraction; photoionization | | | |
| Nicolaides, A. (2003). Singlet hydrocarbon carbenes with high barriers toward isomerization: A computational investigation. *Journal of the American Chemical Society, 125*(30), 9070-9073. doi: 10.1021/ja0299699 | A prerequisite for a stable singlet hydrocarbon carbene is the existence of high barriers toward isomerization. Four derivatives of cyclopentylidene (1-4) with rigid and varying carbon cages are examined computationally at the B3LYP/6-311+G(d,p) level of theory. Singlet ground states are predicted for carbenes 1-4, with DeltaE(ST)'s = 7-22 kcal/mol. The rearrangement paths considered are 1,3-hydrogen shift, 1,2-carbon shift and beta-CC bond-cleavage. Carbenes 3 and 4 lie in relatively shallow potential-energy wells (around 4 and 6 kcal/mol, respectively) and are expected to rearrange via 1,3-hydrogen shifts to cyclopropane derivatives. For 1 and 2, the lowest energy rearrangement path is beta-CC bond-cleavage requiring about 12 and 20 kcal/mol, respectively, placing 2 in the deepest potential-energy | | stable carbenes; density; cyclobutylidene; cyclobutenylidene; rearrangements; migration; hydrogen; carbon; state | | | |

| Publications | Abstract | Keywords | Keywords_plus | Relevant | No relevant | Further comments |
|---|---|---|---|---|---|---|
| | well among the four carbenes. | | | | | |
| Mao, C., Qi, J., & Belcher, A. M. (2003). Building Quantum Dots into Solids with Well- Defined Shapes. *Advanced Functional Materials, 13*(8), 648-656. doi: 10.1002/adfm.200304297 | Quantum dots (QDs) chemically synthesized in solution at a higher temperature (85degreesC) were built in situ into a variety of three-dimensional (3D) close-packed QD ensembles (QD solids) with well-defined shapes: needles, disks, rods, spheres, bundles, stars, ribbons, and transition structures (TSs). Design strategies using a novel cold-treatment (-25 to 0degreesC) process immediately following the synthesis of the QDs provided control over these shapes, independently from the II-VI materials used. Transformation occurred between different shapes by the rearrangement of the QDs within the QD ensembles. The QD solids were characterized by advanced electron microscopy and photoluminescence spectroscopy. The cold treatment strategy is versatile and has been applied to several II-VI QDs (CdS, ZnS, and CdSe) and may be extended to other QD systems and other chemical approaches. | | semiconductor nanocrystals; cdse nanocrystals; optical gain; nanoparticles; superlattices; coalescence; assemblies; ensemble; states | | | |

| Publications | Abstract | Keywords | Keywords_plus | Relevant | No relevant | Further comments |
|---|---|---|---|---|---|---|
| Hichria, A., & Jazirib, S. (2003). Two Carriers in Vertically Coupled Quantum Dots: Magnetic Field Effect. *Journal of nanoscience and nanotechnology*, *3*(3), 263-269. doi: 10.1166/jnn.2003.162 | We study a two-charge-carrier (two holes or two electrons) quantum dot molecule in a magnetic field. In comparison with the electron states in the double quantum dot, the switching between the hole states is achieved by changing both the inter-dot distance and magnetic field. We use harmonic potentials to model the confining of two charge carriers and calculate the energy difference DeltaE between the two lowest energy states with the Hund-Mulliken technique, including the Coulomb interaction. Introducing the Zeeman effect, we note a ground-state crossing, which can be observed as a pronounced jump in the magnetization at a perpendicular magnetic field of a few Tesla. The ground states of the molecule provide a possible realization for a quantum gate. | quantum dots; valence band mixing; harmonic Potentials; Hund-Mulliken technique; switching; phase transition; qubit; magnetization | artificial atoms |  |  |  |
| Bao, M., Yang, H., Sun, Y., & French, P. J. (2003). Modified Reynolds' equation and analytical analysis of squeeze-film air damping of perforated structures. *Journal of Micromechanics and Microengineering, 13*(6), 795. | We modified the Reynolds equation to extend its applications to squeeze-film air damping of perforated plates (i.e., the hole-plates) by adding a term related to the damping effect of gas flow through holes. The modified Reynolds equation (MRE) is generally effective for a perforated hole-plate with a finite thickness and finite lateral dimensions as well as a non-perforated hole-plate. Analytical expressions of damping pressure for long rectangular hole-plates and regular rectangular hole-plates have been found. For MEMS devices with typical dimensions, 'effective damping width' approximation is introduced so that the boundary effect on damping force can be treated easily. The conditions for 'effective damping width' approximation are discussed. Based on the concept of 'effective damping width', damping forces for circular plates and even hole-plates with irregular shapes can be found. The results obtained by the MIZE method match the numerical results obtained by ANSYS/FLOTRAN very well. The comparison between the MIZE results and the experimental results by Kim et al (1999 MEMS '99 pp 296-301) shows that the MIZE results agree with the experimental results much better than the FEM simulation given by Kim et al. |  |  |  |  |  |

| Publications | Abstract | Keywords | Keywords_plus | Relevant | No relevant | Further comments |
|---|---|---|---|---|---|---|
| Alcala, M. D., Criado, J. M., Real, C., Grygar, T., Nejezchleba, M., Subrt, J., & Petrovsky, E. (2004). Synthesis of nanocrystalline magnetite by mechanical alloying of iron and hematite. *Journal of Materials Science, 39*(7), 2365-2370. | The synthesis of magnetite has been studied by mechanical alloying in an inert atmosphere of a stoichiometric mixture of micrometric particle size iron and hematite powders. The final products have been characterised by chemical analysis, SEM, TEM, XRD, Mossbauer spectroscopy as well as specific surface and magnetic measurements. The magnetite obtained in this way exhibits a high magnetic hardness. The formation of a wustite layer on the magnetite core, because of the reaction between magnetite and iron contamination coming from the bowls and grinding balls, tends to decrease the coercive force of magnetite. The formation of this phase would be avoided by controlling the grinding time. | | electrochemical dissolution; gamma-fe2o3 particles; thermal-stability; size; microparticles; nanoparticles; fabrication; voltammetry; iron(iii); evolution | | | |
| Shi, Z., Wang, Q., Ye, W., Li, Y., & Yang, Y. (2006). Synthesis and characterization of mesoporous titanium pyrophosphate as lithium intercalation electrode materials. *Microporous and mesoporous materials, 88*(1), 232-237. doi: 10.1016/j.micromeso.2005.09.013 | Mesoporous titanium pyrophosphates have been synthesized by a sol-gel template method with further calcinations at or below the temperature of 700 degrees C. When calcined at 800 degrees C, crystalline TiP2O7 will be formed accompanied with the break down of meso-structure in the precursor. Mesoporous TiP2O7 shows a similar lithium ion intercalation behavior to that of solid solution in the electrochemical tests. When cycled at high charge/discharge rate, mesoporous TiP2O7 calcined at 700 degrees C delivers a higher specific discharge capacity than that of crystalline TiP2O7, indicating that mesoporous structure is beneficial for improving the transportation and intercalation/deintercalation behavior of lithium ions in the materials, thus improving the charge/discharge performance of the materials at high charge/discharge rate. | mesoporous materials; titanium pyrophosphate; cathode material; lithium ion batteries | cathode materials; ion batteries; nanostructured materials; block-copolymer; performance; progress; oxide; nanotechnology; insertion; storage | | | |
| Matsubara, M., Kortus, J., Parlebas, J. C., & Massobrio, C. (2006). Dynamical identification of a threshold instability in Si-doped heterofullerenes. *Physical Review Letters, 96*(15), 155502. doi: 10.1103/PhysRevLett.96.155502 | We rationalize the origins of a threshold instability and the mechanism of finite temperature fragmentation in highly Si-doped C60-mSim heterofullerenes via a first-principles approach. Cage disruption is driven by enhanced fluctuations of the most internal Si atoms. These are located within fully segregated Si regions neighboring the C-populated part of the cage. The predominance of inner Si atoms over those involved in Si-C bonds marks the transition from thermally stable to unstable C60-mSim systems at m=20. | | metal-fullerene clusters; molecular-dynamics; 1st principles; carbon clusters; mixed clusters; cage; density; boron | | | |

| Publications | Abstract | Keywords | Keywords_plus | Relevant | No relevant | Further comments |
|---|---|---|---|---|---|---|
| Hájek, M., Veselý, J., & Cieslar, M. (2007). Precision of electrical resistivity measurements. *Materials Science and Engineering: A*, *462*(1), 339-342. doi: 10.1016/j.msea.2006.01.175 | The influence of the specimen shape on the electrical resistivity measurements was investigated both theoretically, by means of finite element method modelling, and experimentally, by measuring electrical resistivity of samples of variable shapes. Precision of electrical resistivity measurements using a four-point method for samples of arbitrary shape was critically reviewed. | electrical resistivity; four-point method; finite elements | deformation; order; al | | | |
| Stopa, M., & Marcus, C. M. (2008). Magnetic field control of exchange and noise immunity in double quantum dots. *Nano letters*, *8*(6), 1778-1782. doi: 10.1021/nl801282t | We employ density functional calculated eigenstates as a basis for exact diagonalization studies of semiconductor double quantum dots, with two electrons, through the transition from the symmetric bias regime to the regime where both electrons occupy the same dot. We calculate the singlet-triplet splitting J(epsilon) as a function of bias detuning epsilon and explain its functional shape with a simple, double anticrossing model. A voltage noise suppression "sweet spot," where dJ(epsilon)/d epsilon = 0 with nonzero J(epsilon), is predicted and shown to be tunable with a magnetic field B. | | coulomb-blockade; electron-spin; polarization | | | |
| Ao, Y., Xu, J., & Fu, D. (2009). Study on the effect of different acids on the structure and photocatalytic activity of mesoporous titania. Applied Surface Science, 256(1), 239-245. doi: 10.1016/j.apsusc.2009.08.008 | Nanocrystalline mesoporous titania was synthesized via a combined sol-gel process with surfactant-assisted templating method using cetyltrimethyl ammonium bromide (CTAB) as the structure-directing agent. The process was catalyzed by different acid (hydrochloric acid, nitric acid, sulfuric acid, or phosphoric acid). The prepared samples were characterized by XRD, TEM, BET and FT-IR. The photocatalytic activity of the samples was determined by degradation of phenol in aqueous solution. Results showed that different acid had different effect on the structure and crystal phase of the samples. The sample adjusted by phosphoric acid showed highest surface area and photocatalytic activity. The formation mechanism of the samples catalyzed by different acid was also discussed. | Mesoporous structure; Titania; Sol-gel; Mechanism; Photocatalysis | visible-light irradiation; tio2 particles; sol; degradation; adsorption; nanocrystallites; silica; films; oxide | | | |
| Kher, S., Dixit, A., Rawat, D. N., & Sodha, M. S. (2010). Experimental verification of light induced field emission. *Applied Physics Letters, 96*(4), 044101. doi: 10.1063/1.3293297 | This letter reports the experimental verification of the recently predicted phenomenon that the electric field emission current from a negatively charged surface gets enhanced by incidence of light (even of frequency below the photoelectric threshold) on the cathode. | cathodes; electric fields; field emission; photoelectricity | | | | |
| Ghodake, G., Seo, Y. D., Park, D., & Lee, D. S. (2010). Phytotoxicity of carbon nanotubes assessed by Brassica juncea and Phaseolus mungo. *Journal of Nanoelectronics and Optoelectronics*, *5*(2), 157-160. doi: | Two agronomic plant species were evaluated for the phytotoxicity of multi-walled carbon nanotubes (CNTs) using germination and seedling growth of Brassica juncea and Phaseolus mungo. Both B. juncea and P mungo seeds showed 100% germination with the application of CNTs, which indicated their non-hazardous nature for the germination of the | Phytotoxicity; Germination; Brassica juncea; Phaseolus mungo; Carbon | bulk zno; toxicity; nanoparticles; behavior; tio2; nanomaterials; | | | |

| Publications | Abstract | Keywords | Keywords_plus | Relevant | No relevant | Further comments |
|---|---|---|---|---|---|---|
| 10.1166/jno.2010.1084 | seeds. As the B. juncea seed was grown with CNTs at 10 mu g/ml, 20 mu g/ml, and 40 mu g/ml, the enhancement of root growth was evidenced up to 138%, 202%, and 135%, respectively, compared to the control. In the case of B. juncea, the heights of the shoot and root were not affected at all by the studied CNTs' concentrations; however, phytotoxicity was evidenced at 40 mu g/ml CNTs by optical microscopy of the hairy root system, since a severe reduction in both the number of root hairs and their length was observed. | nanotubes | environment; spinach | | | |
| Hu, W., Zhang, X. B., Cheng, Y. L., Wu, Y. M., & Wang, L. M. (2011). Low-cost and facile one-pot synthesis of pure single-crystalline ε-Cu 0.95 V 2 O 5 nanoribbons: high capacity cathode material for rechargeable Li-ion batteries. *Chemical Communications, 47*(18), 5250-5252. doi: 10.1039/c1cc11184a | Pure single-crystalline epsilon-Cu0.95V2O5 nanoribbons have been successfully synthesized via a facile one-pot solvothermal route using low-cost raw materials. The obtained materials can react electrochemically with 2.64 Li in a reversible fashion and thus greatly expands the range of cathode choices. | | lithiation characteristics; electrochemical properties; electrode material; silicon nanowires; room-temperature; lithium; alpha-cuv2o6; gel | | | |
| Lai, S. L., Tan, W. L., & Yang, K. L. (2011). Detection of DNA targets hybridized to solid surfaces using optical images of liquid crystals. ACS applied materials & interfaces, 3(9), 3389-3395. doi: 10.1021/am200571h | In this paper, we report a method of detecting DNA targets hybridized to a solid surface by using liquid crystals (LC). The detection principle is based on different interference colors of LC supported on surfaces decorated with single-stranded DNA (ssDNA) or double-stranded DNA (dsDNA). However, the contrast between the ssDNA and dsDNA is not obvious, unless DNA-streptavidin complexes are introduced to the dsDNA to increase the surface mass density. Two different approaches of introducing streptavidin to the system are studied and compared. We find that by premixing the biotin-labeled DNA targets with streptavidin prior to the DNA hybridization, branched-streptavidin complexes are formed and clear LC signal can be observed. This LC-based DNA detection principle represents an important step toward the development of a simple, instrument- and fluorophore-free DNA detection method. | liquid crystals; DNA microarray; DNA targets detection; single-stranded DNA; double-stranded DNA; Biotin; Streptavidin | gold surfaces; microarrays; amplification; transitions; technology; proteins; systems; arrays; probes | | | |

| Publications | Abstract | Keywords | Keywords_plus | Relevant | No relevant | Further comments |
|---|---|---|---|---|---|---|
| Wu, W. Q., Xu, Y. F., Rao, H. S., Su, C. Y., & Kuang, D. B. (2013). A double layered TiO2 photoanode consisting of hierarchical flowers and nanoparticles for high-efficiency dye-sensitized solar cells. Nanoscale, 5(10), 4362-4369. doi: 10.1039/c3nr00508a | We report the innovative development of a double layered photoanode made of hierarchical TiO2 flowers (HTFs) as the overlayer and TiO2 nanoparticles (TNPs) as the underlayer, for dye-sensitized solar cells (DSSCs). They were prepared via a mild and simple one-step hydrothermal reaction of TiO2 nanoparticles/FTO glass substrate in an alkaline solution. The underlayer made of TNPs with a small size (20 nm in diameter) serves as a transparent photoanode for efficient dye adsorption. The overlayer consisting of HTFs (3-5 mu m in diameter) embedded by TiO2 nanosheets plays multiple roles in enhancing light-scattering and fast electron transport. DSSCs based on this novel double layered photoanode (5 mu m TNPs + 5 mu m HTFs) exhibit greater than 7.4% power conversion efficiency (PCE), which is higher than that of single layer TNP based photoanodes (6.59%) with similar thickness (similar to 10 mu m), and this is mainly attributed to the superior light scattering ability and fast electron transport of the former. Meanwhile, the thickness of the TNP underlayer has been optimized to further improve the PCE and an excellent PCE of over 9% has been achieved based on a 15 mu m TNP + a 5 mu m HTF double layered photoanode, accompanied by a short-circuit photocurrent density of 17.85 mA cm(-2), an open-circuit voltage of 763 mV and a fill factor of 0.67. | | photovoltaic performance; conversion efficiency; nanowire arrays; morphology; spectroscopy; temperature; electrolyte; candidate; transport; spheres | | | |
| Wang, W. B., Li, X. H., Wen, L., Zhao, Y. F., Duan, H. H., Zhou, B. K., Shi, T. F. Zeng, X. S., Li, N., & Wang, Y. Q. (2014). Optical simulations of P3HT/Si nanowire array hybrid solar cells. *Nanoscale research letters*, *9*(1). doi: 10.1186/1556-276X-9-238 | An optical simulation of poly(3-hexylthiophene) (P3HT)/Si nanowire array (NWA) hybrid solar cells was investigated to evaluate the optical design requirements of the system by using finite-difference time-domain (FDTD) method. Steady improvement of light absorption was obtained with increased P3HT coating shell thickness from 0 to 80 nm on Si NWA. Further increasing the thickness caused dramatic decrease of the light absorption. Combined with the analysis of ultimate photocurrents, an optimum geometric structure with a coating P3HT thickness of 80 nm was proposed. At this structure, the hybrid solar cells show the most efficient light absorption. The optimization of the geometric structure and further understanding of the optical characteristics may contribute to the development for the practical experiment of the promising hybrid solar cells. | P3HT; Si nanowire array; Hybrid solar cells; Finite-difference time-domain (FDTD) method | photovoltaic applications; absorption; enhancement; morphology; shell; layer; glass | | | |

| Publications | Abstract | Keywords | Keywords_plus | Relevant | No relevant | Further comments |
|---|---|---|---|---|---|---|
| Hashimoto, K., Sasaki, F., Hotta, S., & Yanagi, H. (2016). Amplified Emission and Field-Effect Transistor Characteristics of One-Dimensionally Structured 2, 5-Bis (4-biphenylyl) thiophene Crystals. Journal of nanoscience and nanotechnology, 16(4), 3200-3205. doi: 10.1166/jnn.2016.12285 | One-dimensional (1D) structures of 2,5-bis(4-biphenylyl)thiophene (BP1T) crystals are fabricated for light amplification and field-effect transistor (FET) measurements. A strip-shaped 1D structure (10 mu m width) made by photolitography of a vapor-deposited polycrystalline film shows amplified spontaneous emission and lasing oscillations under optical pumping. An FET fabricated with this 1D structure exhibits hole-conduction with a mobility of mu(h) = 8.0 x 10(-3) cm(2)/Vs. Another 1D-structured FET is fabricated with epitaxially grown needle-like crystals of BP1T. This needle-crystal FET exhibits higher mobility of mu(h) = 0.34 cm(2)/Vs. This improved hole mobility is attributed to the single-crystal channel of epitaxial needles while the grain boudaries in the polycrystalline 1D-structure decrease the carrier transport. | One-Dimensional Structure; 5-Bis(4-biphenylyl)thiophene; Amplified Spontaneous Emission; Organic Laser; Field-Effect Transistor | thiophene/phenylene co-oligomer; single-crystals | | | |

Appendix 7. Random sample with answers

| Approach | Publication | No relevant | Relevant | No answer |
|---|---|---|---|---|
| $A_0$ | Lin, T., Zheng, K., Wang, C. L., & Ma, X. Y. (2007). Photoluminescence study of AlGaInP/GaInP quantum well intermixing induced by zinc impurity diffusion. *Journal of Crystal Growth, 309*(2), 140-144. doi: 10.1016/j.jcrysgro.2007.09.029 | 55 | 42 | 1 |
| $A_0$ | Ao, Y., Xu, J., & Fu, D. (2009). Study on the effect of different acids on the structure and photocatalytic activity of mesoporous titania. *Applied Surface Science, 256*(1), 239-245. doi: 10.1016/j.apsusc.2009.08.008 | 54 | 42 | 2 |
| $A_0$ | Hu, W., Zhang, X. B., Cheng, Y. L., Wu, Y. M., & Wang, L. M. (2011). Low-cost and facile one-pot synthesis of pure single-crystalline ε-Cu 0.95 V 2 O 5 nanoribbons: high capacity cathode material for rechargeable Li-ion batteries. *Chemical Communications, 47*(18), 5250-5252. doi: 10.1039/c1cc11184a | 16 | 81 | 1 |
| $A_0$ | Ghalamestani, S. G., Heurlin, M., Wernersson, L. E., Lehmann, S., & Dick, K. A. (2012). Growth of InAs/InP core–shell nanowires with various pure crystal structures. *Nanotechnology, 23*(28), 285601. doi: 10.1088/0957-4484/23/28/285601 | 10 | 86 | 2 |
| $A_0$ | Park, S., Lee, J., Do, I. G., Jang, J., Rho, K., Ahn, S., Maruja, L., Kim S. J., Kim, K. M., Mao, M., Oh, E., Kim, Y. J., Kim, J., & Choi, Y. L. (2014). Aberrant CDK4 amplification in refractory rhabdomyosarcoma as identified by genomic profiling. *Scientific reports, 4*, 3623. doi: 10.1038/srep03623 | 77 | 16 | 5 |
| $A_1$ | Dechow, J., Forchel, A., Lanz, T., & Haase, A. (2000). Fabrication of NMR - Microsensors for nanoliter sample volumes. *Microelectronic engineering, 53*(1), 517-519. | 62 | 32 | 4 |

| Approach | Publication | No relevant | Relevant | No answer |
|---|---|---|---|---|
| A$_1$ | Hilleringmann, U., Vieregge, T., & Horstmann, J. T. (2000). A structure definition technique for 25 nm lines of silicon and related materials. *Microelectronic engineering*, *53*(1-4), 569-572. | 12 | 83 | 3 |
| A$_1$ | Wilson, L. R., Carder, D. A., Steer, M. J., Cockburn, J. W., Hopkinson, M., Chia, C. K., ... & Airey, R. (2002). Strategies for reducing the emission wavelength of GaAs–AlAs quantum cascade lasers. *Physica E: Low-dimensional Systems and Nanostructures, 13*(2), 835-839. | 51 | 43 | 4 |
| A$_1$ | Du, X., & Hlady, V. (2002). Monolayer formation on silicon and mica surfaces rearranged from N-hexadecanoyl-L-alanine supramolecular structures. *The Journal of Physical Chemistry B, 106*(29), 7295-7299. doi: 10.1021/jp0209603 | 40 | 54 | 4 |
| A$_1$ | Mao, C., Qi, J., & Belcher, A. M. (2003). Building Quantum Dots into Solids with Well-Defined Shapes. *Advanced Functional Materials, 13*(8), 648-656. doi: 10.1002/adfm.200304297 | 6 | 89 | 3 |
| A$_1$ | Bao, M., Yang, H., Sun, Y., & French, P. J. (2003). Modified Reynolds' equation and analytical analysis of squeeze-film air damping of perforated structures. *Journal of Micromechanics and Microengineering, 13*(6), 795. | 87 | 9 | 2 |
| A$_1$ | Shi, Z., Wang, Q., Ye, W., Li, Y., & Yang, Y. (2006). Synthesis and characterization of mesoporous titanium pyrophosphate as lithium intercalation electrode materials. *Microporous and mesoporous materials, 88*(1), 232-237. doi: 10.1016/j.micromeso.2005.09.013 | 62 | 33 | 3 |
| A$_1$ | Hájek, M., Veselý, J., & Cieslar, M. (2007). Precision of electrical resistivity measurements. *Materials Science and Engineering: A*, *462*(1), 339-342. doi: 10.1016/j.msea.2006.01.175 | 85 | 12 | 1 |
| A$_1$ | Bergman, L., Rosenholm, J., Öst, A. B., Duchanoy, A., Kankaanpää, P., Heino, J., & Lindén, M. (2008). On the complexity of electrostatic suspension stabilization of functionalized silica nanoparticles for biotargeting and imaging applications. *Journal of Nanomaterials*, *2008*. doi: 10.1155/2008/712514 | 18 | 76 | 4 |
| A$_1$ | Lai, S. L., Tan, W. L., & Yang, K. L. (2011). Detection of DNA targets hybridized to solid surfaces using optical images of liquid crystals. *ACS Applied Materials & Interfaces*, *3*(9), 3389-3395. doi: 10.1021/am200571h | 54 | 39 | 5 |
| A$_1$ | Yin, Z., Zheng, Q., Chen, S. C., Li, J., Cai, D., Ma, Y., & Wei, J. (2015). Solution-derived poly (ethylene glycol)-TiOx nanocomposite film as a universal cathode buffer layer for enhancing efficiency and stability of polymer solar cells. *Nano Research*, 8(2), 456-468. doi: 10.1007/s12274-014-0615-8 | 26 | 69 | 3 |
| A$_1$ | Todai, M., Hagihara, K., Kishida, K., Inui, H., & Nakano, T. (2016). Microstructure and fracture toughness in boron added NbSi 2 (C40)/MoSi 2 (C11 b) duplex crystals. *Scripta Materialia*, 113, 236-240. doi: 10.1016/j.scriptamat.2015.11.004 | 76 | 18 | 4 |
| A$_1$ | Gong, J., & Wilkinson, A. J. (2016). Sample size effects on grain boundary sliding. *Scripta Materialia*, *114*, 17-20. doi: 10.1016/j.scriptamat.2015.11.029 | 75 | 21 | 2 |
| A$_1$ | Hu, D., Wang, H. Y., & Zhu, Q. F. (2016). Design of an ultra-broadband and polarization-insensitive solar absorber using a circular-shaped ring resonator. *Journal of Nanophotonics*, *10*(2). doi: 10.1117/1.JNP.10.026021 | 68 | 28 | 2 |
| A$_1$ | Hashimoto, K., Sasaki, F., Hotta, S., & Yanagi, H. (2016). Amplified Emission and Field-Effect Transistor Characteristics of One-Dimensionally Structured 2, 5-Bis (4-biphenylyl) thiophene Crystals. *Journal of Nanoscience and Nanotechnology*, *16*(4), 3200-3205. doi: 10.1166/jnn.2016.12285 | 38 | 52 | 8 |
| A$_2$ | Alcala, M. D., Criado, J. M., Real, C., Grygar, T., Nejezchleba, M., Subrt, J., & Petrovsky, E. (2004). Synthesis of nanocrystalline magnetite by mechanical alloying of iron and hematite. *Journal of Materials Science, 39*(7), 2365-2370. | 37 | 59 | 2 |

| Approach | Publication | No relevant | Relevant | No answer |
|---|---|---|---|---|
| A₂ | Xiang, C., Yang, Y., & Penner, R. M. (2009). Cheating the diffraction limit: electrodeposited nanowires patterned by photolithography. *Chemical Communications*, (8), 859-873. doi: 10.1039/b815603d | 5 | 92 | 1 |
| A₂ | Sumesh, C. K., Patel, K. D., Pathak, V. M., & Srivastava, R. (2008). Twofold conduction mechanisms in molybdenum diselenide single crystals in the wide temperature range of 300K to 12K. *Chalcogenide Letters*, 5(8), 177-180. | 79 | 17 | 2 |
| A₂ | Borysenko, K. M., Mullen, J. T., Barry, E. A., Paul, S., Semenov, Y. G., Zavada, J. M., Nardelli, M.& Kim, K. W. (2010). First-principles analysis of electron-phonon interactions in graphene. *Physical Review B*, 81(12). doi 10.1103/PhysRevB.81.121412 | 32 | 64 | 2 |
| A₂ | Vikram, S. V., Phase, D. M., & Chandel, V. S. (2010). High-TC phase transition in K2Ti6O13 lead-free ceramic synthesised using solid-state reaction. *Journal of Materials Science: Materials in Electronics*, 21(9), 902-905. doi: 10.1007/s10854-009-0015-0 | 83 | 12 | 3 |
| A₂ | Ghodake, G., Seo, Y. D., Park, D., & Lee, D. S. (2010). Phytotoxicity of carbon nanotubes assessed by Brassica juncea and Phaseolus mungo. *Journal of Nanoelectronics and Optoelectronics*, 5(2), 157-160. doi: 10.1166/jno.2010.1084 | 30 | 66 | 2 |
| A₂ | Westenfelder, B., Meyer, J. C., Biskupek, J., Kurasch, S., Scholz, F., Krill III, C. E., & Kaiser, U. (2011). Transformations of carbon adsorbates on graphene substrates under extreme heat. *Nano letters*, 11(12), 5123-5127. doi 10.1021/nl203224z | 24 | 71 | 3 |
| A₂ | Sherman, E. Y., Ban, Y., Gulyaev, L. V., & Khomitsky, D. V. (2012). Spin Tunneling and Manipulation in Nanostructures. *Journal of Nanoscience and Nanotechnology*, 12(9), 7535-7539. doi 10.1166/jnn.2012.6554 | 10 | 87 | 1 |
| A₂ | Wu, W. Q., Xu, Y. F., Rao, H. S., Su, C. Y., & Kuang, D. B. (2013). A double layered TiO2 photoanode consisting of hierarchical flowers and nanoparticles for high-efficiency dye-sensitized solar cells. *Nanoscale*, 5(10), 4362-4369. doi: 10.1039/c3nr00508a | 12 | 82 | 4 |
| A₂ | Chiacchiarelli, L. M., Rallini, M., Monti, M., Puglia, D., Kenny, J. M., & Torre, L. (2013). The role of irreversible and reversible phenomena in the piezoresistive behavior of graphene epoxy nanocomposites applied to structural health monitoring. *Composites Science and Technology*, 80, 73-79. doi: 10.1016/j.compscitech.2013.03.009 | 34 | 58 | 6 |
| A₂ | Foos, E. E., Yoon, W., Lumb, M. P., Tischler, J. G., & Townsend, T. K. (2013). Inorganic photovoltaic devices fabricated using nanocrystal spray deposition. ACS applied materials & interfaces, 5(18), 8828-8832. doi: 10.1021/am402423f | 22 | 74 | 2 |
| A₂ | Wang, W. B., Li, X. H., Wen, L., Zhao, Y. F., Duan, H. H., Zhou, B. K., Shi, T. F. Zeng, X. S., Li, N., & Wang, Y. Q. (2014). Optical simulations of P3HT/Si nanowire array hybrid solar cells. *Nanoscale Research Letters*, 9(1). doi: 10.1186/1556-276X-9-238 | 18 | 74 | 6 |
| A₂ | Shao Z. Q., Chen, J. W. ,Li Y. Q., & Pan, X.Y. (2014). Thermodynamical properties of a three-dimensional free electron gas confined in a one-dimensional harmonical potential. *Acta Physica Sinica*, 63(24). doi: 10.7498/aps.63.240502 | 79 | 17 | 2 |
| A₂ | Soylemez, S., Udum, Y. A., Kesik, M., Hizliates, C. G., Ergun, Y., & Toppare, L. (2015). Electrochemical and optical properties of a conducting polymer and its use in a novel biosensor for the detection of cholesterol. *Sensors and Actuators B: Chemical*, 212, 425-433. doi 10.1016/j.snb.2015.02.045 | 69 | 27 | 2 |
| A₂ | Dyshin, A. A., Eliseeva, O. V., Bondarenko, G. V., Kolker, A. M., & Kiselev, M. G. (2016). Dispersion of single-walled carbon nanotubes in dimethylacetamide and a dimethylacetamide–cholic acid mixture. *Russian Journal of Physical Chemistry A*, 90(12), 2434-2439. doi 10.1016/j.snb.2015.02.045 | 18 | 78 | 2 |

| Approach | Publication | No relevant | Relevant | No answer |
|---|---|---|---|---|
| A₃ | Gotter, R., Ruocco, A., Butterfield, M. T., Iacobucci, S., Stefani, G., & Bartynski, R. A. (2003). Angle-resolved Auger-photoelectron coincidence spectroscopy (AR-APECS) of the Ge (100) surface. *Physical Review B*, *67*(3), 033303. doi: 10.1103/PhysRevB.67.033303 | 78 | 16 | 4 |
| A₃ | Nicolaides, A. (2003). Singlet hydrocarbon carbenes with high barriers toward isomerization: A computational investigation. *Journal of the American Chemical Society, 125(30)*, 9070-9073. doi: 10.1021/ja0299699 | 79 | 17 | 2 |
| A₃ | Hichria, A., & Jazirib, S. (2003). Two Carriers in Vertically Coupled Quantum Dots: Magnetic Field Effect. *Journal of Nanoscience and Nanotechnology*, *3*(3), 263-269. doi: 10.1166/jnn.2003.162 | 20 | 76 | 2 |
| A₃ | Matsubara, M., Kortus, J., Parlebas, J. C., & Massobrio, C. (2006). Dynamical identification of a threshold instability in Si-doped heterofullerenes. *Physical Review Letters, 96*(15), 155502. doi: 10.1103/PhysRevLett.96.155502 | 43 | 51 | 4 |
| A₃ | Chong, Y. M., Leung, K. M., Ma, K. L., Zhang, W. J., Bello, I., & Lee, S. T. (2006). Growing cubic boron nitride films at different temperatures. *Diamond and related materials, 15*(4), 1155-1160. | 71 | 25 | 2 |
| A₃ | Yang, P., Xu, R., Nanita, S. C., & Cooks, R. G. (2006). Thermal formation of homochiral serine clusters and implications for the origin of homochirality. *Journal of the American Chemical Society, 128*(51), 17074-17086. doi: 10.1021/ja064617d | 64 | 31 | 3 |
| A₃ | Stopa, M., & Marcus, C. M. (2008). Magnetic field control of exchange and noise immunity in double quantum dots. *Nano letters*, *8*(6), 1778-1782. doi: 10.1021/nl801282t | 20 | 77 | 1 |
| A₃ | Gonzalez, J. (2008). Kohn-Luttinger superconductivity in graphene. *Physical Review B*, *78*(20), 205431. doi: 10.1103/PhysRevB.78.205431 | 37 | 57 | 4 |
| A₃ | Kanamadi, C. M., Das, B. K., Kim, C. W., Cha, H. G., Ji, E. S., Kang, D. I., Jadhav, A. P., & Kang, Y. S. (2009). Template assisted growth of cobalt ferrite nanowires. *Journal of Nanoscience and Nanotechnology*, *9*(8), 4942-4947. doi: 10.1166/jnn.2009.1271 | 8 | 87 | 3 |
| A₃ | Kher, S., Dixit, A., Rawat, D. N., & Sodha, M. S. (2010). Experimental verification of light induced field emission. *Applied Physics Letters, 96*(4), 044101. doi: 10.1063/1.3293297 | 78 | 16 | 4 |
| A₃ | Bora, D. K., Braun, A., Erni, R., Fortunato, G., Graule, T., & Constable, E. C. (2011). Hydrothermal treatment of a hematite film leads to highly oriented faceted nanostructures with enhanced photocurrents. *Chemistry of Materials, 23*(8), 2051-2061. doi: 10.1021/cm102826n | 25 | 72 | 1 |
| A₃ | Fang, L., & Li, W. J. (2012). Hydrothermal synthesis of flake-like MnCO3 film under high gravity field and their thermal conversion to hierarchical Mn3O4. *Micro & Nano Letters, 7*(4), 353-356. doi: 10.1049/mnl.2011.0530 | 56 | 39 | 3 |
| A₃ | Howell, S. L., Padalkar, S., Yoon, K., Li, Q., Koleske, D. D., Wierer, J. J., Wang, G. T., & Lauhon, L. J. (2013). Spatial mapping of efficiency of GaN/InGaN nanowire array solar cells using scanning photocurrent microscopy. *Nano Letters, 13*(11), 5123-5128. | 9 | 88 | 1 |
| A₃ | Gu, J., Xiao, P., Huang, Y., Zhang, J., & Chen, T. (2015). Controlled functionalization of carbon nanotubes as superhydrophobic material for adjustable oil/water separation. *Journal of Materials Chemistry A, 3*(8), 4124-4128. doi: 10.1039/c4ta07173e | 15 | 80 | 3 |
| A₃ | Meng, D., Wang, G., San, X., Shen, Y., Zhao, G., Zhang, Y., & Meng, F. (2015). CTAB-assisted hydrothermal synthesis of WO 3 hierarchical porous structures and investigation of their sensing properties. *Journal of Nanomaterials, 16*(1), 364. doi: 10.1155/2015/393205 | 30 | 64 | 4 |

# Appendix 8. Precision by experts' field and subfield of knowledge

*Precision by experts' field of knowledge*

| Approach | Biology | Chemistry | Computer science | Engineering | Materials science | Physics |
|---|---|---|---|---|---|---|
| $A_0$ | 0,67 | 0,55 | 0,80 | 0,60 | 1,00 | 0,54 |
| $A_1$ | 0,60 | 0,45 | 0,73 | 0,66 | 0,62 | 0,45 |
| $A_2$ | 0,74 | 0,60 | 0,87 | 0,77 | 0,77 | 0,60 |
| $A_3$ | 0,72 | 0,49 | 0,87 | 0,72 | 0,79 | 0,57 |

*Precision by experts' nano subfield of knowledge*

| Approach | Nanobiomedicine | Nanobiosystems | Nanoelectronics | Nanoenergy | Nanoengineering | Nanofabrication | Nanomagnetism | Nanomaterials | Nanometrology | Nanooptoelectronics | Nanophotonics | Nanoprocesses | Nanotheoretical physics | Nanotoxicology & Sustainability |
|---|---|---|---|---|---|---|---|---|---|---|---|---|---|---|
| $A_0$ | 0.72 | 0.58 | 0.51 | 0.60 | 0.64 | 0.76 | 0.50 | 0.55 | 0.40 | 0.75 | 0.39 | 0.60 | 0.50 | 0.67 |
| $A_1$ | 0.51 | 0.52 | 0.47 | 0.43 | 0.43 | 0.60 | 0.35 | 0.47 | 0.53 | 0.58 | 0.47 | 0.47 | 0.44 | 0.58 |
| $A_2$ | 0.71 | 0.72 | 0.59 | 0.57 | 0.67 | 0.67 | 0.55 | 0.58 | 0.73 | 0.57 | 0.60 | 0.64 | 0.57 | 0.73 |
| $A_3$ | 0.64 | 0.55 | 0.50 | 0.60 | 0.49 | 0.70 | 0.56 | 0.53 | 0.73 | 0.71 | 0.56 | 0.58 | 0.46 | 0.67 |